\documentclass[11pt,]{article}
\usepackage{lmodern}
\usepackage{amssymb,amsmath}
\usepackage{ifxetex,ifluatex}
\usepackage{verbatim}
\usepackage{lscape}
\usepackage{fixltx2e} % provides \textsubscript
\ifnum 0\ifxetex 1\fi\ifluatex 1\fi=0 % if pdftex
  \usepackage[T1]{fontenc}
  \usepackage[utf8]{inputenc}
\else % if luatex or xelatex
  \ifxetex
    \usepackage{mathspec}
  \else
    \usepackage{fontspec}
  \fi
  \defaultfontfeatures{Ligatures=TeX,Scale=MatchLowercase}
\fi
\IfFileExists{upquote.sty}{\usepackage{upquote}}{}
\IfFileExists{microtype.sty}{%
\usepackage{microtype}
\UseMicrotypeSet[protrusion]{basicmath} % disable protrusion for tt fonts
}{}
\usepackage[margin=1in]{geometry}
\usepackage{hyperref}
\hypersetup{unicode=true,
            pdftitle={M-models},
            pdfauthor={Gonzalo},
            pdfborder={0 0 0},
            breaklinks=true}
\usepackage{graphicx,grffile}
\makeatletter
\def\maxwidth{\ifdim\Gin@nat@width>\linewidth\linewidth\else\Gin@nat@width\fi}
\def\maxheight{\ifdim\Gin@nat@height>\textheight\textheight\else\Gin@nat@height\fi}
\makeatother
\setkeys{Gin}{width=\maxwidth,height=\maxheight,keepaspectratio}
\IfFileExists{parskip.sty}{%
\usepackage{parskip}
}{% else
\setlength{\parindent}{0pt}
\setlength{\parskip}{6pt plus 2pt minus 1pt}
}
\ifx\paragraph\undefined\else
\let\oldparagraph\paragraph
\renewcommand{\paragraph}[1]{\oldparagraph{#1}\mbox{}}
\fi
\ifx\subparagraph\undefined\else
\let\oldsubparagraph\subparagraph
\renewcommand{\subparagraph}[1]{\oldsubparagraph{#1}\mbox{}}
\fi

\let\rmarkdownfootnote\footnote%
\def\footnote{\protect\rmarkdownfootnote}

\usepackage{titling}

\usepackage{float}
\usepackage[svgnames]{xcolor}
\usepackage{booktabs, multirow}
\usepackage{rotating}
\usepackage{booktabs}
\usepackage[font=small,labelfont=bf]{caption}

\usepackage{makecell} %% dividir celdas
\usepackage{mathdots}

\usepackage[round]{natbib}
\def\L{\boldsymbol{L}}

\def\X{\bf X}
\def\Q{\boldsymbol{Q}}
\def\C{\bf W}
\def\M{\boldsymbol{M}}

\def\C{\bf C}
\def\A{\bf A}
\def\D{\boldsymbol{D}}
\def\V{\bf V}

\def\I{\boldsymbol{I}}

\def\U{\boldsymbol{U}}
\def\W{\boldsymbol{W}}

\def\Z{\bf Z}

\def\X{\boldsymbol{X}}
\def\Z{\boldsymbol{Z}}
\def\B{\bf B}

\def\1{\bf 1}
\def\0{\bf 0}
\def\rr{\boldsymbol{r}}

\def\SSigma{\mbox{$\boldsymbol{\Sigma}$}}

\def\ttheta{\mbox{\boldmath{$\theta$}}}

\newfont{\Sc}{eusm10}

\def\aalpha{\mbox{\boldmath{$\alpha$}}}
\def\bbeta{\mbox{$\boldsymbol{\beta}$}}
\def\eepsi{\mbox{\boldmath{$\epsilon$}}}

\def\OOmega{\mbox{\boldmath{$\Omega$}}}

\def\PPhi{\mbox{\boldmath{$\Phi$}}}

\title{A simplified spatial+ approach to mitigate spatial confounding in multivariate spatial areal models}
\date{}
\author{Urdangarin, A.$^{1,2}$, Goicoa, T.$^{1,2}$, Kneib, T.$^3$, Ugarte, M.D.$^{1,2*}$\\
\small {\textit{$^1$ Department of Statistics, Computer Science, and Mathematics, Public University of Navarre, Spain.}} \\
\small {\textit{$^2$ INAMAT$^2$, Public University of Navarre, Spain.}} \\
\small {\textit{$^3$ Georg-August-University of Göttingen, Chairs of Statistics and Econometrics, Germany}} \\
\small {$*$\textbf{Corresponding author:} Mar\'ia Dolores Ugarte, Department of Statistics, Computer Science, and}\\
\small { Mathematics, Public University of Navarre, Campus de Arrosadia, 31006 Pamplona, Spain.} \\
\small {\textbf{E-mail}: lola@unavarra.es }}

\usepackage{listings}
\begin{document}
\maketitle

%%---------------------------------------------------------------------------------------
%% ABSTRACT
%%---------------------------------------------------------------------------------------
\begin{abstract}
Spatial areal models encounter the well-known and challenging problem of spatial confounding. This issue makes it arduous to distinguish between the impacts of observed covariates and spatial random effects. Despite previous research and various proposed methods to tackle this problem, finding a definitive solution remains elusive. In this paper, we propose a simplified version of the spatial+ approach that involves dividing the covariate into two components. One component captures large-scale spatial dependence, while the other accounts for short-scale dependence. This approach eliminates the need to separately fit spatial models for the covariates. We apply this method to analyse two forms of crimes against women, namely rapes and dowry deaths, in Uttar Pradesh, India, exploring their relationship with socio-demographic covariates. To evaluate the performance of the new approach, we conduct extensive simulation studies under different spatial confounding scenarios. The results demonstrate that the proposed method provides reliable estimates of fixed effects and posterior correlations between different responses.
\end{abstract}


%%%%%%%%%%%%%%%%%%
%% INTRODUCTION %%
%%%%%%%%%%%%%%%%%%

\section{Introduction}
Univariate spatial models for areal count data have been a prevailing approach
in smoothing standardized incidence or mortality ratios of chronic diseases.
Though these techniques have been mainly applied to study incidence and
mortality of different types of cancer, other applications exist. A very
interesting one is the study of crimes against women in
India~\citep[see][]{Vicente2020a} where gender based violence is an issue.
Univariate modelling is crucial for visualizing the spatial patterns of crimes
against women. However, adopting a multivariate approach can enhance the
accuracy of estimates and reveals latent correlations between the spatial
patterns of crimes~\citep{Vicente2021}. This is essential for gaining a better
understanding of this complex and multifaceted problem, and ultimately, for
effective prevention.

Despite the substantial growth in research on multivariate spatial models for
areal count data, their practical application is still limited due to the
computational complexity involved in their implementation. While various
approaches exist for constructing multivariate models for disease
mapping~\citep[see for example][]{MacNab2018Test}, this study follows the
corregionalization framework~\citep{Jin2007}. In particular, the research
by~\citet{Martinez-Beneito2013} presents a comprehensive corregionalization
approach that encompasses most of the multivariate methodologies suggested in
the existing literature. However, this approach can be computationally
prohibitive and an alternative reformulation known as
M-models~\citep{Botella-Rocamora2015} is considered.

Incorporating potential risk factors (covariates) into a multivariate spatial
model is important to evaluate the potential relationship between the
covariates and the individual responses. This can contribute to gain knowledge
about crimes against women, a multifaceted  problem affected by social,
religious, or economic characteristics with intricate interactions difficult to
disentangle. However, when both, the covariate and spatial random effects enter
in the model, an important challenge appears. Namely,  how to separate the
fixed effects from the spatial random effects. This is known as \lq\lq spatial
confounding''. When spatial random effects enter in the model, a change in the
fixed effect estimates is observed compared to a simple linear or generalized
linear model (GLM) that does not consider spatial correlation. Although spatial
confounding has usually been considered as a collinearity problem between fixed
effects and spatial random effects~\citep[see for instance][]{Reich2006,Hodges2010,Hughes2013,Hanks2015,Page2017,Adin2021}, there
is not a unique general definition neither a unique solution. In
fact,~\citet{Gilbert2021} identify four different but related phenomena
commonly referred as spatial confounding. In what follows, we consider that the
effect of the unobserved covariates is approximated by spatial random effects
in the model~\citep{Congdon2013,Marques2022}, and we seek procedures that can
simultaneously avoid bias in the estimates of the fixed effects, reduce their
variance, and appropriately smooth the relative risks.

Various procedures have been proposed in the literature to address spatial
confounding in the univariate framework, with restricted spatial regression
(RSR)~\citep{Reich2006} being the most extended approach. RSR effectively deals
with the collinearity between the covariate and spatial random effects by
constraining the spatial random effects to lie in the orthogonal complement of
the space spanned by the covariates. As a result, RSR provides fixed effects
estimates that are equivalent to those obtained using a simple GLM.
Additionally, the variance of the RSR fixed effects estimates falls somewhere
between the variance obtained with the simple GLM and the variance obtained
with the classical disease mapping spatial model~\citep{Reich2006,Hodges2010}.
Consequently, RSR might help prevent variance inflation of the fixed effects
estimates, a concern that is widely recognized in spatial modelling.  However,
recent evidence contradicts some prior beliefs about RSR.~\citet{Khan2020}
demonstrate that for normal responses, the variance of the fixed effects
estimated with RSR is either equal to or lower than the variance obtained from
the model without spatial random effects, leading to overly liberal inference.
These authors also highlight potential issues of RSR for count data.
Additionally,~\citet{Gilbert2021} argue that RSR implicitly assumes the absence
of unobserved covariates that overlap with the observed ones, as it forces the
random effects to be orthogonal to the fixed effects. Furthermore, they support
that the collinearity between the covariate and spatial random effects may not
be a concern, but something expected if we assume the existence of unobserved
covariates.

The spatial+ approach~\citep{Dupont2022} appears to be a promising method among
the various alternatives for addressing spatial confounding. This is a two-step
procedure that consists in removing spatial dependence from the covariate in a
first step by fitting a spatial model to the covariate itself. Then, in the
second step, a classical spatial regression model for the outcome is fitted
replacing the covariate by the residuals obtained in the first step. In the
univariate case,~\citet{Urdangarin2022} conducted simulations to compare the
fixed effects estimates of various approaches, including a simple GLM, the
classical spatial model, the spatial+ approach, as well as other proposals such
as RSR and transformed Gaussian Markov random fields
(TGMRF)~\citep{Prates2015}. Overall, the spatial+ method demonstrated superior
performance in terms of fixed effect estimates. Recent procedures to deal with
spatial confounding include spectral methods~\citep{Guan2022}, and a joint
Gaussian Markov random field model for the covariate and the
response~\citep[see][]{Marques2022}.

This work has two main objectives: firstly, we aim to propose a modified spatial+ approach  that alleviates spatial confounding eliminating the need for separately fitting spatial models for the covariates. Secondly, we seek to estimate the latent correlations between several responses. To deal with these two goals we use M-models incorporating the modified spatial+ method that offer a tailored approach by accommodating the inclusion of distinct covariates for each individual response. Moreover, these models provide the necessary flexibility to effectively address spatial confounding for each response.
To illustrate the procedure, we analyse crimes against women, rapes and dowry
deaths, in the districts of Uttar Pradesh in 2011 and we assess their
relationship with some sociodemographic covariates. To examine how well our
proposal recovers the fixed effects and the correlations between crimes, we
have simulated several scenarios under spatial confounding. The data generating
model includes one observed covariate that has a distinct regression
coefficient for each individual response and additional variability. Model
fitting and inference is carried out from a full Bayes approach, using
integrated nested Laplace approximations~\citep{Rue2009}.

The rest of the article is organized as follows. In Section 2, we introduce a simplified spatial+ approach for fitting spatial models (univariate or multivariate) without the need to fit a separate spatial model to the covariate. Section 3 reviews the multivariate spatial models making emphasis on the M-models. Section 4 discusses some model implementation details and identifiability constraints. In Section 5 we use the procedures to analyse two crimes against women, rape and dowry death, in Uttar Pradesh, India, in the year 2011. Section 6 presents an exhaustive simulation study. Finally, the paper ends with a discussion.

%%%%%%%%%%%%%%%%%%%%%%%%%
%% SIMPLIFIED SPATIAL+ %%
%%%%%%%%%%%%%%%%%%%%%%%%%

\section{Simplified spatial+ approach}

The spatial+ method~\citep{Dupont2022} is a two-step procedure designed to
reduce bias in the fixed effect estimates of univariate spatial models by
eliminating the spatial dependence of the covariates. The first step consists
in fitting a spatial model to the covariate to remove the spatial dependence.
Then, in the second step, the spatial model is fitted replacing the covariate
by the residuals obtained in the first step. Here, we modify the spatial+
procedure to remove the spatial dependence of the covariate in a simpler way,
avoiding fitting a spatial model to the covariate. We also extend the spatial+
approach to the multivariate framework.

Let us consider the following univariate spatial disease mapping model where, conditional on the relative risk $r_i$, the number of counts in a given area $i=1,\ldots,n$, is assumed to follow a Poisson distribution
\begin{gather*}
Y_{i}\arrowvert r_{i} \sim Poisson(\mu_{i}=e_{i}r_{i}), \\
\log \mu_{i}=\log e_{i} + \log r_{i}.
\end{gather*}
Here, $e_{i}$ is the number of expected counts and the log risk is modelled as
\[
\log r_{i}=\alpha+ \beta x_{i}+\theta_{i},\]
where $x_{i}$ is value of the covariate ${\X}$ in the $i$th area. In matrix form
\begin{equation}
\label{eq:univariate_1}
\log \rr=\mathbf{1}_{n}\alpha + \X\beta + \ttheta,
\end{equation}
where $\rr=(r_1,\ldots,r_n)'$, ${\mathbf 1}_{n}$ is a column vector of ones of length
$n$, ${\X}=(x_1,\ldots,x_n)'$, and ${\ttheta}=(\theta_1,\ldots,\theta_n)'$ is the vector of spatial random
effects with precision matrix ${\OOmega}$, which may be different depending on
the spatial prior used for $\ttheta$. In this paper we consider one of the
following priors: intrinsic conditional autoregressive (ICAR)
prior~\citep{Besag1974}, proper conditional autorregresive (PCAR)
prior~\citep[see for example][chapter 4]{Banerjee2015} or BYM2
prior~\citep{Riebler2016}. In more detail, $\ttheta\sim N(\mathbf{0}, \OOmega)$, where the precision
matrix $\OOmega$ depends on the spatial prior chosen:
\begin{itemize}
\item \textbf{ICAR}: $\OOmega=\tau(\D-\W)=\tau \Q$ where $\tau$ is the precision parameter, $\D$ is a diagonal matrix with the number of neighbours of each area in the main diagonal, and $\boldsymbol{W}=(w_{il})$ is a binary adjacency matrix where each entry $w_{il}$ takes value $1$ if areas $i$ and $l$ are neighbours and 0 otherwise, and $w_{ii}=0$.  Note that ${\Q}$ is the usual neighbourhood matrix where the $i$th diagonal element is equal to the number of neighbours of the area $i$, and the off diagonal elements ${\Q}_{il}=-1$ if the  areas $i$ and $l$ share a common border and 0 otherwise.
\item \textbf{PCAR}: $\OOmega=\tau(\D-
\rho \W)$, which provides a valid
 distribution if and only if $1/d_{min}<\rho<1/d_{max}$ being $d_{min}$ and
  $d_{max}$ the minimum and maximum eigenvalues of
  $\D^{-1/2}\W\D^{-1/2}$~\citep{Jin2007}. When $\rho=1$, the PCAR becomes
  the ICAR prior.
\item \textbf{BYM2}: $\OOmega=\tau [(1-\lambda)\I_{n}+\lambda\Q_{*}^{-}]^{-1}$ where $\I_{n}$ is an $n \times n$ identity
 matrix and $\Q_{*}^{-}$ is the Moore--Penrose generalized inverse of the
  scale-transformed precision matrix $\Q$~\citep[for more details about this scale transformation see][]{Riebler2016}. In
  this prior, $0 \leq \lambda \leq 1$ represents the proportion of the
  marginal variance explained by the structured effect.
\end{itemize}

The spatial+ approach assumes that the covariate $\X$ is modelled as
\begin{equation}
\label{eq:spatial+_original}
\X= \boldsymbol{f}(\boldsymbol{s}_1,\boldsymbol{s}_2) + \eepsi
\end{equation}
where $(\boldsymbol{s}_1,\boldsymbol{s}_2)$ are the coordinates (longitude and latitude) of the centroid of the small areas,
${\boldsymbol f}({\boldsymbol s}_1,{\boldsymbol s}_2)=(f(s_{11}, s_{12}), f(s_{21}, s_{22}), \dots, f(s_{n1}, s_{n2}))^{'}$ is a smooth function to be estimated using thin plate splines, and $\eepsi \sim N(\mathbf{0}, \sigma_{\X}^2 \boldsymbol{I}_n)$, where $\sigma_{\X}$ is the standard deviation of the independent and identically distributed errors. The residuals of this model are defined as $\hat{\eepsi}=\X-\hat{\boldsymbol{f}}(\boldsymbol{s}_1,\boldsymbol{s}_2)$ and the spatial model \eqref{eq:univariate_1} is fitted replacing the covariate of interest $\X$ by the residuals $\hat{\eepsi}$.

\citet{Urdangarin2022} suggest an alternative to model
\eqref{eq:spatial+_original} to remove the spatial dependence in the
covariate. It consists in fitting a linear model where the response is now the
covariate $\X$, and some of the eigenvectors of the spatial precision
matrix $\boldsymbol{Q}$ act as regressors. Each eigenvector captures spatial
dependence at different scales, the eigenvectors corresponding to the lowest
non-null eigenvalues being the smoothest ones. Thus, the number of eigenvectors
included as covariates in the model defines the quantity of spatial dependence
removed from $\X$. The simulation study performed by these authors
indicates that with the new approach the fixed effects are recovered better
than using thin plate splines. However this approach also requires fitting a
linear model to the covariate.

In this paper, we introduce a simplified spatial+ approach without the need of fitting a spatial model to the covariate. This simplified spatial+ approach is a simpler and more straightforward procedure. Consider the spectral decomposition of the spatial precision matrix, $\boldsymbol{Q}=\boldsymbol{U}\boldsymbol{\Delta} \boldsymbol{U}^{\prime}$, where $\boldsymbol{U}=(\boldsymbol{U}_{1}, \dots, \boldsymbol{U}_{n})$ is an orthogonal matrix whose columns are the eigenvectors of $\boldsymbol{Q}$ and $\boldsymbol{\Delta}$ is a $n \times n$ diagonal matrix with the eigenvalues of $\boldsymbol{Q}$ in the main diagonal. Note that for connected graphs $\boldsymbol{U}_{n}=\mathbf{1}_{n}$ up to a normalizing constant and $\boldsymbol{\Delta} _{nn}=0$. Since the eigenvectors of $\boldsymbol{Q}$ form a basis of $\mathbb{R}^n$ and contain spatial information at different scales, we now express the covariate $\boldsymbol{X}$ as a linear combination of the eigenvectors $\boldsymbol{U}_{i}$, $i=1,\ldots,n$, that is,
\[
\boldsymbol{X}=a_{1}\boldsymbol{U}_{1} +\cdots+ a_{n}\boldsymbol{U}_{n}.
\]
Assuming that the collinearity between the fixed and random effects primarily
arises from the eigenvectors associated with the lowest non-null eigenvalues,
that is, presuming that the eigenvectors linked to the high eigenvalues remain
unconfounded with the covariate~\citep[equivalent to the unconfoundedness at high frequencies assumption for identifiability in the work by][]{Guan2022}, we
split the covariate into two parts
\begin{equation}
\label{eq:zeta_zetaprima}
\boldsymbol{X}=\boldsymbol{Z} + \boldsymbol{Z}^{*},
\end{equation}
where $\boldsymbol{Z}^{*}=a_{n-k}\U_{n-k}+\cdots+a_{n}\boldsymbol{U}_{n}$ comprises large-scale eigenvectors associated with the lowest eigenvalues, $\boldsymbol{Z}=a_{1}\boldsymbol{U}_{1}+\cdots + a_{n-(k+1)}\boldsymbol{U}_{n-(k+1)}$ contains the rest of eigenvectors, and $k$ is the number of large-scale eigenvectors  assigned to $\boldsymbol{Z}^{*}$. Finally, the spatial model \eqref{eq:univariate_1} is fitted replacing the covariate $\boldsymbol{X}$  by its spatially decorrelated part $\boldsymbol{Z}$ as
\begin{equation}
\label{eq:u-spatplus2}
\log \rr=\mathbf{1}_{n}\alpha + \Z\beta + \ttheta.
\end{equation}

We have verified that the spatial+ proposed in~\citet{Urdangarin2022} and the
simplified spatial+ proposed here yield to almost identical results in
univariate models.

In the next section, we introduce multivariate disease mapping models and extend the modified spatial+ approach to this framework.

%%%%%%%%%%%%%%%%%%%%%%%%%
%% MULTIVARIATE MODELS %%
%%%%%%%%%%%%%%%%%%%%%%%%%

\section{Multivariate ecological spatial areal models}

Multivariate spatial areal models handle two types of dependencies: within-response dependence, often characterized by spatial dependence, and between-response correlations, which typically lack of a specific structure and pose a challenge. Multivariate models, in general, have the potential to provide more reliable risk estimates compared to traditional univariate models. In this paper, we employ multivariate ecological regression models to examine potential linear associations between the responses and a covariate of interest.

Let now $Y_{ij}$ and $e_{ij}$ denote, respectively, the number of observed and expected cases in the $i$th small area $(i=1, \dots, n)$ and $j$th crime $(j=1, \dots, J)$, and assume that $Y_{ij}$ follows a Poisson distribution conditioned on the relative risk $r_{ij}$, that is
\begin{gather*}
Y_{ij}\arrowvert r_{ij} \sim Poisson(\mu_{ij}=e_{ij}r_{ij}), \\
\log \mu_{ij}=\log e_{ij} + \log r_{ij}.
\end{gather*}
The log-risk is modelled as
\begin{equation}
 \label{eq:multivariate_1}
\log r_{ij}=\alpha_{j}+ \beta_{j}x_{i}+\theta_{ij}
\end{equation}
where $\alpha_{j}$ is the intercept of $j$th crime, $\beta_{j}$ is a crime-specific regression coefficient related to the covariate $\boldsymbol{X}=(x_1, \dots, x_{n})^{\prime}$ and $\theta_{ij}$ is the spatial effect of area $i$ and crime $j$. Model \eqref{eq:multivariate_1} can be expressed in matrix form as
\begin{equation}
\label{eq:multivariate_2}
\log \rr=(\I_{J}\otimes \boldsymbol{1}_{n})\aalpha + (\I_{J}\otimes \X)\bbeta + \ttheta
\end{equation}
where $\rr=(\rr_{1}^{\prime}, \dots, \rr_{J}^{\prime})^{\prime}$ is the vector of relative risks with $\rr_{j}=(r_{1j}, \dots, r_{nj})^{\prime}$ for $j=1, \dots, J$, $\I_{J}$ is a $J \times J$ identity matrix, $\mathbf{1}_{n}$ is a vector of ones of length $n$ and $\otimes$ refers to the Kronecker product of two matrices. The vectors of crime-specific intercepts and regression coefficients are denoted by $\aalpha=(\alpha_{1}, \dots, \alpha_{J})^{\prime}$ and $\bbeta=(\beta_{1}, \dots, \beta_{J})^{'}$ respectively. Finally, $\ttheta=(\ttheta_{1}^{\prime}, \dots, \ttheta_{J}^{\prime})^{\prime}$ where $\ttheta_{j}=(\theta_{1j}, \dots, \theta_{nj})^{\prime}$ is the vector of spatial random effects of the $j$th crime. The within and between-crime dependence is introduced through the precision/covariance matrix of the spatial random effect. In particular the following prior distribution with Gaussian kernel $p(\ttheta)\propto exp\left(-\dfrac{{\tau}}{2}{\ttheta}'\boldsymbol{\Omega}_{\boldsymbol{\theta}}{\ttheta}\right)$ is  assumed for ${\ttheta}$ with precision matrix
\begin{equation}
\label{eq:precision}
\boldsymbol{\Omega}_{\boldsymbol{\theta}}=(\boldsymbol{M}^{-1} \otimes \boldsymbol{I}_{n})Blockdiag(\boldsymbol{\Omega}_{1}, \dots, \boldsymbol{\Omega}_{J})(\boldsymbol{M}^{-1} \otimes \boldsymbol{I}_{n})^{\prime},
\end{equation}
where $\M$ is any square root of the $J \times J$ symmetric and positive definite between-crime covariance matrix $\SSigma_{b}$, that is $\SSigma_{b}=\M^{\prime}\M$, and $\boldsymbol{\Omega}_{j}$  is the spatial precision matrix for the $j$ crime.  The elements of the between-crime covariance matrix $\SSigma_{b}$ are all unknown and have to be estimated. Details about how the between-crime dependence is introduced in Model \eqref{eq:multivariate_2} through the matrix ${\M}$ are provided below. Here $\boldsymbol{\Omega}_{j}$ can be the precision matrix of the ICAR, PCAR or BYM2 priors.  For the ICAR prior, the precision matrix is the same for all crimes and $\boldsymbol{\Omega}_{\boldsymbol{\theta}}$ has a separable structure, $\boldsymbol{\Omega}_{\boldsymbol{\theta}}= \SSigma_{b}^{-1} \otimes (\tau[\D-\W])$. Likewise, for the PCAR and BYM2 priors,  if $\rho_{j}=\rho$ and $\lambda_{j}=\lambda$ for $j=1,\dots, J$, we have the separable precision matrices $\boldsymbol{\Omega}_{\boldsymbol{\theta}}= \SSigma_{b}^{-1} \otimes (\tau[\D-\rho \W)])$ and $\boldsymbol{\Omega}_{\boldsymbol{\theta}}= \SSigma_{b}^{-1} \otimes (\tau[(1-\lambda)\I_{n}+\lambda\Q_{*}^{-}]^{-1})$, respectively.
%\QUERY[3]

In the following, we review how to introduce spatial dependencies within crimes
as well as the correlation between spatial patterns of different crimes through
the M-models~\citep{Botella-Rocamora2015}. In this paper, the multivariate
models include a single covariate, but the proposed methodology can be extended
to the case of multiple covariates.

To better understand how the M-models deal with the within and between-crime
dependence, we rearrange the spatial effects into the matrix $\boldsymbol{\Theta}=(\boldsymbol{\theta}_{1}, \dots, \boldsymbol{\theta}_{J})=\{\theta_{ij}: i=1, \dots, n;\; j=1, \dots, J\}$.
Following~\citet{Botella-Rocamora2015}, $\boldsymbol{\Theta}$ can be expressed as
\begin{equation}
\label{eq:M-model}
\boldsymbol{\Theta}=\PPhi \boldsymbol{M}
\end{equation}
where the columns of the matrix $\PPhi=(\PPhi_{1}, \dots, \PPhi_{J})$ are independent and follow a
spatially correlated distribution to deal with spatial dependence within
crimes, namely $\PPhi_{j} \sim N(\mathbf{0}, \OOmega_{j})$ for each column $j=1, \dots, J$. Then, the spatial
covariance matrix of column $j$ is  $\SSigma_{j}=\boldsymbol{\Omega}_{j}^{-1}$ and it is usually
known as the within $j$th crime covariance matrix.  The $J \times J$
matrix $\boldsymbol{M}$ induces dependence between the columns of $\PPhi$,
that is, it induces correlation between the spatial patterns of different
crimes. A typical election for ${\M}$ is to consider the lower triangular
matrix of the Cholesky decomposition of the between-crime covariance matrix
$\SSigma_{b}$, however the effects of this election depend on the ordering of
the crimes in the vector $\ttheta$ as pointed out
by~\citet{Martinez-Beneito2013} (see Appendix \ref{appenA} for details). To avoid this
undesired situation, $\boldsymbol{M}$ is any nonsingular matrix without any other
restriction that satisfies the condition
$\SSigma_{b}=\M^{\prime}\M$~\citep{Botella-Rocamora2015}. Using that
\[
\text{vec}({\A}{\B}{\C})=({\C}'\otimes {\A})\text{vec}({\B}),
\]
\citep[see for example][p. 345]{Harville2008} where the $\text{vec}$ operator stacks the columns of a matrix one under the other and ${\A}$, ${\B}$, and ${\C}$ are conformable matrices for multiplication, it is obtained that
\[
\text{vec}(\boldsymbol{\Theta})={\ttheta}=(\boldsymbol{M}'\otimes {\I}_n)\text{vec}(\PPhi).
\]

Then, the covariance matrix of ${\ttheta}$ is
\begin{eqnarray}
\label{eq_covglobal}
\text{Cov}({\ttheta})={\SSigma}_{\ttheta}&=&(\boldsymbol{M} \otimes \boldsymbol{I}_{n})'\text{Cov}(\text{vec}(\PPhi))(\boldsymbol{M} \otimes \boldsymbol{I}_{n})\nonumber\\
&=& (\boldsymbol{M} \otimes \boldsymbol{I}_{n})'Blockdiag({\SSigma}_{1}, \dots, {\SSigma}_{J})(\boldsymbol{M} \otimes \boldsymbol{I}_{n}).
\end{eqnarray}
 Clearly, the inverse of this matrix is the precision matrix $\boldsymbol{\Omega}_{\boldsymbol{\theta}}$ in Equation~\eqref{eq:precision}.

To make the spatial random effects identifiable, appropriate sum-to-zero
constraints must be contemplated. Here, we will consider the constraints
proposed by~\citet{Goicoa2018} for each crime. In addition, we fix the
precision parameters of the spatial precision matrices at 1 to make the
between-crime covariance matrix $\SSigma_{b}$
identifiable~\citep{Martinez-Beneito2013}.
Note that in the case of separable covariance structures, we set the parameter $\tau=1$ as any change in scale in the within-crime covariance matrix can be compensated for by an appropriate change in the scale of the between-crime covariance matrix $\SSigma_{b}$. Regarding non separable covariance structures, the between-crime covariance matrix $\SSigma_{b}$ will be identifiable if the precision parameters $\tau_j$ are fixed at 1. More precisely, setting $\tau_1=\cdots=\tau_J$ (not necessarily equal to 1) only identifies $\SSigma_{b}$ up to scaling. In any case, fixing the precision parameters at 1 is not restrictive as the elements of the matrix ${\M}$ control the degree of smoothing within each crime (see Appendix \ref{appenB} for details).

 By estimating the entries of the matrix $\M$, the covariance structure between the spatial patterns of different crimes can be estimated as
$\SSigma_{b}=\boldsymbol{M}^{\prime}\boldsymbol{M}$. Note that the entries of $\boldsymbol{M}$ can be interpreted as
coefficients in the regression of the log-relative risks on the columns of
$\boldsymbol{\Phi}$. Hence, they can be treated as fixed effects and assigning a
zero-centred normal prior with large fixed variance is a reasonable choice.
This is equivalent to assume a Wishart prior to $\SSigma_{b}$, i.e.~$\SSigma_{b}=\boldsymbol{M}^{\prime}\boldsymbol{M} \sim Wishart(J, \sigma^2 \boldsymbol{I}_{J})$~\citep[see][for more details ]{Botella-Rocamora2015}.

Finally, to deal with spatial confounding, the multivariate M-model \eqref{eq:multivariate_2} is fitted replacing the covariate $\boldsymbol{X}$  by its spatially decorrelated part $\boldsymbol{Z}$ as
\begin{equation}\label{eq:M-spatplus2}
\log \rr=(\I_{J}\otimes \boldsymbol{1}_{n})\aalpha + (\I_{J}\otimes \Z)\bbeta + \ttheta.
\end{equation}
 It is worth mentioning that Model \eqref{eq:M-spatplus2} allows removing a different number of eigenvectors for each crime. That is
\begin{eqnarray*}
\log \rr=(\I_{J}\otimes \boldsymbol{1}_{n})\aalpha + \left(
\begin{array}{ccc}
     {\Z}_1 &\cdots &{\boldsymbol 0} \\
     \vdots & \ddots & \vdots \\
     {\boldsymbol 0} & \cdots &{\Z}_J \\
\end{array}
\right)
\bbeta + \ttheta,
\end{eqnarray*}
where ${\Z}_j$ is the part of the covariate that we retain for crime $j=1\ldots,J$.

\section{Model implementation}

In this paper, models are fitted using the Integrated nested Laplace
approximation (INLA) approach~\citep{Rue2009}. INLA is designed for approximate
Bayesian inference avoiding the convergence issues of MCMC techniques and
saving computing time. Hence it has become very popular for performing Bayesian
inference with a wide range of hierarchical models. The \texttt{R} package
\texttt{R-INLA}~\citep{Lindgren2015} has many models directly available and
allows to implement other models using \texttt{rgeneric} or \texttt{cgeneric}
constructions. The M-models are not directly available in \texttt{R-INLA}, so
here we will use the \texttt{rgeneric} construction as in~\citet{Vicente2020b}.

The matrix ${\M }$ involves $J \times J$ parameters when only
$J\times(J+1)/2$ parameters are needed to determine the convariance matrix
$\SSigma_{b}=\M^{\prime}\M $. Thus, to avoid overparameterization, when we consider the Wishart
prior on the covariance matrix, $\SSigma_{b}=\M^{\prime}\M \sim Wishart(J, \sigma^2 \I_{J})$, we use the Bartlett decomposition
of Wishart-distributed matrices~\citep[see for example][]{Pena2022}. More
precisely, if $\SSigma_{b}\sim Wishart(\nu,  {\V})$, the Bartlett decomposition of $\SSigma_{b}$ is
 $\SSigma_{b}={\L}\A\A^{\prime}{\L}'$. Here ${\L}$ is the Cholesky factor of ${\V}$ and
\begin{equation}
\A=
\begin{bmatrix}
c_1 & 0 &0 & \cdots & 0\\
n_{21} & c_2 & 0 & \cdots & 0 \\
n_{31} & n_{32} & c_3 & \cdots & 0 \\
\vdots & \vdots & \vdots & \ddots & \vdots \\
n_{J1} & n_{J2} & n_{J3} & \cdots & c_{J}
\end{bmatrix}
,
\end{equation}
where the diagonal and non-diagonal elements are independently distributed as $c_{j} \sim \chi_{\nu-j+1}$ and $n_{jl}\sim N(0,1)$ for $j,l=1, \ldots, J$ with $j>l$. Using the  Bartlett decomposition we only use $J\times(J+1)/2$  parameters instead of $J \times J$ parameters. In our case,  ${\V}=\sigma^{2}{\I}_{J}$ and then ${\L}=\sigma{\I}_{J}$. Finally, to avoid dependence on the ordering of the crimes of $\ttheta$,  instead of estimating $\M=\sigma \A^{\prime}$ (which is upper triangular), we first compute $\SSigma_{b}=\sigma^{2} \A\A^{\prime}$ and then we estimate $\boldsymbol{M}=(\boldsymbol{H} diag(\sqrt{\kappa_1}, \dots, \sqrt{\kappa_{J}}))^{\prime}$, where $\kappa_1, \dots, \kappa_{J}$ are the eigenvalues of the between-crime covariance matrix $\SSigma_{b}$ and the columns of $\boldsymbol{H}$ are the corresponding eigenvectors.
Using this ${\M}$ instead of the Cholesky square root in the global
covariance matrix \eqref{eq_covglobal}, we avoid dependence on the ordering of
the crimes in $\ttheta$. Details about the implementation of the Bartlett
decomposition can be found in~\citet{vicente2022}. Finally, we note that the
parameter $\sigma^2$ can be absorbed by the elements of ${\A}$ and we
fix it at 1 for identifiability issues. We run the analysis with different
values of $\sigma^2$ and the results did not change.

Finally, we have considered a uniform distribution Unif(0,1) for the
hyperparameters $\rho_{j}$ and $\lambda_{j}$ of the PCAR and BYM2 spatial
priors. Note that the Unif(0,1) prior on the parameters $\rho_{j}$ is used to
consider only positive spatial correlations~\citep[see for example][Chap. 4, p.147]{MartBotell2019}. A normal distribution with mean 0 and precision 0.001
are given to crime-specific intercepts $\alpha_{j}$ and regression coefficients
$\beta_{j}$. All the models are fitted using  \texttt{R} version 4.2.3
and \texttt{R-INLA} package version 22.12.16 (dated 2022-12-23) with the
simplified Laplace strategy. The full code and data to reproduce results are
available at
\url{https://github.com/spatialstatisticsupna/Multivariate_confounding}.

%%%%%%%%%%%%%%%%%%%%%%%%
%% REAL DATA ANALYSES %%
%%%%%%%%%%%%%%%%%%%%%%%%
\section{Illustration: joint analysis of rapes and dowry deaths in Uttar Pradesh in 2011}

In this section we analyse jointly rapes and dowry deaths, two forms of crimes against women in the 70 districts of Uttar Pradesh in 2011. The main objective is to assess whether there is linear association between a covariate of interest and the relative risks of each crime using the multivariate models with the spatial+ approach defined in the previous sections.

Uttar Pradesh is the most populated state of India and it is located in the
north of the country. Despite underreporting, the elevated number of rapes in
India is a matter of concern~\citep[see][for the temporal evolution of rapes in Uttar Pradesh]{vicente2018small}. Dowry death is deeply-rooted in the marriage
system in India and it is related to the dowry. The dowry represents the goods
that the bride's family offers to the groom's family before the marriage.
Disputes related to the dowry are frequent and the use of violence against the
bride to obtain a higher dowry is common. Sadly, this violence often ends in
the dead of the woman. This is known as dowry death. Any death related to
dowry, such as, murder or suicide of the woman within the first seven years of
marriage is considered a dowry death~\citep{Vicente2020a}.  Although in 1961
the Dowry Prohibition Act was approved, the tradition of dowry is still
widespread.

\begin{figure}[h]
\centering
\includegraphics[width=\textwidth]{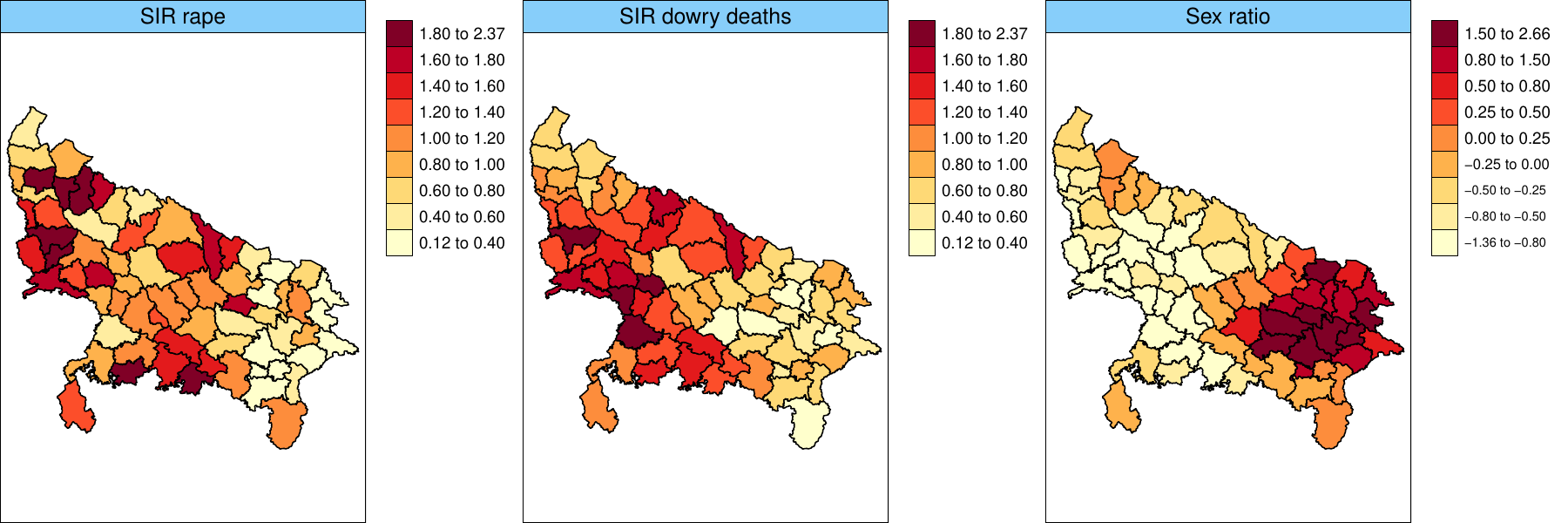}
\caption{Standardized incidence ratio for rapes (left), dowry deaths (center), and the spatial pattern of sex ratio (right) in Uttar Pradesh in 2011.}
\label{fig:SIR_UttarPradesh_2011}
\end{figure}

The minimum values for rapes and dowry deaths per district in 2011 are 2 and 6 respectively, whereas the maximum values are 89 and 95. So, the number of rapes and dowry deaths per district in 2011 is highly variable. Figure  \ref{fig:SIR_UttarPradesh_2011} displays the standardized incidence ratio (SIR) of rapes and dowry deaths in 2011. Some similarities between both spatial patterns can be observed with a Pearson correlation coefficient of 0.4471. Thus, it is sensible to analyse both crimes jointly. Figure \ref{fig:SIR_UttarPradesh_2011} also shows the spatial pattern of the  (standardized) covariate sex ratio $(\X_1)$ defined as the number of females per 1000 males. Here, the goal is to assess the potential linear association between the sex ratio and the relative risks of rapes and dowry deaths. However, this task may not be straightforward due to the evident spatial pattern of the covariate sex ratio, and the addition of spatial random effects as a proxy of unobserved covariates can induce bias in the fixed effects estimates (spatial confounding).

Here we fit the M-models \eqref{eq:multivariate_2} and \eqref{eq:M-spatplus2} introduced in Section 2. First, the covariate sex ratio is expressed as a linear combination of the eigenvectors of the precision matrix $\Q$ (see Equation~\eqref{eq:zeta_zetaprima}). Then we remove ${\Z}^{*}$, the part of the linear combination containing the large scale eigenvectors associated with the $k$ lowest eigenvalues. The number of large-scale eigenvectors chosen depends on the dimensions of $\Q$. Here, $k$ is between $5$ and $20$, which represent 7\% and 29\% of all eigenvectors. Table
\ref{notation_models} presents the notation of the M-models used. All the M-models are fitted considering ICAR, PCAR and BYM2 priors for the spatial random effects.

\begin{table}[htbp]
\centering
\caption{Different proposals for the multivariate M-models fitted to the data. The column Covariate refers to the covariate included in the M-model, $k$ indicates the number of large-scale eigenvectors of $\Q$ that comprises $\Z^{*}$ and $n-(k+1)$ is the number of the rest of eigenvectors comprised in $\Z$.}
\begin{tabular}{lllll}
\hline
Name & M-model & Covariate & $k$ & $n-(k+1)$  \\
\hline
M-Spatial & (\ref{eq:multivariate_2})& $\X_1$ & -- & --\\
M-SpatPlus64 & (\ref{eq:M-spatplus2})& $\Z$   & 5  & 64 \\
M-SpatPlus59 & (\ref{eq:M-spatplus2})& $\Z$   & 10 & 59 \\
M-SpatPlus54 & (\ref{eq:M-spatplus2})& $\Z$   & 15 & 54 \\
M-SpatPlus49 & (\ref{eq:M-spatplus2})& $\Z$   & 20 & 49 \\
\hline
\end{tabular}
\label{notation_models}
\end{table}%
Table \ref{UttarPradesh_betas} provides posterior means, posterior standard errors
and 95\% credible intervals of sex ratio for rapes ($\beta_1$) and dowry
deaths ($\beta_2$) with different M-models. A negative linear association is
observed between sex ratio and both crimes. For rapes, the 95\% credible
intervals always contain 0 irrespective of the models and the spatial prior,
indicating a non-significant relationship between rapes and sex ratio. For
dowry deaths, we observed a reduction of the posterior mean in the spatial+
models with respect to the spatial model. However, 95\% credible intervals
exclude $0$ with the M-Spatial, the M-SpatPlus64 and the M-SpatPlus59
models. In general, multivariate models with the spatial+ approach reduce the
posterior standard deviations in comparison to the usual multivariate spatial
model. Overall, the three spatial priors lead to similar fixed effects
estimates, though some differences appears in the multivariate spatial models
without the spatial+. The results are in line with previous literature as dowry
death is the only crime against women that seems to be associated with sex
ratio~\citep[see for example][]{mukherjee2001crimes}.

\begin{table}[ht]
	\centering
	\caption{Posterior means, standard errors and $95\%$ credible intervals of $\beta_1$ (rapes) and $\beta_2$ (dowry deaths) for Uttar Pradesh data in 2011.}
    \resizebox{\textwidth}{!}{
	\begin{tabular}{llllllllll}
		\hline
         & \multicolumn{1}{l}{Model} & \multicolumn{4}{l}{$\beta_1$} & \multicolumn{4}{l}{$\beta_2$} \\
         \cmidrule(lr){3-6} \cmidrule(lr){7-10}
		 &  & \multicolumn{1}{l}{Mean} & \multicolumn{1}{l}{SD} & \multicolumn{2}{l}{$95\%$ CI} & \multicolumn{1}{l}{Mean} & \multicolumn{1}{l}{SD} & \multicolumn{2}{l}{$95\%$ CI} \\
		\hline
		\multirow{5}[2]{*}{ICAR} & M-Spatial & -0.1560 & 0.1050 & -0.3630 & 0.0500 & -0.1920 & 0.0600 & -0.3080 & -0.0720 \\
		& M-SpatPlus64 & -0.0750 & 0.0680 & -0.2090 & 0.0590 & -0.0940 & 0.0410 & -0.1730 & -0.0130 \\
		& M-SpatPlus59 & -0.0750 & 0.0640 & -0.2010 & 0.0500 & -0.0810 & 0.0390 & -0.1580 & -0.0040 \\
		& M-SpatPlus54 & -0.0720 & 0.0580 & -0.1860 & 0.0420 & -0.0430 & 0.0370 & -0.1150 & 0.0290 \\
		& M-SpatPlus49 & -0.0870 & 0.0550 & -0.1960 & 0.0210 & -0.0460 & 0.0350 & -0.1150 & 0.0230 \\
		\hline
		\multirow{5}[2]{*}{PCAR} & M-Spatial & -0.2270 & 0.0950 & -0.4070 & -0.0330 & -0.2520 & 0.0570 & -0.3600 & -0.1350 \\
		& M-SpatPlus64 & -0.0750 & 0.0690 & -0.2100 & 0.0600 & -0.0970 & 0.0430 & -0.1830 & -0.0130 \\
		& M-SpatPlus59 & -0.0770 & 0.0650 & -0.2060 & 0.0520 & -0.0840 & 0.0410 & -0.1640 & -0.0030 \\
		& M-SpatPlus54 & -0.0720 & 0.0590 & -0.1890 & 0.0450 & -0.0440 & 0.0390 & -0.1200 & 0.0320 \\
		& M-SpatPlus49 & -0.0870 & 0.0570 & -0.1990 & 0.0240 & -0.0460 & 0.0380 & -0.1200 & 0.0270 \\
		\hline
		\multirow{5}[2]{*}{BYM2} &  M-Spatial & -0.1800 & 0.0990 & -0.3720 & 0.0150 & -0.2500 & 0.0590 & -0.3620 & -0.1310 \\
		& M-SpatPlus64 & -0.0680 & 0.0670 & -0.2000 & 0.0650 & -0.1100 & 0.0430 & -0.1940 & -0.0260 \\
		& M-SpatPlus59 & -0.0720 & 0.0640 & -0.1990 & 0.0550 & -0.0960 & 0.0410 & -0.1780 & -0.0150 \\
		& M-SpatPlus54 & -0.0710 & 0.0610 & -0.1900 & 0.0480 & -0.0490 & 0.0390 & -0.1250 & 0.0270 \\
		& M-SpatPlus49 & -0.0930 & 0.0590 & -0.2080 & 0.0220 & -0.0540 & 0.0380 & -0.1280 & 0.0200 \\
		\hline
	\end{tabular}}
    \label{UttarPradesh_betas}
\end{table}

Table \ref{UttarPradesh_correlations} presents the posterior median of the correlation between rapes and dowry deaths with their 95\% credible intervals. The estimated correlations are about 0.3 but slight discrepancies are noted in the estimates depending on the model and the spatial prior. Multivariate spatial+ models with the PCAR prior point towards a significant correlation. However, the rest of models indicate that the correlation is not significant. Given that the lower bounds of the credible intervals are very close to zero, we might conclude that the correlation is on the verge of significance. Finally, to compare the models in terms of goodness of fit and complexity, we have computed the Deviance Information Criterion (DIC) and the Watanabe-Akaike Information Criterion (WAIC) (see
Table \ref{UttarPradesh_DIC_WAIC}). M-models with spatial+ improve slightly the fit compared to the M-Spatial model, but differences are negligible. This is expected as the spatial+ approach can be seen as a kind of reparameterization of the spatial model.

\begin{table}[htbp]
	\centering
	\caption{Posterior medians and $95\%$ credible intervals of estimated correlations between rapes and dowry deaths for Uttar Pradesh data in 2011.}
    \resizebox{\textwidth}{!}{
    \begin{tabular}{llllllllll}
          \hline
          Model & \multicolumn{3}{l}{ICAR} & \multicolumn{3}{l}{PCAR} & \multicolumn{3}{l}{BYM2} \\
           \cmidrule(lr){2-4} \cmidrule(lr){5-7}\cmidrule(lr){8-10}
          & \multicolumn{1}{l}{Median} & \multicolumn{2}{l}{$95\%$ CI} & \multicolumn{1}{l}{Median} & \multicolumn{2}{l}{$95\%$ CI} & \multicolumn{1}{l}{Median} & \multicolumn{2}{l}{$95\%$ CI} \\
          \hline
          M-Spatial & 0.3066 & -0.0677 & 0.6087 & 0.3127 & -0.0508 & 0.6177 & 0.2502 & -0.1483 & 0.5703 \\
          M-SpatPlus64 & 0.3376 & -0.0090 & 0.6275 & 0.3698 & 0.0302 & 0.6503 & 0.3096 & -0.0722 & 0.6499 \\
          M-SpatPlus59 & 0.3320 & -0.0151 & 0.6313 & 0.3601 & 0.0245 & 0.6361 & 0.3086 & -0.0021 & 0.6079 \\
          M-SpatPlus54 & 0.3347 & -0.0056 & 0.6097 & 0.3713 & 0.0448 & 0.6304 & 0.2622 & -0.0688 & 0.5869 \\
          M-SpatPlus49 & 0.3222 & -0.0185 & 0.6229 & 0.3657 & 0.0405 & 0.6245 & 0.2581 & -0.0368 & 0.5464 \\
          \hline
    \end{tabular}}
  \label{UttarPradesh_correlations}
\end{table}%
To summarize, different fixed effects are estimated for both crimes depending
on the model fitted to the data. For rapes, the posterior mean of the sex ratio
coefficient $\beta_1$ is negative but the 95\% credible intervals do not
point towards a significant negative association. For dowry deaths, the
conclusions depend on the number of large-scale eigenvectors excluded in the
M-models with spatial+. According to~\citet{Urdangarin2022}, in univariate
spatial models with spatial+,  removing 14\% to 21\% of eigenvectors could lead
to unbiased fixed effect estimates. Here, M-SpatPlus59 and M-SpatPlus54 are the
equivalent models to removing 14\% and 21\% of eigenvectors. The 95\% credible
interval of M-SpatPlus59 points towards a significant negative association
between sex ratio and dowry deaths. Nevertheless, caution is recommended when
reaching conclusions.

We have also run the models including other socioeconomic covariates. The results are omitted to save space. Among all the covariates analysed, only the covariate \lq\lq number of burglaries''  shows a significant positive linear association with rapes, but it does not show a significant association with dowry deaths.  We have fitted all the previous M-models including both, sex ratio and burglaries as covariates but overall, the fixed effect estimates for each response do not change. The M-models proposed here allow removing different quantities of spatial dependence from each covariate, as well as the inclusion of specific covariates for a particular response. For example, the covariate burglary might be considered only for the analysis of rapes, while the sex ratio might be included only for dowry deaths.

\begin{table}[ht]
	\centering
	\caption{DIC and WAIC for Uttar Pradesh data in 2011.}
    %\resizebox{\textwidth}{!}{
	\begin{tabular}{lllllll}
		\hline
		Model & \multicolumn{2}{l}{ICAR} & \multicolumn{2}{l}{PCAR} & \multicolumn{2}{l}{BYM2} \\
        \cmidrule(lr){2-3} \cmidrule(lr){4-5}\cmidrule(lr){6-7}
         & \multicolumn{1}{l}{DIC} & \multicolumn{1}{l}{WAIC} & \multicolumn{1}{l}{DIC} & \multicolumn{1}{l}{WAIC} & \multicolumn{1}{l}{DIC} & \multicolumn{1}{l}{WAIC} \\
        \hline
        M-Spatial & 959.1602 & 949.0371 & 961.2425 & 948.8055 & 959.7526 & 946.1454 \\
        M-SpatPlus64 & 958.7318 & 945.9487 & 960.0558 & 944.6206 & 960.2536 & 943.0920 \\
        M-SpatPlus59 & 958.8249 & 945.6563 & 960.0255 & 944.4730 & 960.3539 & 942.8094 \\
        M-SpatPlus54 & 959.3916 & 945.3981 & 960.7766 & 944.4437 & 960.7177 & 943.4793 \\
        M-SpatPlus49 & 959.3160 & 945.5091 & 960.9914 & 944.7533 & 960.3658 & 943.0917 \\
		\hline	
	\end{tabular}
    \label{UttarPradesh_DIC_WAIC}
\end{table}

%%%%%%%%%%%%%%%%%%%%%%
%% SIMULATION STUDY %%
%%%%%%%%%%%%%%%%%%%%%%
\section{Simulation study}

In this section, we conduct two simulation studies, named Simulation study 1 and Simulation study 2, to evaluate the performance of the simplified spatial+ method in multivariate models in the presence of spatial confounding. For both simulation studies, we have employed the geographical configuration of Uttar Pradesh consisting of $70$ connected districts. Two related responses, denoted as crime 1 and crime 2 from now on, are generated. In Simulation study 1 we examine if the spatial+ method recovers the true fixed effects for each of the crimes when there is spatial confounding. Simulation Study 2 focuses on evaluating how well the multivariate models with the spatial+ technique simultaneously estimate the correlation between crimes and the fixed effects.

\textbf{Simulation study 1:} the data generating model includes the standardized covariate sex ratio of the real data analysis, denoted as $\X_1$, and two additional covariates, $\X_2$ and $\X_3$. Both $\X_2$ and $\X_3$ play the role of unobserved covariates in the fitted model for crimes 1 and 2 respectively, and hence they can induce spatial confounding. The correlation between $\X_2$ and $\X_3$, as well as the correlations between $\X_1$ and $\X_2$ and $\X_1$ and $\X_3$ are fixed. The fitted models do not include $\X_2$ and $\X_3$. In more detail, the counts for crime 1 and crime 2 are simulated as follows,
\begin{gather} \label{eq:simu_data_1}
\log \rr=(\I_{J}\otimes \boldsymbol{1}_{n})\aalpha + (\I_{J}\otimes \X_{1})\bbeta + \left(
                                                                                      \begin{array}{cc}
                                                                                        {\X}_2 & \boldsymbol{0} \\
                                                                                        \boldsymbol{0} & {\X}_3 \\
                                                                                      \end{array}
                                                                                    \right)
\bbeta^{*}\\
\boldsymbol{Y}^{l}\arrowvert\boldsymbol{r} \sim Poisson(\boldsymbol{\mu}=\boldsymbol{er})
\end{gather}
where $\aalpha=(\alpha_1, \alpha_2)^{\prime}=(-0.12, -0.03)^{\prime}$, $\bbeta=(\beta_1, \beta_2)^{'}=(-0.15, -0.20)^{'}$, $\bbeta^{*}=(\beta_1^{*}, \beta_2^{*})'=(-0.30,-0.30)'$, $l=1,\ldots, L$, and $\boldsymbol{e}$ is the vector of expected cases of the real case study. Two scenarios are simulated depending on the correlations between $\X_1$, $\X_2$ and $\X_3$. Scenario 1 addresses moderate-high correlations between the covariates and Scenario 2 deals with moderate-low correlations.

\begin{itemize}
\item \textbf{Scenario 1:} the correlations are $cor(\X_1, \X_2)=0.5$, $cor(\X_1, \X_3)=0.7$ and $cor(\X_2, \X_3)=0.7$. Here spatial confounding might be a major concern.
\item \textbf{Scenario 2:} the correlations are $cor(\X_1, \X_2)=0.3$, $cor(\X_1, \X_3)=0.5$ and $cor(\X_2, \X_3)=0.3$. In this case, spatial confounding might not be so severe.
\end{itemize}

\begin{figure}[h]
\centering
\includegraphics[width=0.8\textwidth]{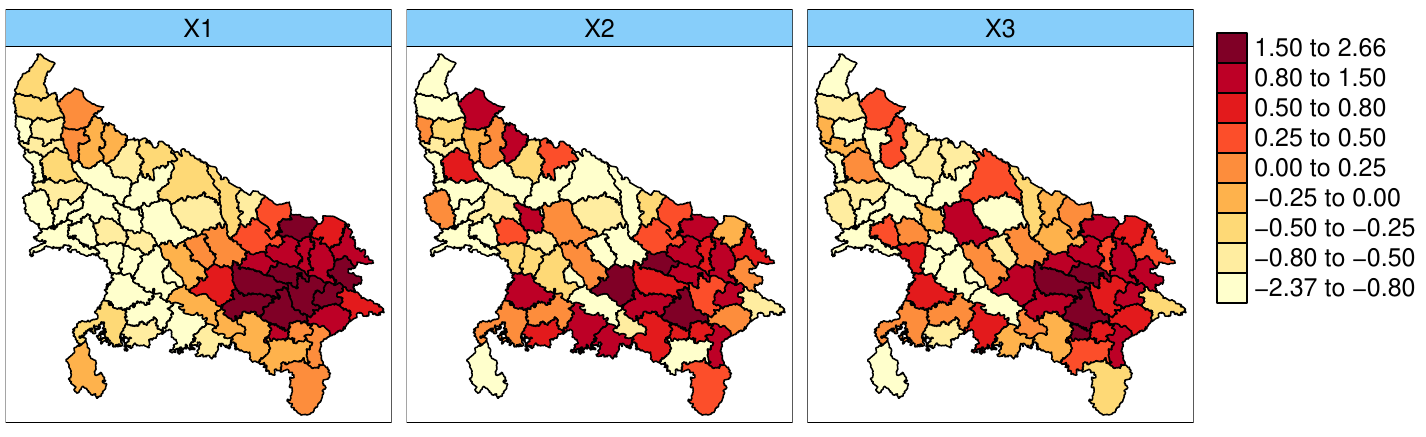}
\includegraphics[width=0.8\textwidth]{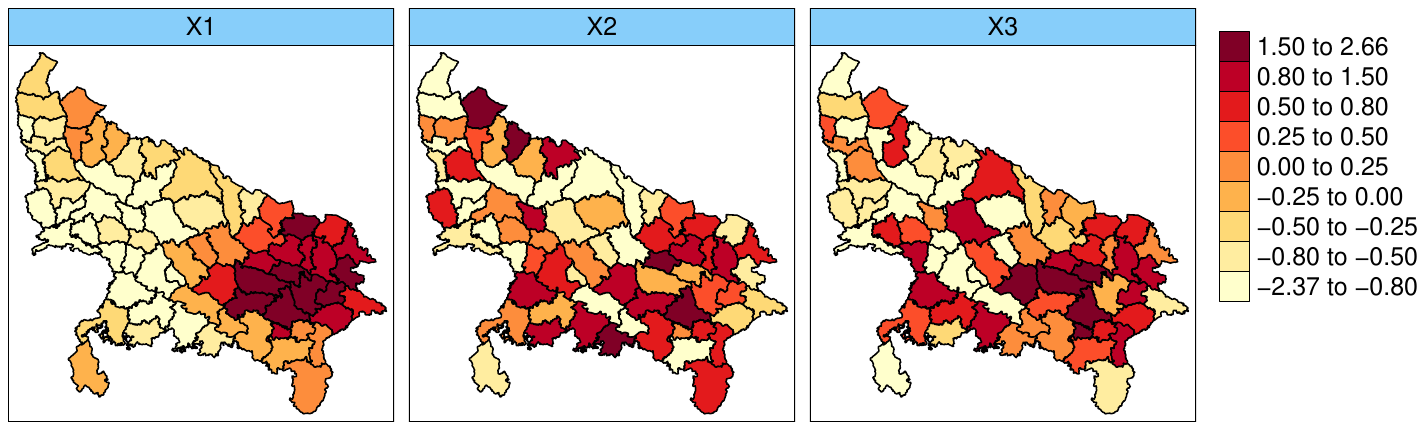}
\caption{From left to right, standardized sex ratio covariate in 2011, standardized simulated covariate $\X_2$ and standardized simulated covariate $\X_3$ in Simulation Study 1. The top row corresponds to Scenario 1 and the bottom row to Scenario 2.}
\label{fig:SimuStudy1}
\end{figure}

Figure \ref{fig:SimuStudy1} displays the standardized sex ratio covariate, along with the simulated covariates $\X_2$ and $\X_3$. The first row illustrates the spatial patterns of these covariates for Scenario 1, while the second row presents the patterns for Scenario 2. For each of the scenarios a total of $L=300$ counts data sets are generated. To complete the simulation study, Supplementary material A presents the details about the data generating process of some additional scenarios, Scenarios 3, 4, 5 and 6, and their results. (Note that in Scenarios 5 and 6, $cor(\X_1, \X_2)=0$ and hence, confounding should not be an issue for crime 1.)

\textbf{Simulation study 2:} the aim of this simulation study is to see how well multivariate models with the spatial+ method estimate the fixed effects but also the correlation between crimes. The objective is to see whether the modified spatial+ method introduces some bias in the estimation of the between-crime correlations. Here, the data generating process is quite different to Simulation study 1. First, a vector of spatial effects, $\ttheta=(\ttheta'_1, \ttheta'_2)^{\prime}$ is generated as $\ttheta \sim N(\mathbf{0}, \boldsymbol{\Omega}_{\boldsymbol{\theta}})$ where the precision matrix $\boldsymbol{\Omega}_{\boldsymbol{\theta}}$ follows the separable structure $\boldsymbol{\Omega}_{\boldsymbol{\theta}}= \SSigma_{b}^{-1} \otimes (\D-\W)$. The $2 \times 2$  between-crime covariance matrix $\SSigma_{b}$ is conveniently expressed as
\[
{\SSigma}_{b}=
\begin{pmatrix}
 \sigma_1 & 0 \\
 0 & \sigma_{2}
\end{pmatrix}
\begin{pmatrix}
 1 & \rho \\
 \rho & 1
\end{pmatrix}
\begin{pmatrix}
 \sigma_1 & 0 \\
 0 & \sigma_{2}
\end{pmatrix}\]
where $\sigma^2_1=0.9$, $\sigma^2_2=0.2$ are the variances obtained with the M-Spatial model in the real data analysis and $\rho$ is the pre-defined linear correlation between crime 1 and crime 2. In a second step, a covariate $\X^{*}_1$ is simulated such that its linear correlation with the spatial effects of crime 1, $\ttheta_1$, and the spatial effects of crime 2, $\ttheta_2$, is fixed. In more detail, data are generated as,
\begin{gather} \label{eq:simu_data_2}
\log \rr=(\I_{J}\otimes \boldsymbol{1}_{n})\aalpha + (\I_{J}\otimes \X^{*}_1)\bbeta + \ttheta\\
\boldsymbol{Y}^{l}\arrowvert\boldsymbol{r} \sim Poisson(\boldsymbol{\mu}=\boldsymbol{er})
\end{gather}
where $\ttheta=(\ttheta'_1, \ttheta'_2)^{\prime}$, $l=1,\ldots, L$ and $\boldsymbol{e}$ is the vector of expected cases of the real case study. Now, $\aalpha=(\alpha_1, \alpha_2)^{\prime}=(0.12, 0.03)^{\prime}$ and $\bbeta=(\beta_1, \beta_2)^{'}=(0.15, 0.20)^{'}$ are considered. Two scenarios are simulated depending on the correlation between crimes, $\rho$, and the correlations between the simulated $\X^{*}_1$ and the spatial effects $\ttheta_1$ and $\ttheta_2$. Scenario 1 addresses moderate-high correlations and Scenario 2 deals with low-moderate correlations.

\begin{itemize}
\item \textbf{Scenario 1:} the correlations are $\rho=0.7$, $cor(\X^{*}_1, \ttheta_1)=0.5$ and $cor(\X^{*}_1, \ttheta_2)=0.7$. Here spatial confounding might be a major concern and the crimes are highly correlated.
\item \textbf{Scenario 2:} the correlations are $\rho=0.3$, $cor(\X^{*}_1, \ttheta_1)=0.3$ and $cor(\X^{*}_1, \ttheta_2)=0.5$. In this case, spatial confounding might not be so severe and the correlation between crimes is low.
\end{itemize}

\begin{figure}[h]
\centering
\includegraphics[width=0.8\textwidth]{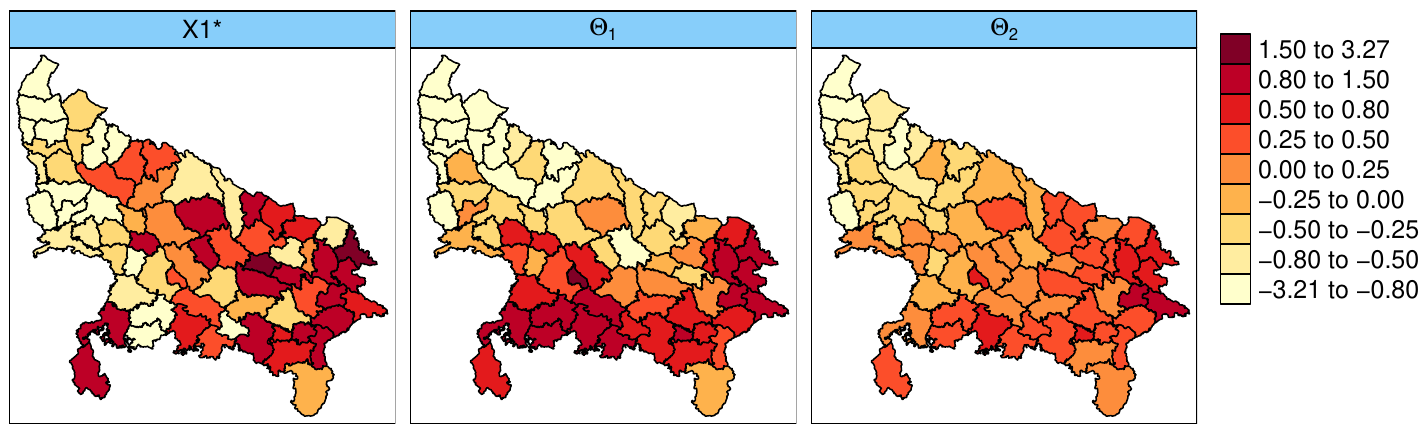}
\includegraphics[width=0.8\textwidth]{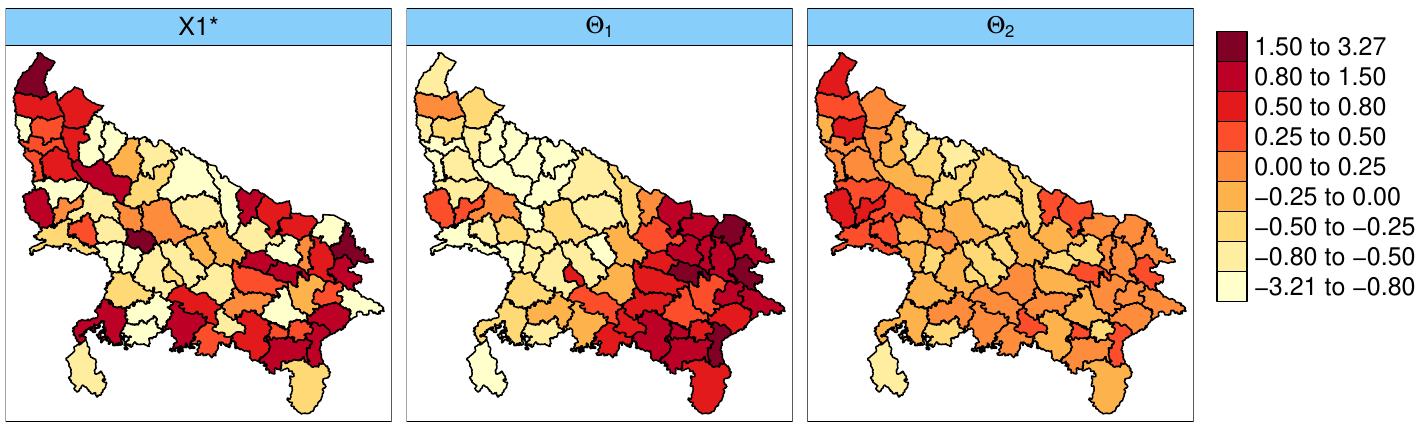}
\caption{From left to right, standardized simulated covariate $\X^{*}_1$, spatial effects $\ttheta_1$ for crime 1 and spatial effects $\ttheta_2$ for crime 2 in Simulation Study 2. The top row corresponds to Scenario 1 and the bottom row to Scenario 2.}
\label{fig:SimuStudy2}
\end{figure}

Figure \ref{fig:SimuStudy2} shows the standardized simulated covariate $\X^{*}_1$ and the spatial effects $\ttheta_1$ and $\ttheta_2$ for each of the crimes. The first row displays the spatial patterns for Scenario, 1 while the second row depicts those for Scenario 2. For each of the scenarios a total of $L=300$ counts data sets are generated. Supplementary material B presents the details about the data generating process of some additional scenarios, Scenarios 3, 4 and 5, and their results.

The simulated data are analysed using the same models as those employed in the real data analysis. All the models are fitted assigning ICAR, PCAR and BYM2 priors to the spatial random effects. The objective is to examine in more detail how the simplified spatial+ method recovers the true fixed effects in different scenarios and if the between-crime correlation estimates are affected by the method.

%%%%%%%%%%%%%%%%%%%%%%%%%%%%%%%%%
%% SIMULATION STUDY 1: RESULTS %%
%%%%%%%%%%%%%%%%%%%%%%%%%%%%%%%%%
\subsection{Simulation study 1: Results}
The main goal of this simulation study is to evaluate the fixed effect estimates of the spatial+ method in multivariate models. Intuitively, the correlation between the unobserved covariates should be captured by the between-crime correlation matrix, so we will also look at those parameters.

\begin{figure}[htbp]
\centering
\includegraphics[width=1\textwidth]{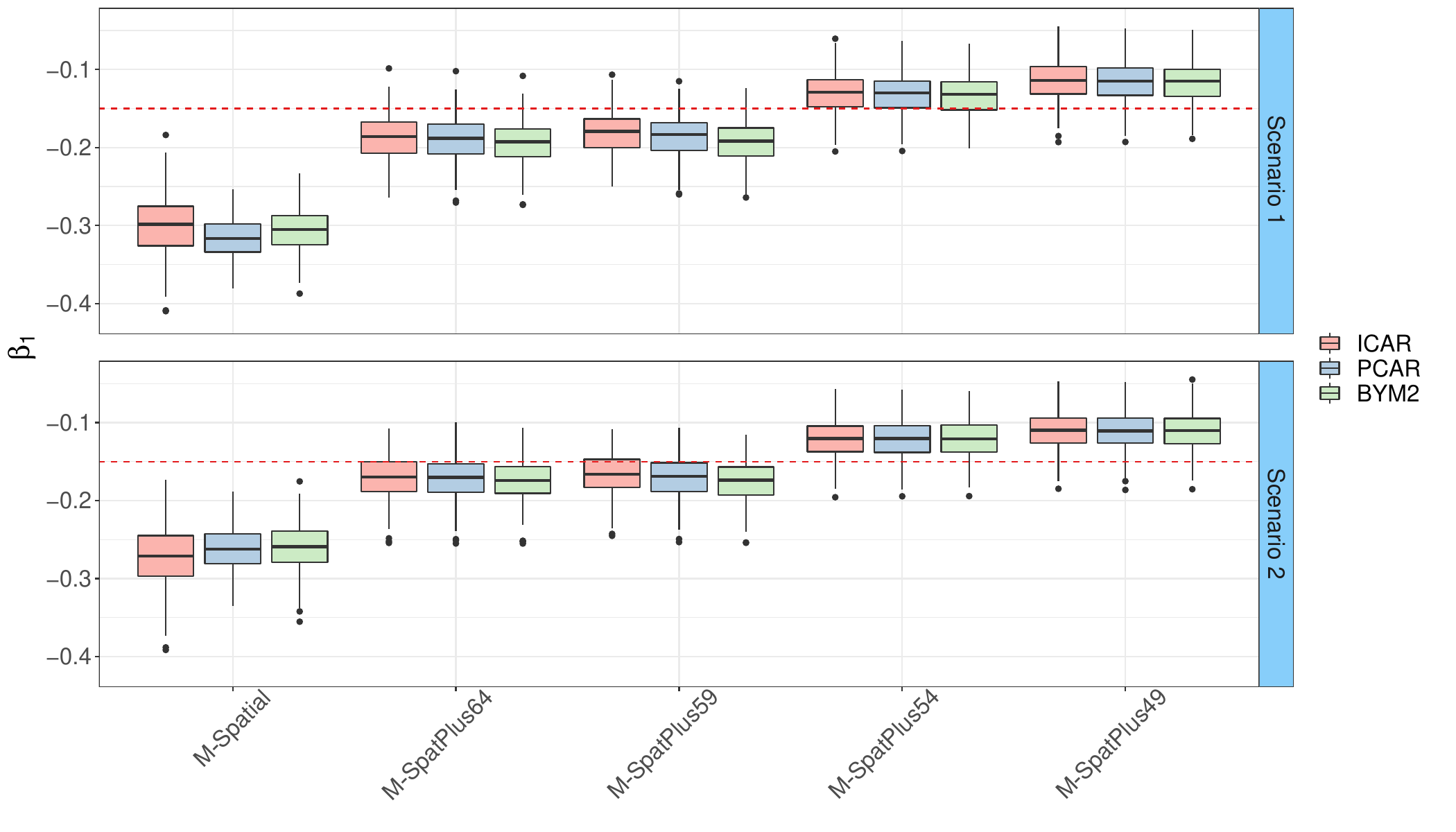}
\caption{Boxplots of the estimated means of $\beta_1$ based on 300 simulated datasets for Simulation Study 1, Scenario 1 (top row) and Scenario 2 (bottom row). Each color represents a different prior given to the columns of $\PPhi$, namely, red for ICAR, blue for PCAR and green for BYM2.}
\label{SimuStudy1_boxplots_rapes}
\end{figure}

\begin{figure}[ttbp]
\centering
\includegraphics[width=1\textwidth]{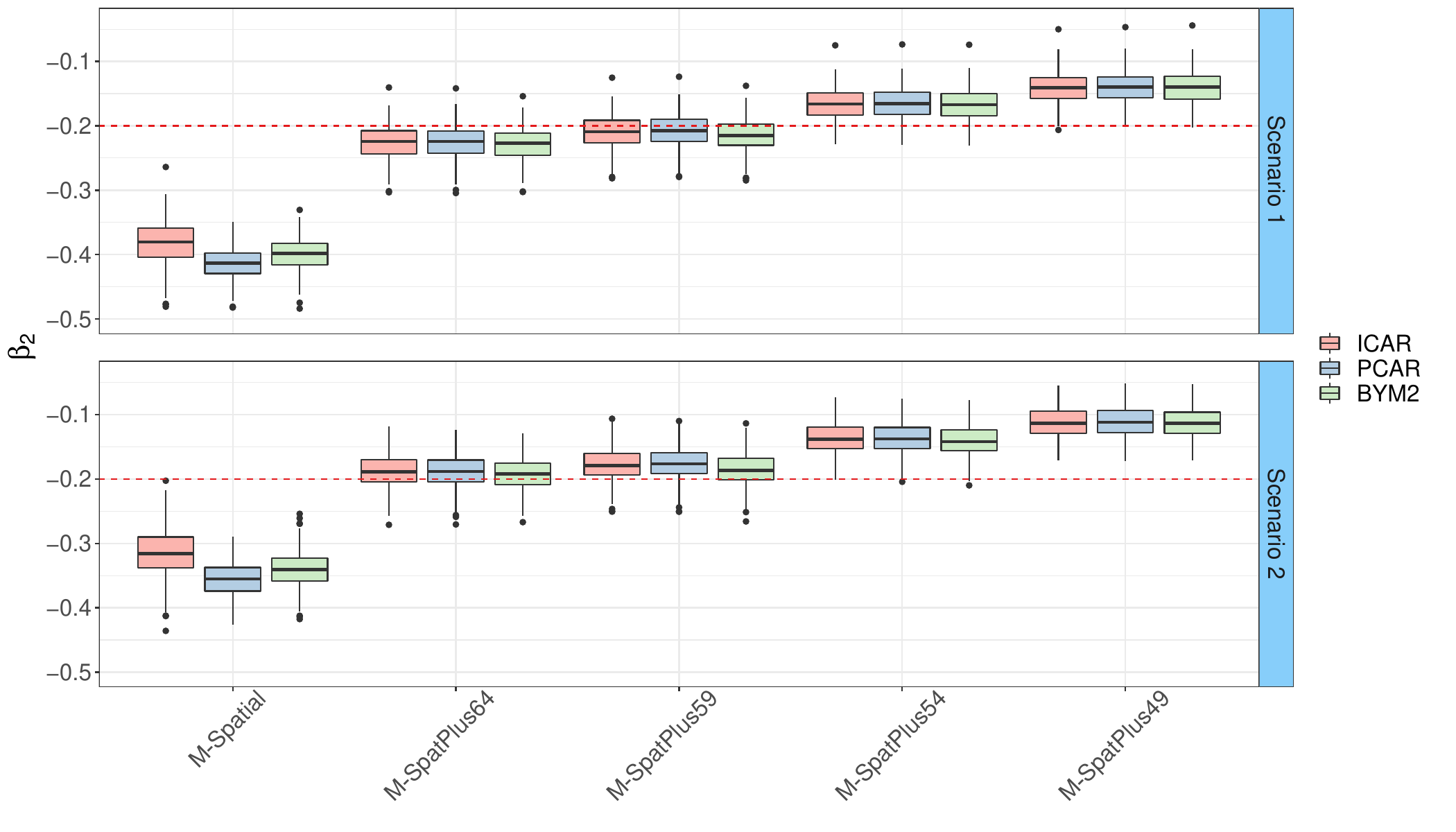}
\caption{Boxplots of the estimated means of $\beta_2$ based on 300 simulated datasets for Simulation Study 1, Scenario 1 (top row) and Scenario 2 (bottom row). Each color represents a different prior given to the columns of $\PPhi$, namely, red for ICAR, blue for PCAR and green for BYM2.}
\label{SimuStudy1_boxplots_dowry}
\end{figure}

Tables \ref{SimuStudy1_mean_sd_beta1} and \ref{SimuStudy1_mean_sd_beta2} provide the average over the 300 simulated datasets of the posterior means and posterior standard deviations of the regression coefficients $\beta_1$ and $\beta_2$ respectively. To visually examine the fixed effect estimates of different models, Figures \ref{SimuStudy1_boxplots_rapes} and \ref{SimuStudy1_boxplots_dowry}  display boxplots of the posterior means of $\beta_1$ and $\beta_2$ over the 300 simulated datasets for Scenarios 1 and 2 respectively. In general, the M-Spatial model provides highly biased fixed effects estimates whereas some of the M-models with the spatial+ method recover quite well the true values of the regression coefficients. The number of large-scale eigenvectors of $\boldsymbol{Q}$ that should be omitted in $\Z$ and hence assigned to $\Z^{*}$ (see Equation~\ref{eq:zeta_zetaprima}) depends on the relationship between the observed and unobserved covariates. In the scenarios addressed here, we contemplate correlations of 0.5 and 0.3 (moderate and low) between sex ratio and the unobserved covariate in crime 1 and correlations of 0.7 and 0.5 (high and moderate) in crime 2. Supplementary material A addresses correlations of 0.5, 0.3 and 0.0 (moderate, low and no correlation) between sex ratio and the unobserved covariate ${\X}_2$ in crime 1 and correlation of 0.7 (high) between sex ratio and the unobserved covariate ${\X}_3$ in crime 2.
%\QUERY[4]

\begin{table}[htbp]
	\centering
	\caption{Posterior means and standard deviations of $\beta_1$ based on 300 simulated datasets for Simulation study 1 and Scenarios 1 and 2.}
    \resizebox{\textwidth}{!}{
	\begin{tabular}{llcllllll}
		\hline
        & Model & \text{True value} & \multicolumn{2}{l}{ICAR} & \multicolumn{2}{l}{PCAR} & \multicolumn{2}{l}{BYM2} \\
        \cmidrule(lr){4-5} \cmidrule(lr){6-7}\cmidrule(lr){8-9}
        & &  & \multicolumn{1}{l}{Mean} & \multicolumn{1}{l}{SD} & \multicolumn{1}{l}{Mean} & \multicolumn{1}{l}{SD} & \multicolumn{1}{l}{Mean} & \multicolumn{1}{l}{SD}  \\
		\hline
		\multirow{5}[2]{*}{Scenario 1} & M-Spatial & \multirow{5}[2]{*}{-0.1500} & -0.2996 & 0.0674 & -0.3162 & 0.0457 & -0.3048 & 0.0534 \\
		& M-SpatPlus64 & & -0.1869 & 0.0446 & -0.1895 & 0.0429 & -0.1942 & 0.0433 \\
		& M-SpatPlus59 & & -0.1818 & 0.0419 & -0.1863 & 0.0407 & -0.1932 & 0.0415 \\
		& M-SpatPlus54 & & -0.1303 & 0.0407 & -0.1315 & 0.0402 & -0.1332 & 0.0420 \\
		& M-SpatPlus49 & & -0.1141 & 0.0399 & -0.1152 & 0.0398 & -0.1162 & 0.0419 \\
		\hline
        \multirow{5}[2]{*}{Scenario 2} & M-Spatial & \multirow{5}[2]{*}{-0.1500} & -0.2707 & 0.0733 & -0.2619 & 0.0477 & -0.2590 & 0.0579 \\
        & M-SpatPlus64 & & -0.1692 & 0.0481 & -0.1710 & 0.0449 & -0.1746 & 0.0452 \\
        & M-SpatPlus59 & & -0.1648 & 0.0453 & -0.1695 & 0.0431 & -0.1748 & 0.0440 \\
		& M-SpatPlus54 & & -0.1209 & 0.0431 & -0.1208 & 0.0422 & -0.1211 & 0.0443  \\
		& M-SpatPlus49 & & -0.1096 & 0.0419 & -0.1110 & 0.0416 & -0.1109 & 0.0442 \\
        \hline
	\end{tabular}}
    \label{SimuStudy1_mean_sd_beta1}
\end{table}

\begin{table}[htbp]
	\centering
	\caption{Posterior means and standard deviations of $\beta_2$ based on 300 simulated datasets for Simulation study 1 and Scenarios 1 and 2.}
    \resizebox{\textwidth}{!}{
	\begin{tabular}{llcllllll}
		\hline
        & Model & \text{True value} & \multicolumn{2}{l}{ICAR} & \multicolumn{2}{l}{PCAR} & \multicolumn{2}{l}{BYM2} \\
        \cmidrule(lr){4-5} \cmidrule(lr){6-7}\cmidrule(lr){8-9}
        & &  & \multicolumn{1}{l}{Mean} & \multicolumn{1}{l}{SD} & \multicolumn{1}{l}{Mean} & \multicolumn{1}{l}{SD} & \multicolumn{1}{l}{Mean} & \multicolumn{1}{l}{SD}  \\
		\hline
		\multirow{5}[2]{*}{Scenario 1} & M-Spatial & \multirow{5}[2]{*}{-0.2000} & -0.3828 & 0.0596 & -0.4140 & 0.0407 & -0.3996 & 0.0462 \\
		& M-SpatPlus64 & & -0.2258 & 0.0429 & -0.2253 & 0.0454 & -0.2282 & 0.0438 \\
		& M-SpatPlus59 & & -0.2094 & 0.0408 & -0.2075 & 0.0432 & -0.2143 & 0.0423  \\
		& M-SpatPlus54 & & -0.1663 & 0.0396 & -0.1655 & 0.0421 & -0.1677 & 0.0424 \\
		& M-SpatPlus49 & & -0.1412 & 0.0393 & -0.1400 & 0.0417 & -0.1410 & 0.0426 \\
		\hline
        \multirow{5}[2]{*}{Scenario 2} & M-Spatial & \multirow{5}[2]{*}{-0.2000} & -0.3152 & 0.0690 & -0.3556 & 0.0443 & -0.3403 & 0.0508 \\
        & M-SpatPlus64 & & -0.1881 & 0.0465 & -0.1885 & 0.0477 & -0.1922 & 0.0459  \\
        & M-SpatPlus59 & & -0.1779 & 0.0438 & -0.1764 & 0.0452 & -0.1857 & 0.0443 \\
		& M-SpatPlus54 & & -0.1368 & 0.0418 & -0.1369 & 0.0439 & -0.1403 & 0.0445 \\
		& M-SpatPlus49 & & -0.1130 & 0.0411 & -0.1116 & 0.0434 & -0.1132 & 0.0447 \\
        \hline
	\end{tabular}}
    \label{SimuStudy1_mean_sd_beta2}
\end{table}

In summary, when the correlation between $\X_1$ and unobserved covariates becomes smaller, lower number of eigenvectors should be omitted in $\Z$. If too many eigenvectors are excluded from $\Z$, the M-models with the spatial+ provide biased fixed effects. Nevertheless, in the case of crime 1, a greater number of large-scale eigenvectors should be excluded from $\Z$ compared to crime 2. This discrepancy could be explained because the number of counts for crime 1 is smaller than for crime 2. The M-models with the spatial+ approach provide the same fixed effects estimates irrespective of the prior (ICAR, PCAR or BYM2) given to the spatial random effects. For the M-Spatial model, slight differences are observed in the mean estimates depending on the prior chosen.

Examining Tables \ref{SimuStudy1_mean_sd_beta1} and \ref{SimuStudy1_mean_sd_beta2} we see that the M-SpatPlus models provide smaller posterior standard deviations than the M-Spatial model. The reduction is about 30\%. However, drawing a definitive conclusion regarding whether the posterior standard deviation accurately measures the variability of the fixed effect estimates based on this information is not straightforward. Table A.3 and A.4 in Supplementary material A provide the true simulated standard error ($s.e._{sim}$) and estimated standard error ($s.e._{est}$) for $\beta_1$ and $\beta_2$ computed as $ s.e._{sim}=\displaystyle \sqrt{\frac{1}{300} \sum_{l=1}^{300}\left(\hat{\beta}_{j}^l-\overline{\hat{\beta}}_{j}\right)^2}$ and $s.e._{est}=\displaystyle \frac{1}{300} \sum_{l=1}^{300} sd(\hat{\beta}_{j}^l)$ for $j=1,2$. Here $\hat{\beta}_{j}^{l}$ is the posterior mean of $\beta_{j}$ estimated in simulation $l$ and $\overline{\hat{\beta}}_{j}$ is the average of all the posterior mean estimates across the simulations. Finally, $sd(\hat{\beta}_{j}^l)$ is the posterior standard deviation of $\beta_{j}$ in simulation $l$.
Overestimation of the standard error occurs when the estimated standard error exceeds the simulated standard error. Conversely, if the estimated standard error is lower than the simulated standard error, it results in an underestimation of the standard error of the fixed effects. Here, both the M-Spatial and the M-SpatPlus models tend to overestimate the standard error of the fixed effects.

To complete the information about fixed effects, Table \ref{SimuStudy1_95coverage} provides the empirical coverage of credible intervals at 95\% nominal value for $\beta_1$ and $\beta_2$. In general, the empirical coverages of $\beta_1$ and $\beta_2$ are relatively low when using the M-Spatial model. This can be attributed to the large bias introduced by the method. On the contrary, the M-models with the spatial+ approach that remove an appropriate number of eigenvectors exhibit credible intervals with certain overcoverage. This could be explained because the spatial+ approach reduces bias but still inflates the variance. If too many eigenvectors are removed, the coverage is poor because the bias increases.

\begin{table}[htbp]
	\centering
	\caption{Empirical $95\%$ coverage probabilities of the true value of $\beta_{1}$ and $\beta_{2}$ based on $300$ simulated datasets for Simulation study 1 and Scenarios 1 and 2.}
    \resizebox{\textwidth}{!}{
	\begin{tabular}{llllllll}
        \hline
		& Model & \multicolumn{3}{l}{$\beta_{1}$} & \multicolumn{3}{l}{$\beta_{2}$} \\
		\cmidrule(lr){3-5} \cmidrule(lr){6-8}
		&  & \multicolumn{1}{l}{ICAR} & \multicolumn{1}{l}{PCAR} & \multicolumn{1}{l}{BYM2} & \multicolumn{1}{l}{ICAR} & \multicolumn{1}{l}{PCAR} & \multicolumn{1}{l}{BYM2}  \\
		\hline
		\multirow{5}[2]{*}{Scenario 1} & M-Spatial & 35.6667 & 0.0000 & 5.6667 & 3.0000 & 0.0000 & 0.0000 \\
		& M-SpatPlus64 & 96.3333 & 94.6667 & 93.0000 & 98.0000 & 98.6667 & 98.6667 \\
		& M-SpatPlus59 & 96.3333 & 95.3333 & 93.6667 & 99.0000 & 100.0000 & 99.0000 \\
		& M-SpatPlus54 & 97.6667 & 98.3333 & 100.0000 & 96.3333& 97.3333 & 97.0000 \\
		& M-SpatPlus49 & 95.6667 & 95.6667 & 96.6667 & 77.3333 & 80.3333 & 85.6667 \\
		\hline
        \multirow{5}[2]{*}{Scenario 2} & M-Spatial & 72.6667 & 26.6667 & 53.3333 & 70.0000 & 0.6667 & 9.6667 \\
		& M-SpatPlus64 & 99.0000& 98.6667 & 98.6667 & 100.0000 & 100.0000 & 100.0000 \\
		& M-SpatPlus59 & 99.6667 & 99.0000 & 98.6667 & 99.3333 & 99.6667 & 100.0000 \\
		& M-SpatPlus54 & 98.3333 & 97.6667 & 99.3333 & 75.6667 & 84.0000 & 88.6667 \\
		& M-SpatPlus49 & 95.0000 & 95.3333 & 97.0000 & 42.3333& 47.0000 & 52.3333 \\
        \hline
	\end{tabular}}
	\label{SimuStudy1_95coverage}
\end{table}

Table \ref{SimuStudy1_correlations} displays the average over the $300$ simulated data sets of the posterior medians and 95\% credible intervals of the between crime correlation in each scenario. Here, we contemplate correlations of 0.7 (Scenario 1) and 0.3 (Scenario 2) between the unobserved covariates $\X_2$ and $\X_3$. In both scenarios, the results vary depending on the model and the prior chosen. In general, higher correlations between crimes are estimated when the BYM2 prior is assigned to the spatial random effects. When a high number of large-scale eigenvectors are excluded from the observed covariate, the M-SpatPlus models exhibit higher correlations between crimes. This probably occurs because the spatial dependence removed from the covariate is accounted for by the spatial effects. As a result, the spatial patterns of both crimes may become more similar, potentially leading to an increased correlation between the two crimes. Moreover, the correlation between crimes estimated with the M-models depends on the correlation between $\X_2$ and $\X_3$ defined in the data generating model. Overall, the estimated between-crime correlations are close to the correlations between $\X_2$ and $\X_3$. It is worth noting that the M-models with the spatial+ approach and the BYM2 prior tend to overestimate the correlations. Finally, when the correlation is low, the credible intervals are wider.

\begin{table}[htbp]
	\centering
	\caption{Posterior medians and $95\%$ credible intervals of estimated correlations between crime 1 and crime 2 based on $300$ simulated datasets for Simulation study 1 and Scenarios 1 and 2.}
    \resizebox{\textwidth}{!}{
    \begin{tabular}{lllllllllll}
          \hline
          &    Model    & \multicolumn{3}{l}{ICAR} & \multicolumn{3}{l}{PCAR} & \multicolumn{3}{l}{BYM2} \\
          \cmidrule(lr){3-5} \cmidrule(lr){6-8}\cmidrule(lr){9-11}
          &      & \multicolumn{1}{l}{Median} & \multicolumn{2}{l}{$95\%$ CI} & \multicolumn{1}{l}{Median} & \multicolumn{2}{l}{$95\%$ CI} & \multicolumn{1}{l}{Median} & \multicolumn{2}{l}{$95\%$ CI} \\
          \hline
          \multirow{5}[2]{*}{Scenario 1} & M-Spatial & 0.7287 & 0.3607 & 0.9305 & 0.6256 & 0.2464 & 0.8739 & 0.5857 & 0.1678 & 0.8694 \\
          & M-SpatPlus64 & 0.7277 & 0.4032 & 0.9191 & 0.6313 & 0.3150 & 0.8598 & 0.8099 & 0.5331 & 0.9439 \\
          & M-SpatPlus59 & 0.7229 & 0.3931 & 0.9178 & 0.6237 & 0.3140 & 0.8541 & 0.8164 & 0.5461 & 0.9459 \\
          & M-SpatPlus54 & 0.7830 & 0.5067 & 0.9380 & 0.6852 & 0.3879 & 0.8868 & 0.8691 & 0.6472 & 0.9631 \\
          & M-SpatPlus49 & 0.7958 & 0.5360 & 0.9410 & 0.7045 & 0.4168 & 0.8942 & 0.8808 & 0.6749 & 0.9664 \\
          \hline
          \multirow{5}[2]{*}{Scenario 2} & M-Spatial & 0.1535 & -0.2631 & 0.5401 & 0.1710 & -0.2115 & 0.5273 & 0.2759 & -0.2240 & 0.6615 \\
          & M-SpatPlus64 & 0.2138 & -0.1782 & 0.5729 & 0.2325 & 0.0381 & 0.4566 & 0.6084 & 0.2456 & 0.8345 \\
          & M-SpatPlus59 & 0.1989 & -0.1966 & 0.5643 & 0.2194 & 0.0362 & 0.4368 & 0.5921 & 0.2223 & 0.8228 \\
          & M-SpatPlus54 & 0.3065 & -0.0621 & 0.6327 & 0.3038 & 0.0830 & 0.5458 & 0.7018 & 0.3673 & 0.8864 \\
          & M-SpatPlus49 & 0.3371 & -0.0224 & 0.6506 & 0.3378 & 0.0963 & 0.5970 & 0.7243 & 0.4087 & 0.8919 \\
          \hline
    \end{tabular}}
  \label{SimuStudy1_correlations}
\end{table}%
To conclude the simulation study, we look at model selection criteria and the relative risks estimates. According to
Table \ref{SimuStudy1_DIC_WAIC}, these criteria do not clearly select any model, though models with the BYM2 prior present slightly lower values, particularly if we compare it with the ICAR prior. Table A.8 in Supplementary material A provides MARB and MRRMSE of the relative risks. M-Spatial and M-SpatPlus models perform similarly in terms of the relative risks estimates and differences between spatial priors are negligible.

\begin{table}[ht]
	\centering
	\caption{DIC and WAIC based on 300 simulated data sets for Simulation Study 1 and Scenarios 1 and 2.}
    \resizebox{\textwidth}{!}{
	\begin{tabular}{lllllllll}
		\hline
		& Model & \multicolumn{2}{l}{ICAR} & \multicolumn{2}{l}{PCAR} & \multicolumn{2}{l}{BYM2} \\
        \cmidrule(lr){3-4} \cmidrule(lr){5-6}\cmidrule(lr){7-8}
        &  & \multicolumn{1}{l}{DIC} & \multicolumn{1}{l}{WAIC} & \multicolumn{1}{l}{DIC} & \multicolumn{1}{l}{WAIC} & \multicolumn{1}{l}{DIC} & \multicolumn{1}{l}{WAIC} \\
        \hline
        \multirow{5}[2]{*}{Scenario 1} & M-Spatial & 946.0498 & 952.7725 & 939.1256 & 937.2348 & 934.8325 & 930.7239 \\
		& M-SpatPlus64 &  944.8425 & 947.5755 & 941.3371 & 937.9480 & 937.8670 & 930.4969 \\
		& M-SpatPlus59 & 945.1169 & 947.9748 & 941.2409 & 937.7502 & 937.9337 & 930.6553 \\
		& M-SpatPlus54 & 945.8632 & 947.4914 & 942.4370 & 937.5271 & 940.5496 & 932.8373 \\
		& M-SpatPlus49 & 946.5776 & 947.8275 & 943.0707 & 937.6770 & 941.4065 & 933.3797  \\
        \hline
        \multirow{5}[2]{*}{Scenario 2 } & M-Spatial & 961.6077 & 962.9980 & 953.7916 & 946.1904 & 949.2563 & 939.0299 \\
		& M-SpatPlus64 & 960.3800 & 958.9302 & 953.9252 & 945.8736 & 950.2834 & 937.4415  \\
		& M-SpatPlus59 & 960.3221 & 959.2081 & 953.7474 & 946.2881 & 949.9048 & 937.1384 \\
		& M-SpatPlus54 & 960.7991 & 957.1269 & 954.7606 & 944.6926 & 952.5815 & 938.9499 \\
		& M-SpatPlus49 & 961.0325 & 956.6160 & 955.0791 & 944.1604 & 953.0834 & 939.0144 \\
		\hline	
	\end{tabular}}
    \label{SimuStudy1_DIC_WAIC}
\end{table}

%%%%%%%%%%%%%%%%%%%%%%%%%%%%%%%%%
%% SIMULATION STUDY 2: RESULTS %%
%%%%%%%%%%%%%%%%%%%%%%%%%%%%%%%%%
\subsection{Simulation study 2: Results}

The primary objective of this second simulation study is to evaluate how well the proposed multivariate models estimate the correlations between crimes and the fixed effects. In the data simulation, the dependence between crimes is introduced by the covariance matrix ${\SSigma}_{b}$ where $\rho$ defines the correlation between the spatial patterns of crime 1 and crime 2. Hence, we will compare the estimated correlations with the true $\rho$ used in the data generating process. Moreover, we will also examine the fixed effect estimates as in Simulation Study 1.

\begin{table}[htbp]
	\centering
	\caption{Posterior medians and $95\%$ credible intervals of estimated correlations between crime 1 and crime 2 based on $300$ simulated datasets for Simulation study 2 and Scenario 1 ($\rho=0.7$) and Scenario 2 ($\rho=0.3$).}
    \resizebox{\textwidth}{!}{
    \begin{tabular}{lllllllllll}
          \hline
          &    Model    & \multicolumn{3}{l}{ICAR} & \multicolumn{3}{l}{PCAR} & \multicolumn{3}{l}{BYM2} \\
          \cmidrule(lr){3-5} \cmidrule(lr){6-8}\cmidrule(lr){9-11}
          &      & \multicolumn{1}{l}{Median} & \multicolumn{2}{l}{$95\%$ CI} & \multicolumn{1}{l}{Median} & \multicolumn{2}{l}{$95\%$ CI} & \multicolumn{1}{l}{Median} & \multicolumn{2}{l}{$95\%$ CI} \\
          \hline
          \multirow{7}[2]{18mm}{Scenario 1 ($\rho=0.7$)} &  M-Spatial & 0.6991 & 0.4669 & 0.8547 & 0.6825 & 0.4611 & 0.8404 & 0.7328 & 0.4952 & 0.8854 \\
            & M-SpatPlus64 & 0.6951 & 0.4738 & 0.8456 & 0.6889 & 0.4795 & 0.8359 & 0.7184 & 0.4928 & 0.8664 \\
          & M-SpatPlus59 & 0.6783 & 0.4499 & 0.8347 & 0.6745 & 0.4624 & 0.8264 & 0.7008 & 0.4718 & 0.8543 \\
          & M-SpatPlus54 & 0.6954 & 0.4775 & 0.8438 & 0.6883 & 0.4841 & 0.8338 & 0.7177 & 0.4954 & 0.8642 \\
          & M-SpatPlus49 & 0.7062 & 0.4926 & 0.8516 & 0.6977 & 0.4996 & 0.8379 & 0.7271 & 0.5102 & 0.8698 \\
          & M-SpatPlus44 & 0.7192 & 0.5137 & 0.8586 & 0.7081 & 0.5116 & 0.8467 & 0.7398 & 0.5326 & 0.8752 \\
          & M-SpatPlus39 & 0.7221 & 0.5197 & 0.8599 & 0.7086 & 0.5148 & 0.8462 & 0.7454 & 0.5369 & 0.8806 \\
          \hline
          \multirow{7}[2]{18mm}{Scenario 2 ($\rho=0.3$)} & M-Spatial & 0.4529 & 0.1511 & 0.6942 & 0.3180 & 0.0074 & 0.6090 & 0.5037 & 0.1921 & 0.7490 \\
          & M-SpatPlus64 & 0.4873 & 0.1998 & 0.7148 & 0.3865 & 0.1048 & 0.6429 & 0.5533 & 0.2664 & 0.7764 \\
          & M-SpatPlus59 & 0.5058 & 0.2339 & 0.7195 & 0.4128 & 0.1378 & 0.6579 & 0.5699 & 0.2971 & 0.7799 \\
          & M-SpatPlus54 & 0.4969 & 0.2232 & 0.7131 & 0.4042 & 0.1293 & 0.6528 & 0.5596 & 0.2856 & 0.7728 \\
          & M-SpatPlus49 & 0.5155 & 0.2469 & 0.7254 & 0.4233 & 0.1469 & 0.6687 & 0.5761 & 0.3092 & 0.7808 \\
          & M-SpatPlus44 & 0.5439 & 0.2913 & 0.7417 & 0.4510 & 0.1761 & 0.6892 & 0.6068 & 0.3498 & 0.7997 \\
          & M-SpatPlus39 & 0.5588 & 0.3129 & 0.7501 & 0.4629 & 0.1801 & 0.7040 & 0.6226 & 0.3742 & 0.8097 \\
          \hline
    \end{tabular}}
  \label{SimuStudy2_correlations}
\end{table}%
Table \ref{SimuStudy2_correlations} displays the averages over the $300$ simulated data sets of the posterior median and 95\% credible intervals of the between-crime correlations in each scenario. Here, the data generating process assumes $\rho=0.7$ (Scenario 1), $\rho=0.3$ (Scenario 2), that is, high and low correlations between the spatial patterns of the crimes. For the high correlation scenario, all the M-models effectively capture $\rho$, but for low correlation, $\rho$ is overestimated. Indeed, in line with the Simulation Study 1 for the low correlation scenario, the estimates of $\rho$ increase in the M-SpatPlus models as more large-scale eigenvectors are removed. Additionaly, unlike in Simulation Study 1, the correlation parameter $\rho$ is significant as the credible intervals are not too wide. This is probably because the generating model is the same as the fitted model. Supplementary material B examines a scenario (Scenario 3) in which the data generating model assumes a moderate correlation $\rho=0.5$ between the spatial patterns. In this case, the estimated correlation coefficient $\rho$ is also overestimated (See Table B.12). In general, the overestimation of $\rho$ is larger with the BYM2 prior than with the ICAR and PCAR priors.

\begin{table}[htbp]
	\centering
	\caption{Posterior means and standard deviations of $\beta_1$ based on 300 simulated datasets for Simulation study 2 and Scenarios 1 and 2.}
    \resizebox{\textwidth}{!}{
	\begin{tabular}{llcllllll}
		\hline
        & Model & \text{True value} & \multicolumn{2}{l}{ICAR} & \multicolumn{2}{l}{PCAR} & \multicolumn{2}{l}{BYM2} \\
        \cmidrule(lr){4-5} \cmidrule(lr){6-7}\cmidrule(lr){8-9}
        & & & \multicolumn{1}{l}{Mean} & \multicolumn{1}{l}{SD} & \multicolumn{1}{l}{Mean} & \multicolumn{1}{l}{SD} & \multicolumn{1}{l}{Mean} & \multicolumn{1}{l}{SD}  \\
		\hline
		\multirow{7}[2]{*}{Scenario 1} & M-Spatial & \multirow{7}[2]{*}{0.1500} & 0.2701 & 0.0815 & 0.2766 & 0.0837 & 0.2760 & 0.0846 \\
        & M-SpatPlus64 & & 0.1878 & 0.0612 & 0.1871 & 0.0619 & 0.1895 & 0.0630 \\
        & M-SpatPlus59 & & 0.1914 & 0.0585 & 0.1915 & 0.0593 & 0.1961 & 0.0605 \\
        & M-SpatPlus54 & & 0.1709 & 0.0580 & 0.1705 & 0.0587 & 0.1731 & 0.0599 \\
        & M-SpatPlus49 & & 0.1580 & 0.0563 & 0.1579 & 0.0571 & 0.1600 & 0.0584 \\
        & M-SpatPlus44 & & 0.1405 & 0.0547 & 0.1407 & 0.0555 & 0.1425 & 0.0574 \\
        & M-SpatPlus39 & & 0.1260 & 0.0517 & 0.1260 & 0.0526 & 0.1289 & 0.0549 \\
		\hline
        \multirow{7}[2]{*}{Scenario 2} & M-Spatial & \multirow{7}[2]{*}{0.1500} & 0.2688 & 0.0685 & 0.2726 & 0.0712 & 0.2733 & 0.0708 \\
        & M-SpatPlus64 & & 0.2406 & 0.0656 & 0.2400 & 0.0681 & 0.2407 & 0.0675 \\
        & M-SpatPlus59 & & 0.2160 & 0.0640 & 0.2140 & 0.0665 & 0.2140 & 0.0656 \\
        & M-SpatPlus54 & & 0.2173 & 0.0625 & 0.2157 & 0.0650 & 0.2164 & 0.0642 \\
        & M-SpatPlus49 & & 0.1996 & 0.0612 & 0.1982 & 0.0636 & 0.1985 & 0.0631 \\
        & M-SpatPlus44 & & 0.1700 & 0.0599 & 0.1685 & 0.0624 & 0.1683 & 0.0624 \\
        & M-SpatPlus39 & & 0.1454 & 0.0567 & 0.1445 & 0.0593 & 0.1441 & 0.0599 \\
        \hline
	\end{tabular}}
    \label{SimuStudy2_mean_sd_beta1}
\end{table}

\begin{figure}[t]
\centering
\includegraphics[width=1\textwidth]{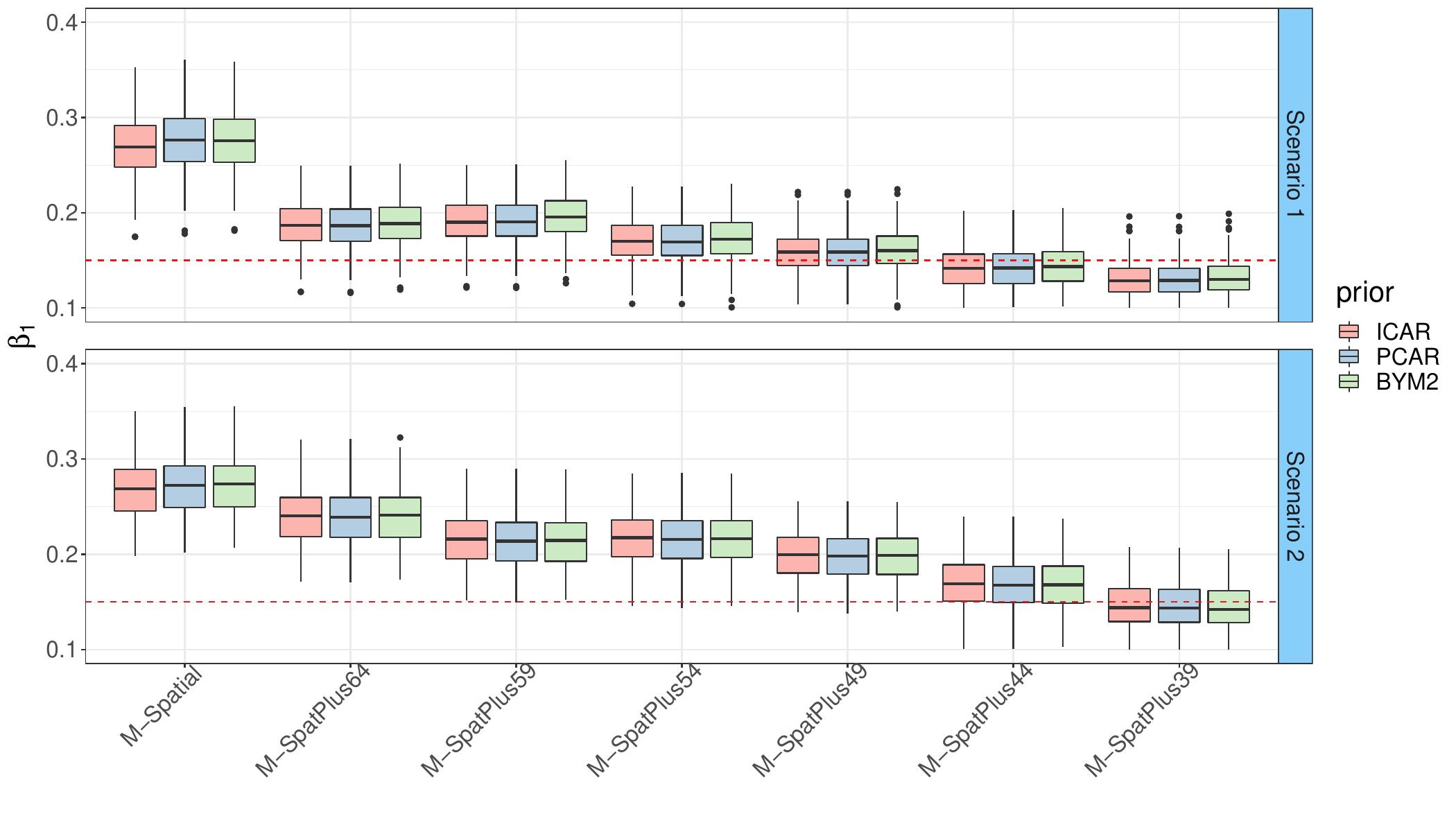}
\caption{Boxplots of the estimated means of $\beta_1$ based on 300 simulated datasets for Simulation Study 2, Scenario 1 (top row) and Scenario 2 (bottom row). Each color represents a different prior given to the columns of $\PPhi$, namely, red for ICAR, blue for PCAR and green for BYM2.}
\label{SimuStudy2_boxplots_rapes}
\end{figure}

\begin{figure}[htbp]
\centering
\includegraphics[width=1\textwidth]{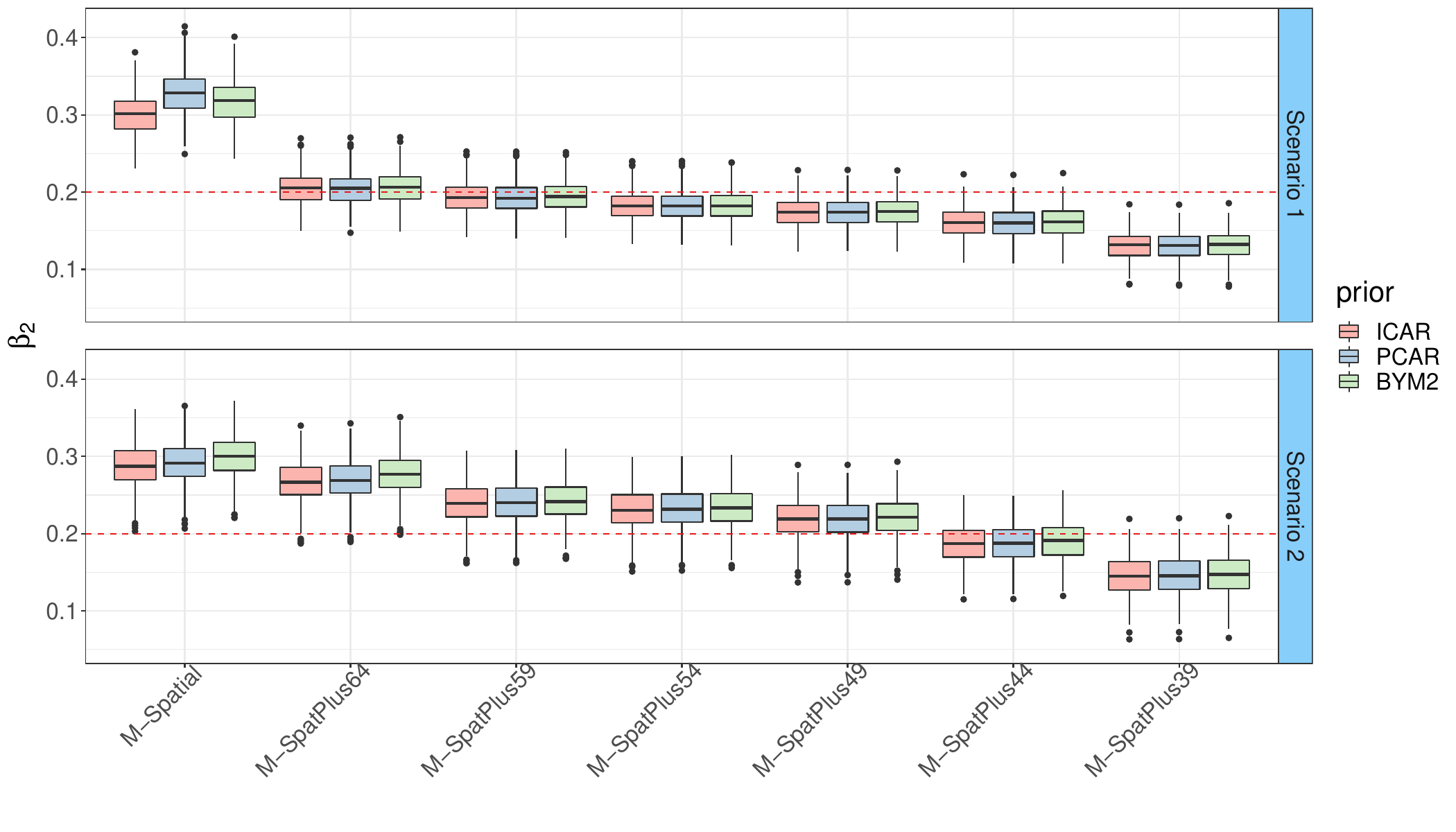}
\caption{Boxplots of the estimated means of $\beta_2$ based on 300 simulated datasets for Simulation Study 2, Scenario 1 (top row) and Scenario 2 (bottom row). Each color represents a different prior given to the columns of $\PPhi$, namely, red for ICAR, blue for PCAR and green for BYM2.}
\label{SimuStudy2_boxplots_dowry}
\end{figure}

\begin{table}[!htbp]
	\centering
	\caption{Posterior means and standard deviations of $\beta_2$ based on 300 simulated datasets for Simulation study 2 and Scenarios 1 and 2.}
    \resizebox{\textwidth}{!}{
	\begin{tabular}{llcllllll}
		\hline
        & Model & \text{True value} & \multicolumn{2}{l}{ICAR} & \multicolumn{2}{l}{PCAR} & \multicolumn{2}{l}{BYM2} \\
        \cmidrule(lr){4-5} \cmidrule(lr){6-7}\cmidrule(lr){8-9}
        & & & \multicolumn{1}{l}{Mean} & \multicolumn{1}{l}{SD} & \multicolumn{1}{l}{Mean} & \multicolumn{1}{l}{SD} & \multicolumn{1}{l}{Mean} & \multicolumn{1}{l}{SD}  \\
		\hline
		\multirow{7}[2]{*}{Scenario 1} & M-Spatial & \multirow{7}[2]{*}{0.2000} & 0.3018 & 0.0476 & 0.3284 & 0.0519 & 0.3179 & 0.0501 \\
        & M-SpatPlus64 & & 0.2055 & 0.0380 & 0.2051 & 0.0404 & 0.2069 & 0.0398 \\
        & M-SpatPlus59 & & 0.1938 & 0.0370 & 0.1935 & 0.0393 & 0.1947 & 0.0388 \\
        & M-SpatPlus54 & & 0.1829 & 0.0367 & 0.1829 & 0.0390 & 0.1830 & 0.0385 \\
        & M-SpatPlus49 & & 0.1741 & 0.0355 & 0.1739 & 0.0378 & 0.1745 & 0.0374 \\
        & M-SpatPlus44 & & 0.1605 & 0.0353 & 0.1599 & 0.0375 & 0.1613 & 0.0375 \\
        & M-SpatPlus39 & & 0.1304 & 0.0341 & 0.1302 & 0.0364 & 0.1312 & 0.0367 \\
		\hline
        \multirow{7}[2]{*}{Scenario 2} & M-Spatial & \multirow{7}[2]{*}{0.2000} & 0.2876 & 0.0381 & 0.2914 & 0.0393 & 0.2997 & 0.0399\\
        & M-SpatPlus64 & & 0.2674 & 0.0371 & 0.2694 & 0.0382 & 0.2771 & 0.0392 \\
        & M-SpatPlus59 & & 0.2394 & 0.0380 & 0.2403 & 0.0389 & 0.2423 & 0.0401 \\
        & M-SpatPlus54 & & 0.2314 & 0.0374 & 0.2323 & 0.0383 & 0.2337 & 0.0395 \\
        & M-SpatPlus49 & & 0.2189 & 0.0368 & 0.2190 & 0.0376 & 0.2213 & 0.0389 \\
        & M-SpatPlus44 & & 0.1876 & 0.0378 & 0.1880 & 0.0384 & 0.1905 & 0.0404 \\
        & M-SpatPlus39 & & 0.1454 & 0.0373 & 0.1458 & 0.0379 & 0.1472 & 0.0404 \\
        \hline
	\end{tabular}}
    \label{SimuStudy2_mean_sd_beta2}
\end{table}
Regarding the fixed effects,
Table \ref{SimuStudy2_mean_sd_beta1} and
Table \ref{SimuStudy2_mean_sd_beta2} show the average over the 300 simulated datasets of the posterior means and posterior standard deviations of the regression coefficients $\beta_1$ (crime 1) and $\beta_2$ (crime 2) in Scenarios 1 and 2 respectively. Figures \ref{SimuStudy2_boxplots_rapes} and \ref{SimuStudy2_boxplots_dowry} provide the boxplots of the posterior means of $\beta_1$ and $\beta_2$ over the 300 simulated data sets. Looking at these tables and figures, some differences can be observed compared to Simulation Study 1. Here, in the M-SpatPlus models, the number of eigenvectors that must be omitted from the observed covariate to recover the fixed effects does not appear to depend on the linear correlation between $\X^{*}_1$ and the spatial effects used in the data generating model, but rather on the spatial pattern of the covariate. The covariate and both spatial patterns of Scenario 1 have quite similar structure: the values of the districts in the northwest are low, and as we move to the middle southeast districts, the values gradually increase. Conversely, in Scenario 2 the structures are more complex. The spatial patterns show contrasting directions, and somehow a combination of both spatial patterns can be observed in $\X^{*}_1$. In the high correlation scenario (Scenario 1), removing 20 large-scale eigenvectors from the covariate seems enough to recover $\beta_1$ (see
Table \ref{SimuStudy2_mean_sd_beta1}) and between 5 to 15 eigenvectors for $\beta_2$ (see
Table \ref{SimuStudy2_mean_sd_beta2}).
For crime 2, fewer eigenvectors are likely sufficient, possibly due to the larger number of simulated counts.
In contrast, when examining the low correlation scenario (Scenario 2), the removal of 20 large-scale eigenvectors in the M-SpatPlus models leads to an overestimation of the fixed effects, particularly for crime 1 (see Table
\ref{SimuStudy2_mean_sd_beta1}). To address this issue, we have fitted the M-SpatPlus model removing more eigenvectors from the covariate. Specifically, we have remove 25 and 30 large-scale eigenvectors (M-SpatPlus44 and M-SpatPlus39 respectively in the tables). Interestingly, when 30 large-scale eigenvectors were removed, the M-SpatPlus estimates of $\beta_1$ are satisfactory. There are no considerable differences between the ICAR, PCAR and BYM2 priors in terms of the fixed effect estimates.

Similarly to Simulation Study 1, we have compared the true simulated standard error ($s.e._{sim}$) and the estimated standard error ($s.e._{est}$) for $\beta_1$ and $\beta_2$ (see Tables B.15 and B.16 of the Supplementary material B) and we provide the empirical coverage of credible intervals at 95\% nominal value for $\beta_1$ and $\beta_2$ (Table \ref{SimuStudy2_95coverage}). The overestimation of the standard deviations of $\beta_1$ is quite apparent (see Table B.15). The estimated standard deviation is approximately three times the simulated estandard error. This overestimation is reflected in the 95\% coverage probabilities of M-SpatPlus, where nearly all the methods lead to an overcoverage. The reason is that the M-SpatialPlus models reduce bias, but still the variance seems inflated, hence the excess of coverage. In the case of $\beta_2$ the overestimation of the standard deviations is lower (see Table B.16) and more reasonable 95\% coverage probabilities are obtained for the M-SpatPlus models. As in Simulation Study 1, the empirical coverage of $\beta_1$ and $\beta_2$ with the usual multivariate M-Spatial model is low because the overestimation of the standard deviation does not compensate for the large bias.

To finish, we examine model selection criteria and MARB and MRRMSE of the relative risks estimates.
Table \ref{SimuStudy2_DIC_WAIC} provides the DIC and WAIC for Scenarios 1 and 2. Again, model selection criteria do not effectively differentiate between the models and spatial priors. Table B.19 provides MARB and MRRMSE of the relative risks. All the M-models have similar bias and error concerning the relative risks estimates.

\begin{table}[htbp]
	\centering
	\caption{Empirical $95\%$ coverage probabilities of the true value of $\beta_{1}$ and $\beta_{2}$ based on $300$ simulated datasets for Simulation study 2 and Scenarios 1 and 2.}
    \resizebox{\textwidth}{!}{
	\begin{tabular}{llllllll}
        \hline
		& Model & \multicolumn{3}{l}{$\beta_{1}$} & \multicolumn{3}{l}{$\beta_{2}$} \\
		\cmidrule(lr){3-5} \cmidrule(lr){6-8}
		&  & \multicolumn{1}{l}{ICAR} & \multicolumn{1}{l}{PCAR} & \multicolumn{1}{l}{BYM2} & \multicolumn{1}{l}{ICAR} & \multicolumn{1}{l}{PCAR} & \multicolumn{1}{l}{BYM2}  \\
		\hline
		\multirow{7}[2]{*}{Scenario 1} & M-Spatial & 86.0000 & 84.6667 & 85.0000 & 40.3333 & 19.6667 & 25.3333 \\
        & M-SpatPlus64 & 100.0000 & 100.0000 & 100.0000 & 100.0000 & 100.0000 & 100.0000 \\
        & M-SpatPlus59 & 100.0000 & 100.0000 & 100.0000 & 100.0000 & 100.0000 & 100.0000 \\
        & M-SpatPlus54 & 100.0000 & 100.0000 & 100.0000 & 100.0000 & 100.0000 & 100.0000 \\
        & M-SpatPlus49 & 100.0000 & 100.0000 & 100.0000 & 99.3333 & 99.3333 & 99.3333 \\
        & M-SpatPlus44 & 100.0000 & 100.0000 & 100.0000 & 92.3333 & 95.3333 & 95.6667 \\
        & M-SpatPlus39 & 100.0000 & 100.0000 & 100.0000 & 47.0000 & 57.6667 & 61.0000 \\
		\hline
        \multirow{7}[2]{*}{Scenario 2} & M-Spatial & 70.0000 & 70.6667 & 69.3333 & 31.6667 & 28.0000 & 20.6667 \\
        & M-SpatPlus64 & 91.0000 & 92.6667 & 92.6667 & 57.3333 & 56.6667 & 51.0000 \\
        & M-SpatPlus59 & 98.6667 & 99.0000 & 99.0000 & 91.6667 & 91.3333 & 92.3333 \\
        & M-SpatPlus54 & 98.3333 & 99.0000 & 99.0000 & 95.0000 & 95.0000 & 95.3333 \\
        & M-SpatPlus49 & 100.0000 & 100.0000 & 100.0000 & 98.3333 & 98.6667 & 98.6667 \\
        & M-SpatPlus44 & 100.0000 & 100.0000 & 100.0000 & 99.0000 & 99.3333 & 99.6667 \\
		& M-SpatPlus39 & 100.0000 & 100.0000 & 100.0000 & 76.3333 & 77.6667 & 87.0000 \\		
        \hline
	\end{tabular}}
	\label{SimuStudy2_95coverage}
\end{table}

\begin{table}[t]
	\centering
	\caption{DIC and WAIC based on 300 simulated data sets for Simulation Study 2 and Scenarios 1 and 2.}
    \resizebox{\textwidth}{!}{
	\begin{tabular}{lllllllll}
		\hline
		& Model & \multicolumn{2}{l}{ICAR} & \multicolumn{2}{l}{PCAR} & \multicolumn{2}{l}{BYM2} \\
        \cmidrule(lr){3-4} \cmidrule(lr){5-6}\cmidrule(lr){7-8}
        &  & \multicolumn{1}{l}{DIC} & \multicolumn{1}{l}{WAIC} & \multicolumn{1}{l}{DIC} & \multicolumn{1}{l}{WAIC} & \multicolumn{1}{l}{DIC} & \multicolumn{1}{l}{WAIC} \\
        \hline
        \multirow{7}[2]{*}{Scenario 1} & M-Spatial & 966.1754 & 951.4779 & 967.5752 & 950.1012 & 967.0064 & 948.7606\\
        & M-SpatPlus64 & 966.8528 & 949.4872 & 967.9400 & 948.4048 & 967.9372 & 947.5738 \\
        & M-SpatPlus59 & 967.6719 & 950.0178 & 968.8122 & 949.0177 & 968.8697 & 948.4177 \\
        & M-SpatPlus54 & 967.8177 & 949.9382 & 968.9508 & 948.8622 & 969.1185 & 948.4952 \\
        & M-SpatPlus49 & 968.4432 & 950.9564 & 969.5225 & 949.7588 & 969.7227 & 949.5140 \\
        & M-SpatPlus44 & 969.1746 & 951.9225 & 970.1753 & 950.5237 & 970.4424 & 950.4116 \\
        & M-SpatPlus39 & 971.8599 & 955.0197 & 972.7043 & 953.2348 & 973.0639 & 953.3253  \\
        \hline
        \multirow{7}[2]{*}{Scenario 2 } & M-Spatial & 969.8757 & 952.8217 & 970.7773 & 949.9615 & 970.5796 & 949.5215  \\
        & M-SpatPlus64 & 969.9933 & 952.1133 & 970.6760 & 949.0728 & 970.6719 & 948.8739 \\
        & M-SpatPlus59 & 971.2157 & 951.1453 & 972.0652 & 948.4785 & 972.2841 & 948.9999 \\
        & M-SpatPlus54 & 971.8931 & 951.8211 & 972.7483 & 949.2491 & 973.0684 & 949.9883 \\
        & M-SpatPlus49 & 973.0603 & 953.3003 & 973.8526 & 950.6847 & 974.2231 & 951.4681 \\
        & M-SpatPlus44 & 975.0822 & 954.8057 & 975.8658 & 952.0949 & 976.2864 & 952.9832 \\
        & M-SpatPlus39 & 978.8893 & 959.4039 & 979.5481 & 956.4251 & 979.9821 & 957.2993  \\
		\hline	
	\end{tabular}}
    \label{SimuStudy2_DIC_WAIC}
\end{table}

%%%%%%%%%%%%%%%%
%% DISCUSSION %%
%%%%%%%%%%%%%%%%
\clearpage
\section{Discussion}
The inclusion of the covariates in spatial models, whether they are univariate or multivariate, faces the well known problem of spatial confounding, that is, the impossibility of disentangling the fixed effects and the spatial random effects. A correct estimation of the fixed effects is crucial to gain knowledge about complex diseases like cancer or intricate socio-demographic phenomena, such as crimes against women in India.

To deal with spatial confounding, different methods have been proposed in the literature, but we focus on the spatial+ approach as it has proven to work well at least in univariate spatial models. In this paper we propose a modified and simpler version of the method that avoids fitting a spatial model to the covariate. Here, the covariate is expressed as a linear combination of eigenvectors of the spatial precision matrix, and from this linear combination we discard a pre-determined number of large-scale eigenvectors which we assume to be the spatially confounded part of the covariate. With this modified spatial+ approach we avoid fitting a model to the covariate and we directly consider a \lq\lq spatially decorrelated'' version of it. The method can be used in univariate as well as in multivariate models, but in this work we focus on the latter. The method is simple and flexible, allowing for the removal of spatial structure from the covariate differently for each of the response variables.

To illustrate the proposal, two crimes against women, namely rapes and dowry deaths, are analysed jointly in the 70 districts of Uttar Pradesh in 2011. In particular, we have focused on the linear association between the covariate sex ratio and the crimes. All the multivariate models indicate that there is no significant linear association between sex ratio and rapes, whereas for dowry deaths the strength of the association depends on the number of removed eigenvectors from the covariate. Additionally, all the models estimate a rather low positive correlation between rapes and dowry deaths that is on the verge of statistical significance. Our proposal requires consideration about the number of eigenvectors to remove from the linear combination of the covariate. To provide some guidelines, we have conducted two simulation studies. In Simulation Study 1, spatial confounding is induced including two covariates in the data generating model for each crime. Namely, the sex ratio ($\X_1$) for both crimes, $\X_2$ for crime 1 and $\X_3$ for crime 2. The covariates $\X_2$ and $\X_3$ play the role of unobserved covariates (they are replaced by spatial random effects in the fit) and are correlated, hence they induce dependence between crimes. In Simulation study 2, we have generated the spatial effects $\ttheta_1$ and $\ttheta_2$ for crimes 1 and 2 respectively from a multivariate ICAR prior with fixed correlation parameter. These spatial effects are responsible for causing correlation between crimes. Then, the observed covariate $\X^{*}_{1}$, is simulated to be correlated with both the spatial effects, $\ttheta_1$ and $\ttheta_2$, and different scenarios are simulated depending on the correlations between the covariates and the spatial effects. Overall, the findings of the simulation studies are interesting. The M-models with the original covariate without the spatial+ method do not recover the true fixed effects in any of the simulated scenarios. This is particularly striking in Scenario 2 where the multivariate ICAR is the data generating model. On the contrary, M-models with the spatial+ method provide fairly unbiased estimates and the estimates do not depend on the prior given to the spatial random effects. The key lies in determining beforehand the amount of spatial dependence that we should remove from the covariate. In Simulation Study 1, where the observed covariate (sex ratio) has a clear and gradual spatial pattern, the greater the correlation between the observed and the unobserved covariates, the larger the number of eigenvectors to remove. To be more specific, we should remove between 7\% and 20\% of the eigenvectors corresponding to the lowest eigenvalues. It is worth noting that in crime 1 we have to remove more eigenvectors than in crime 2, probably because the number of counts is smaller.

In Simulation study 2, results are revealing. It seems that the number of
eigenvectors to remove depends more on the spatial dependence of the observed
covariate than on the correlation between the observed and unobserved parts of
the model. In scenario 1, where the spatial pattern of the simulated observed
covariate is more or less smooth and the correlation with the unobserved part
is moderate-high, we need to remove about $28\%$ of the eigenvectors in crime 1
and between $7\%$ and $14\%$ of the eigenvectors for crime 2.  In scenario 2, where
the spatial pattern of the covariate is not so smooth and the correlation with
the unobserved part is low-moderate, we need to remove between $35\%$ and $43\%$ of
the eigenvectors (this seems to be in line with recent results about
confounding at high frequencies by~\citealp{dupont2023}). As in Simulation Study
1, we need to remove more eigenvectors in crime 1 than in crime 2.

To sum up, if the spatial pattern of the observed covariate is smooth, removing $7\%$ to $20\%$ of the eigenvectors should be a good choice. If the spatial pattern of the covariate is not so smooth, then we should remove $35\%$ to $43\%$ eigenvectors. We find these guidelines useful as they depend on the observed covariate rather than on the unobserved ones.

In terms of the between-crime correlation estimates, all models recover quite well high correlations, though low-moderate correlations might be overestimated. Moreover, the overestimation is larger if we remove more eigenvectors than needed. Additionally, the overestimation is larger with the BYM2 spatial prior.

The multivariate models employing the spatial+ approach address bias in the estimates of fixed effects. However, there seems to be an issue with inflated standard errors, leading to   overcoverage of the credible intervals. Further research is necessary to adequately adjust the standard errors. Additionally, model selection criteria such as DIC or WAIC do not differentiate between the models. This is somewhat expected, considering that the spatial+ approach can be seen as a reparameterized spatial model.

Finally, it is worth noting that the proposed modified spatial+ approach is valid for univariate and multivariate models. In general we do not expect big differences in the estimation of the fixed effects in univariate and multivariate models if a true relationship between the responses and the covariate exists.  However, we think that in complex phenomena such as crimes against women (or cancer, where risk factors only explain a small percentage of cases) multivariate models are more convenient to deal with the problem as a whole and not with each part
individually.

%%%%%%%%%%%%%%%%%%%%%
%% Aknowledgements %%
%%%%%%%%%%%%%%%%%%%%%
\section*{Declaration of competing interest}
The authors declare that they have no known competing financial interests or personal relationship that could have appeared to influence the work reported in this paper.

\section*{Acknowledgements}
This work has been supported
by Project PID2020-113125RB-I00/MCIN/AEI/10.13039/501100011033. We extend our gratitude to the Associate Editor and the reviewer for their insightful comments, which greatly contributed to enhancing the final version of this article.

\clearpage
%%---------------------------------------------------------------------------------------
%% APPENDIX
%%---------------------------------------------------------------------------------------
\setcounter{section}{0} %\setcounter{equation}{0}
\renewcommand{\thesection}{\Alph{section}}

\setcounter{figure}{0}
\renewcommand\thefigure{\thesection.\arabic{figure}}

\setcounter{table}{0}
\renewcommand\thetable{\thesection.\arabic{table}}

\setcounter{equation}{0}
\renewcommand\theequation{\thesection.\arabic{equation}}

\section{Appendix}

\subsection{Appendix A}\label{appenA}
In this Appendix we show that choosing the Cholesky square root ${\tilde{\SSigma}_b}$ of the between-crime covariance matrix ${\SSigma}_b$ has effects that are not independent of the column ordering of $\boldsymbol{\Theta}$, that is, the crime ordering in the analysis.

Consider $J=2$, the real case study in this work.  If $\M=\left(
	\begin{array}{cc}
		m_1 & m_2 \\
		m_3 & m_4 \\
	\end{array}
	\right)
	$ is any nonsingular matrix that satisfies the condition $\SSigma_{b}=\boldsymbol{M}^{'}\boldsymbol{M}$,  then the corresponding covariance matrix of $\ttheta=(\ttheta_1^{'},\ttheta_{2}^{'})^{'}$ is

\begin{eqnarray}\label{eq1_ap}
	\text{Cov}({\ttheta}) &=& (\boldsymbol{M} \otimes \boldsymbol{I}_{n})'Blockdiag(\boldsymbol{\SSigma}_{1},  \boldsymbol{\SSigma}_{2})(\boldsymbol{M} \otimes \boldsymbol{I}_{n}) \\\nonumber
	&=& \left(
	\begin{array}{cc}
		m_1^2{\SSigma}_{1}+m_3^2{\SSigma}_{2} & m_1m_2{\SSigma}_{1}+m_3m_4{\SSigma}_{2} \\
		m_1m_2{\SSigma}_{1}+m_3m_4{\SSigma}_{2} & m_2^2{\SSigma}_{1}+m_4^2{\SSigma}_{2} \\
	\end{array}
	\right).
	\end{eqnarray}
If we want to choose $\M$ to be the Cholesky square root, we need to impose the restriction $m_3=0$, and then ${\tilde{\SSigma}_b}={\M}'$. However, the effect of the restriction is not independent of the crime ordering because for $\ttheta=(\ttheta_1^{\prime}, \ttheta_2^{\prime})^{\prime}$, the restriction implies cross-covariances that only depend on $\SSigma_1$, whereas if we interchange the order, that is, $\ttheta^{*}=(\ttheta_2^{\prime}, \ttheta_1^{\prime})^{\prime}$ cross-covariances depend only on $\SSigma_2$. To avoid this dependence on the crime ordering,
rather than defining $\M$ as the Cholesky square root of $\SSigma_{b}$, we allow it to vary more freely although we use the Bartlett decomposition to avoid overparameterization as outlined in Section 4.

\subsection{Appendix B}\label{appenB}
Here we show that the precision parameters $\tau_j=1/\sigma^2_j$ are fixed at 1 (or $\sigma_j=1$) for identifiability issues. We remark that for identifiability issues we refer to the situation where there are several unknown quantities but it is impossible to determine each quantity separately because different combinations lead to the same result, that is, the estimation problem does not have a unique solution. Here, given a covariance structure, $\text{Cov}({\ttheta})$, the quantities ${\SSigma}_b$, $\sigma_1$, and $\sigma_2$ are not identifiable.

In the case of separable covariance structures, the reason to fix the precision parameter $\tau$ (or $\sigma^2=1/\tau$) equal to 1 is clear. The within-crime covariance matrix ${\SSigma}_w$ (the inverse of the spatial precision matrix) is the same for all crimes and the global covariance structure is
\[
\text{Cov}({\ttheta})={\SSigma}_b\otimes {\SSigma}_w.
\]
Then, a change in the scale of ${\SSigma}_w$ can be compensated for by an appropriate change in the scale of ${\SSigma}_b$.

In the case of non-separable covariance structures, suppose
\begin{eqnarray}\label{eq2_ap}
	\text{Cov}({\ttheta}) &=& (\boldsymbol{M} \otimes \boldsymbol{I}_{n})'Blockdiag(\sigma^2_1\boldsymbol{\SSigma}_{1},  \sigma^2_2\boldsymbol{\SSigma}_{2})(\boldsymbol{M} \otimes \boldsymbol{I}_{n}) \\\nonumber
	&=& \left(
	\begin{array}{cc}
		m_1^{2}\sigma^2_1{\SSigma}_{1}+m_3^{2}\sigma^2_2{\SSigma}_{2} & m_1m_2\sigma^2_1{\SSigma}_{1}+m_3m_4\sigma^2_2{\SSigma}_{2} \\
		m_1m_2\sigma^2_1{\SSigma}_{1}+m_3m_4\sigma^2_2{\SSigma}_{2} & m_2^{2}\sigma^2_1{\SSigma}_{1}+m_4^{2}\sigma^2_2{\SSigma}_{2} \\
	\end{array}
	\right).
\end{eqnarray}
for some $\sigma_1$, $\sigma_2$ and $\SSigma_{b}=\boldsymbol{M}^{'}\boldsymbol{M}$, with $\M=\left(
	\begin{array}{cc}
		m_1 & m_2 \\
		m_3 & m_4 \\
	\end{array}
	\right)
	$.
Then, Eq.~\eqref{eq2_ap} does not have a unique solution. If $\sigma_1$, $\sigma_2$ and $\SSigma_{b}=\boldsymbol{M}^{\prime}\boldsymbol{M}$ is one solution, then so is any combination $\breve{\sigma}_1$, $\breve{\sigma}_2$ and $\breve{\SSigma}_{b}=\breve{\boldsymbol{M}}^{\prime}\breve{\boldsymbol{M}}$ for which $\breve{m}_1 \breve{\sigma}_1=m_1 \sigma_1$, $\breve{m}_2 \breve{\sigma}_1=m_2 \sigma_1$, $\breve{m}_3 \breve{\sigma}_2=m_3 \sigma_2$, and $\breve{m}_4 \breve{\sigma}_2=m_4 \sigma_2$. One such example is the choice $\breve{\sigma}_{i}=\sigma_i^2$ for $i=1,2$ and $\breve{m}_1=m_1/\sigma_1$, $\breve{m}_2=m_2/\sigma_1$, $\breve{m}_3=m_3/\sigma_2$ and $\breve{m}_4=m_4/\sigma_2$.

However, if we fix $\sigma_1=\sigma_2=1$, then $\SSigma_{b}$ is identifiable as
\begin{eqnarray*}
	\text{Cov}({\ttheta}) &=& \left(
	\begin{array}{cc}
		m_1^{2}{\SSigma}_{1}+m_3^{2}{\SSigma}_{2} & m_1m_2{\SSigma}_{1}+m_3m_4{\SSigma}_{2} \\
		m_1m_2{\SSigma}_{1}+m_3m_4{\SSigma}_{2} & m_2^{2}{\SSigma}_{1}+m_4^{2}{\SSigma}_{2} \\
	\end{array}
	\right)
\end{eqnarray*}
uniquely determines
$$\SSigma_{b}=\M'\M=\left(
                      \begin{array}{cc}
                        m_1^2+m_3^2 & m_1m_2+m_3m_4 \\
                        m_1m_2+m_3m_4 & m_2^2+m_4^2 \\
                      \end{array}
                    \right).
$$
Note that setting $\sigma_1=\sigma_2$ (i.e.~not necessarily equal to 1) only identifies $\SSigma_{b}$ up to scaling. That is, the overall scaling can be done by scaling $\sigma_1=\sigma_2$ or scaling $\SSigma_{b}$. This implies that fixing $\sigma_1=\sigma_2=1$ is not so relevant as looking at Equations (\ref{eq1_ap}) and (\ref{eq2_ap}), where the diagonal elements represent the spatial dependence within each crime, the overall smoothing can be controlled by $\sigma_j$  or by the elements of the matrix ${\M}$. Finally, we remark that the matrix ${\M}$ is not identifiable (even if $\sigma_1=\sigma_2=1$) because there are several choices of square roots for the matrix $\SSigma_{b}$.

%%---------------------------------------------------------------------------------------
%% REFERENCES
%%---------------------------------------------------------------------------------------
\clearpage
\bibliographystyle{apalike-ejor}
\bibliography{Urdangarin_et_al_2024_arXiv}
\nocite{}

\clearpage
%%---------------------------------------------------------------------------------------
%% APPENDIX
%%---------------------------------------------------------------------------------------
\setcounter{section}{0} %\setcounter{equation}{0}
\renewcommand{\thesection}{\Alph{section}}

\setcounter{figure}{0}
\renewcommand\thefigure{\thesection.\arabic{figure}}

\setcounter{table}{0}
\renewcommand\thetable{\thesection.\arabic{table}}

%%---------------------------------------------------------------------------------------
%% SUPPLEMENTARY MATERIAL: SIMULATION STUDY 1
%%---------------------------------------------------------------------------------------
\section{Supplementary material: Simulation study 1}
This section includes some additional scenarios to complement the analysis of Simulation study 1. The data is generated following the generating models (11)-(12) with $\aalpha=(\alpha_1, \alpha_2)^{'}=(-0.12, -0.03)^{'}$, $\bbeta=(\beta_1, \beta_2)^{'}=(-0.15, -0.20)^{'}$, $\bbeta^{*}=(\beta^{*}_{1}, \beta^{*}_{2})^{'}=(-0.30, -0.30)^{'}$ and $\X_1$ is the standardized sex ratio covariate. Below we define the correlations between $\X_1$, $\X_2$ and $\X_3$ chosen in each scenario:
\begin{itemize}
\item \textbf{Scenario 3:} $cor(\X_1, \X_2)=0.5$, $cor(\X_1, \X_3)=0.7$ and $cor(\X_2, \X_3)=0.5$ are chosen. This is similar to Scenario 1 but choosing lower correlation between $\X_2$ and $\X_3$.
\item \textbf{Scenario 4:} $cor(\X_1, \X_2)=0.3$, $cor(\X_1, \X_3)=0.7$ and $cor(\X_2, \X_3)=0.3$ are chosen.
\item \textbf{Scenario 5:} $cor(\X_1, \X_2)=0.0$, $cor(\X_1, \X_3)=0.7$ and $cor(\X_2, \X_3)=0.7$ are chosen.
\item \textbf{Scenario 6:} $cor(\X_1, \X_2)=0.0$, $cor(\X_1, \X_3)=0.7$ and $cor(\X_2, \X_3)=0.3$ are chosen.
\end{itemize}

\begin{figure}[h]
\centering
\includegraphics[width=0.8\textwidth]{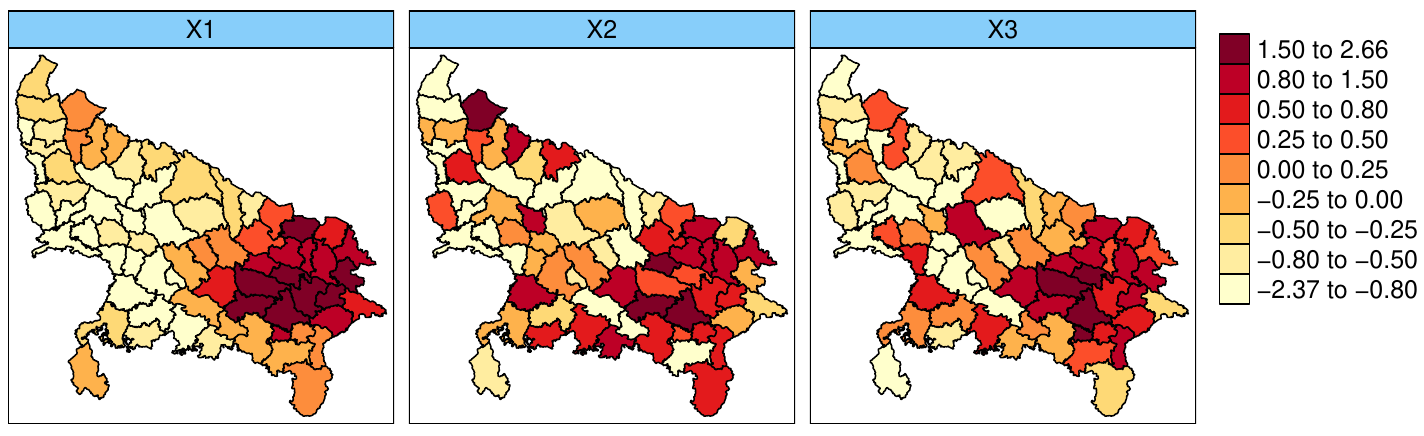}
\includegraphics[width=0.8\textwidth]{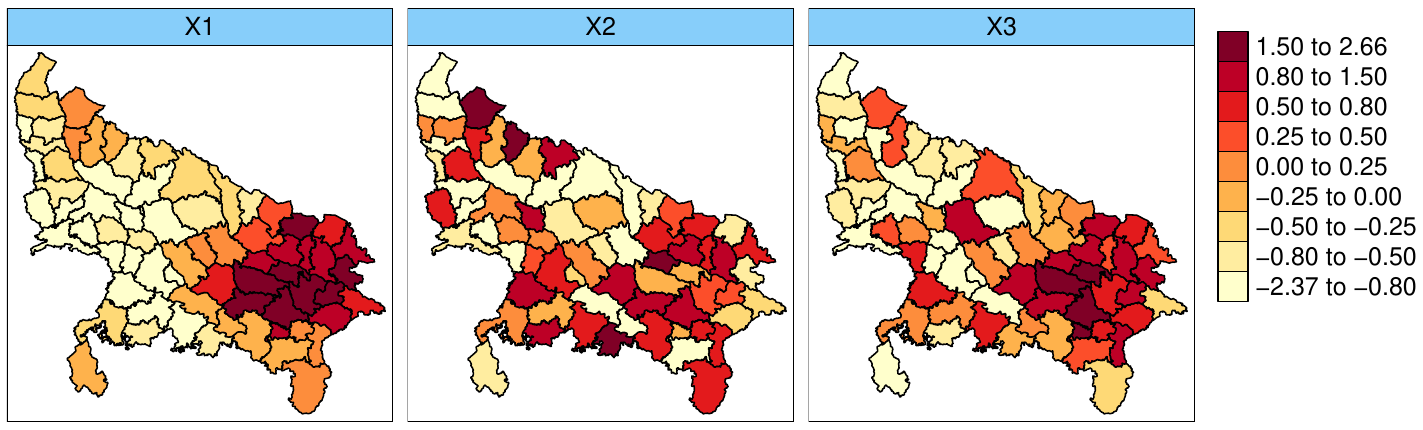}
\includegraphics[width=0.8\textwidth]{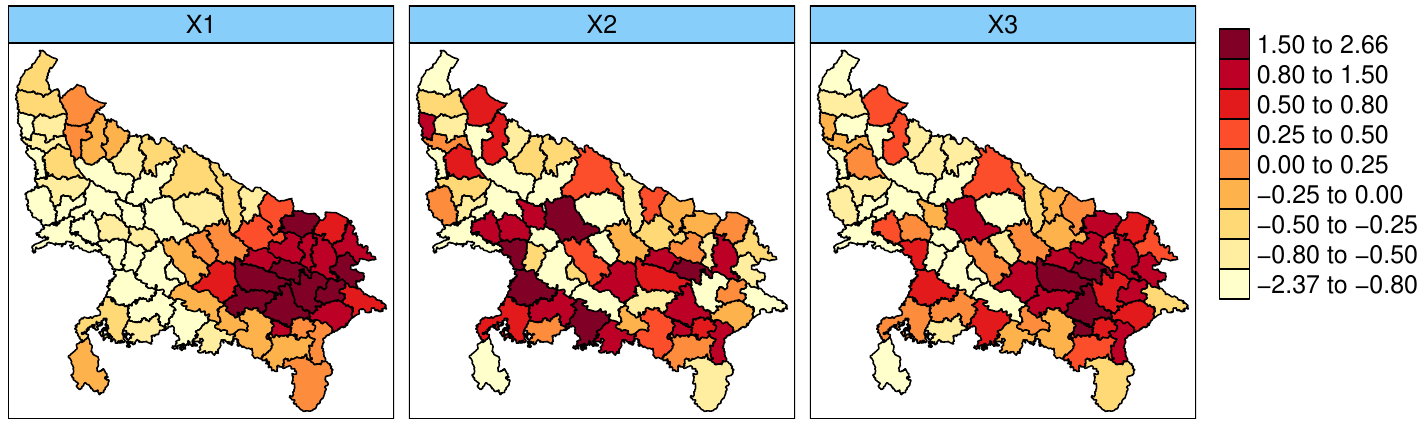}
\includegraphics[width=0.8\textwidth]{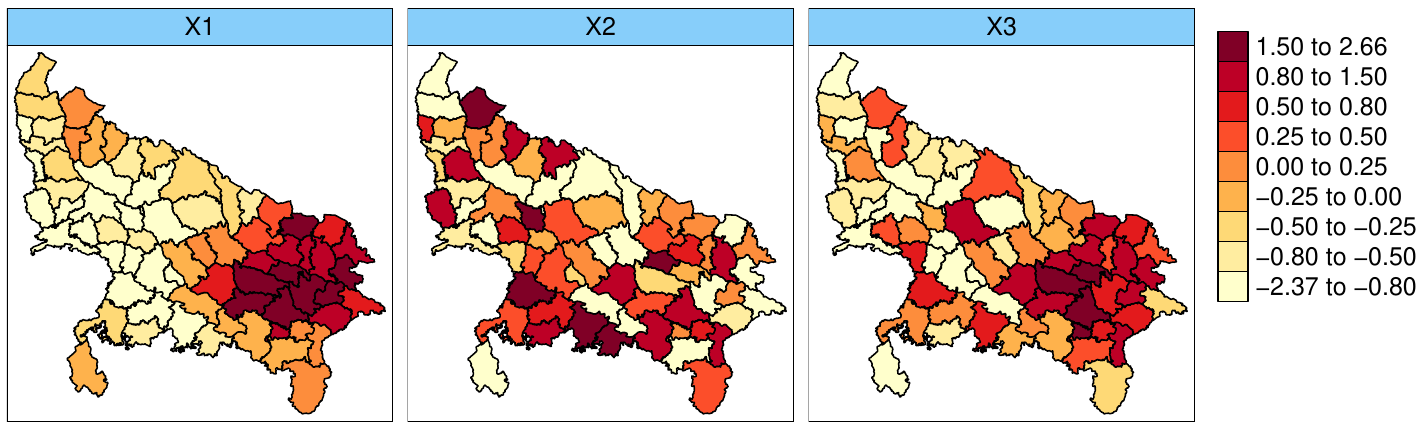}
\caption{From left to right, standardized sex ratio covariate in 2011, standardized simulated covariate $\X_2$ and standardized simulated covariate $\X_3$ in Simulation Study 1. The first row corresponds to Scenario 3, the second to Scenario 4, the third to Scenario 5 and the last row to Scenario 6.}
\label{fig:SimuStudy1_sup}
\end{figure}

\clearpage
\begin{table}[ht]
	\centering
	\caption{Posterior means and standard deviations of $\beta_1$ based on 300 simulated datasets for Simulation study 1 and Scenarios 3, 4, 5 and 6.}
	\begin{tabular}{lcccccccc}
		\hline
		& & & \multicolumn{2}{c}{ICAR} & \multicolumn{2}{c}{PCAR} & \multicolumn{2}{c}{BYM2} \\
        \hline
        & Model & True value & Mean & SD & Mean & SD & Mean & SD  \\
		\hline
		\multirow{5}[2]{*}{Scenario 3} & M-Spatial & \multirow{5}[2]{*}{-0.1500} & -0.3207 & 0.0684 & -0.3169 & 0.0457 & -0.3130 & 0.0552 \\
        & M-SpatPlus64 & & -0.1988 & 0.0456 & -0.2002 & 0.0423 & -0.2034 & 0.0432 \\
        & M-SpatPlus59 & & -0.1915 & 0.0430 & -0.1951 & 0.0406 & -0.1995 & 0.0417 \\
        & M-SpatPlus54 & & -0.1432 & 0.0416 & -0.1434 & 0.0403 & -0.1439 & 0.0422 \\
        & M-SpatPlus49 & & -0.1288 & 0.0407 & -0.1297 & 0.0399 & -0.1301 & 0.0421 \\
		\hline
        \multirow{5}[2]{*}{Scenario 4} & M-Spatial & \multirow{5}[2]{*}{-0.1500} & -0.2721 & 0.0734 & -0.2604 & 0.0476 & -0.2596 & 0.0583 \\
        & M-SpatPlus64 & & -0.1700 & 0.0482 & -0.1715 & 0.0438 & -0.1742 & 0.0446 \\
        & M-SpatPlus59 & & -0.1653 & 0.0454 & -0.1694 & 0.0418 & -0.1740 & 0.0432 \\
        & M-SpatPlus54 & & -0.1225 & 0.0430 & -0.1223 & 0.0407 & -0.1222 & 0.0429 \\
        & M-SpatPlus49 & & -0.1117 & 0.0418 & -0.1126 & 0.0401 & -0.1125 & 0.0427 \\
        \hline
        \multirow{5}[2]{*}{Scenario 5} & M-Spatial & \multirow{5}[2]{*}{-0.1500} & -0.1067 & 0.0773 & -0.1598 & 0.0487 & -0.1361 & 0.0567 \\
        & M-SpatPlus64 & & -0.0657 & 0.0489 & -0.0711 & 0.0488 & -0.0772 & 0.0463 \\
        & M-SpatPlus59 & & -0.0681 & 0.0461 & -0.0742 & 0.0464 & -0.0853 & 0.0448 \\
        & M-SpatPlus54 & & -0.0386 & 0.0421 & -0.0408 & 0.0423 & -0.0447 & 0.0433 \\
        & M-SpatPlus49 & & -0.0255 & 0.0407 & -0.0265 & 0.0408 & -0.0280 & 0.0428 \\
        \hline
        \multirow{5}[2]{*}{Scenario 6} & M-Spatial & \multirow{5}[2]{*}{-0.1500} & -0.1630 & 0.0737 & -0.1711 & 0.0482 & -0.1614 & 0.0580 \\
        & M-SpatPlus64 & & -0.1046 & 0.0476 & -0.1091 & 0.0445 & -0.1130 & 0.0453 \\
        & M-SpatPlus59 & & -0.1078 & 0.0447 & -0.1165 & 0.0422 & -0.1232 & 0.0437 \\
        & M-SpatPlus54 & & -0.0697 & 0.0417 & -0.0711 & 0.0401 & -0.0725 & 0.0426 \\
        & M-SpatPlus49 & & -0.0609 & 0.0404 & -0.0624 & 0.0390 & -0.0630 & 0.0421 \\
        \hline
	\end{tabular}
\end{table}

\begin{table}[ht]
	\centering
	\caption{Posterior means and standard deviations of $\beta_2$ based on 300 simulated datasets for Simulation study 1 and Scenarios 3, 4, 5 and 6.}
	\begin{tabular}{lcccccccc}
		\hline
		&  & & \multicolumn{2}{c}{ICAR} & \multicolumn{2}{c}{PCAR} & \multicolumn{2}{c}{BYM2} \\
        \hline
        & Model & \text{True value} & Mean & SD & Mean & SD & Mean & SD  \\
		\hline
		\multirow{5}[2]{*}{Scenario 3} & M-Spatial & \multirow{5}[2]{*}{-0.2000} & -0.3863 & 0.0593 & -0.4138 & 0.0401 & -0.4029 & 0.0454 \\
        & M-SpatPlus64 & & -0.2299 & 0.0424 & -0.2290 & 0.0442 & -0.2309 & 0.0435 \\
        & M-SpatPlus59 & & -0.2137 & 0.0403 & -0.2112 & 0.0421 & -0.2162 & 0.0420 \\
        & M-SpatPlus54 & & -0.1693 & 0.0392 & -0.1687 & 0.0417 & -0.1706 & 0.0422 \\
        & M-SpatPlus49 & & -0.1446 & 0.0389 & -0.1433 & 0.0414 & -0.1442 & 0.0424 \\
		\hline
        \multirow{5}[2]{*}{Scenario 4} & M-Spatial & \multirow{5}[2]{*}{-0.2000} & -0.3819 & 0.0597 & -0.4113 & 0.0403 & -0.4001 & 0.0453 \\
        & M-SpatPlus64 & & -0.2272 & 0.0426 & -0.2267 & 0.0452 & -0.2287 & 0.0440 \\
        & M-SpatPlus59 & & -0.2121 & 0.0405 & -0.2101 & 0.0430 & -0.2155 & 0.0425 \\
        & M-SpatPlus54 & & -0.1669 & 0.0394 & -0.1664 & 0.0423 & -0.1685 & 0.0432 \\
        & M-SpatPlus49 & & -0.1419 & 0.0391 & -0.1405 & 0.0420 & -0.1415 & 0.0434 \\
        \hline
         \multirow{5}[2]{*}{Scenario 5} & M-Spatial & \multirow{5}[2]{*}{-0.2000} & -0.3730 & 0.0607 & -0.4119 & 0.0402 & -0.3939 & 0.0455 \\
        & M-SpatPlus64 & & -0.2206 & 0.0454 & -0.2240 & 0.0477 & -0.2282 & 0.0440 \\
        & M-SpatPlus59 & & -0.2033 & 0.0430 & -0.2066 & 0.0456 & -0.2156 & 0.0426 \\
        & M-SpatPlus54 & & -0.1614 & 0.0415 & -0.1618 & 0.0443 & -0.1660 & 0.0437 \\
        & M-SpatPlus49 & & -0.1360 & 0.0411 & -0.1353 & 0.0439 & -0.1372 & 0.0443 \\
        \hline
        \multirow{5}[2]{*}{Scenario 6} & M-Spatial & \multirow{5}[2]{*}{-0.2000} & -0.3818 & 0.0590 & -0.4128 & 0.0401 &  -0.4005 & 0.0453 \\
        & M-SpatPlus64 & & -0.2253 & 0.0428 & -0.2259 & 0.0461 & -0.2285 & 0.0437 \\
        & M-SpatPlus59 & & -0.2089 & 0.0407 & -0.2082 & 0.0438 & -0.2150 & 0.0423 \\
        & M-SpatPlus54 & & -0.1655 & 0.0396 & -0.1650 & 0.0432 & -0.1677 & 0.0433 \\
        & M-SpatPlus49 & & -0.1394 & 0.0393 & -0.1382 & 0.0430 & -0.1394 & 0.0438 \\
        \hline
	\end{tabular}
\end{table}

\begin{figure}[h]
\centering
\includegraphics[width=1.1\textwidth]{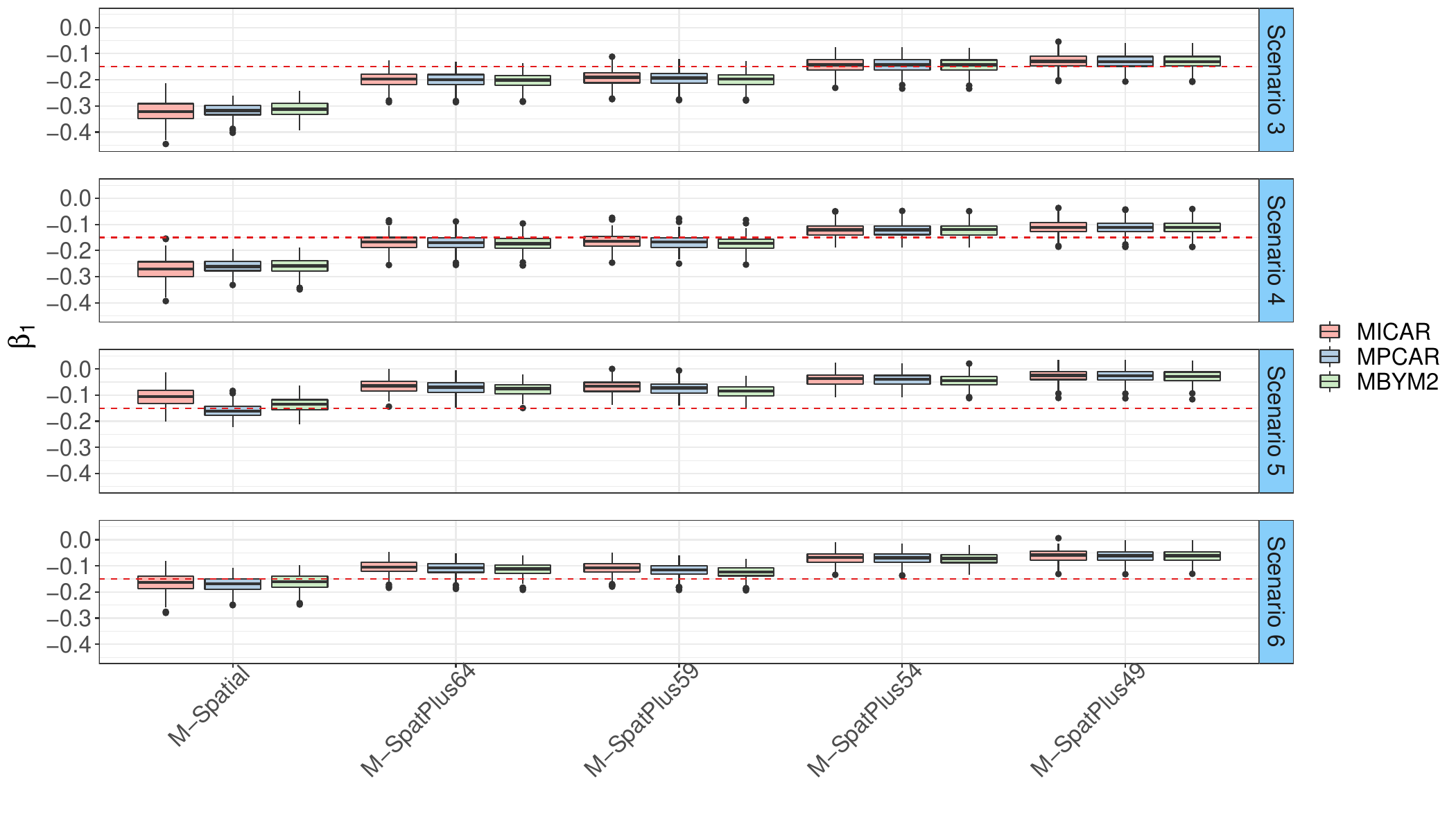}
\caption{Boxplots of the estimated means of $\beta_1$ based on 300 simulated datasets for Simulation Study 1, Scenarios 3, 4, 5 and 6 (ordered from top to bottom). Each color represents a different prior given to the columns of $\PPhi$, namely, red for ICAR, blue for PCAR and green for BYM2.}
\label{SimuStudy1_boxplots_rapes_sup}
\end{figure}

\begin{figure}[h]
\centering
\includegraphics[width=1.1\textwidth]{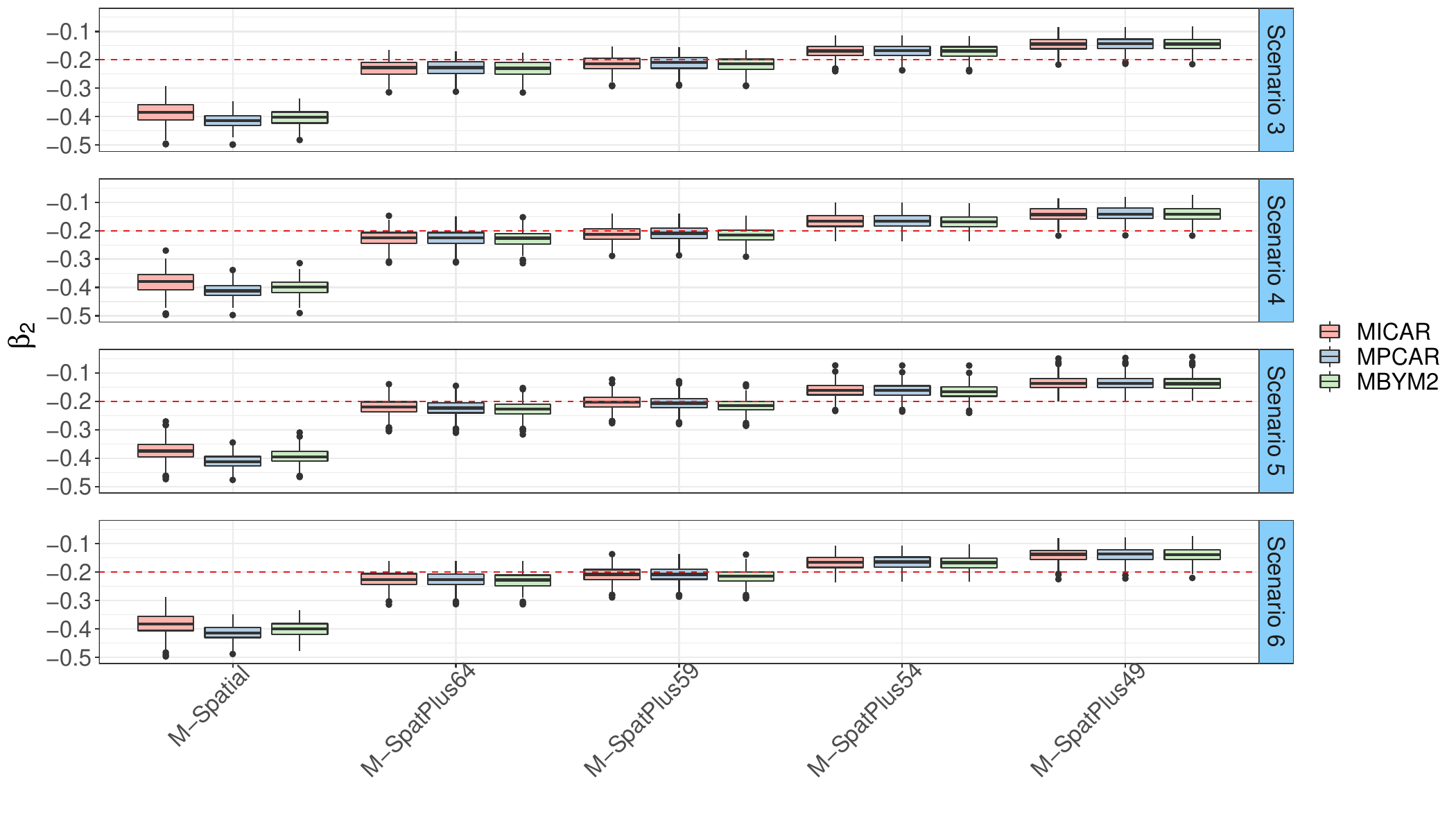}
\caption{Boxplots of the estimated means of $\beta_2$ based on 300 simulated datasets for Simulation Study 1, Scenarios 3, 4, 5 and 6 (ordered from top to bottom). Each color represents a different prior given to the columns of $\PPhi$, namely, red for ICAR, blue for PCAR and green for BYM2.}
\label{SimuStudy1_boxplots_dowry_sup}
\end{figure}

\begin{table}[htbp]
	\centering
	\caption{Estimated standard errors ($s.e._{est}$) and simulated standard errors ($s.e._{sim}$) for $\beta_{1}$ based on $300$ simulated datasets for Simulation study 1 and Scenarios 1 to 6.}
\begin{tabular}{lccccccc}
        \hline
		& & \multicolumn{2}{c}{ICAR} & \multicolumn{2}{c}{PCAR} & \multicolumn{2}{c}{BYM2} \\
		\hline
		& Model & $s.e._{est}$ & $s.e._{sim}$ & $s.e._{est}$ & $s.e._{sim}$ & $s.e._{est}$ & $s.e._{sim}$  \\
		\hline
		\multirow{5}[2]{*}{Scenario 1} &  M-Spatial & 0.0674 & 0.0390 & 0.0457 & 0.0263 & 0.0534 & 0.0289 \\
		& M-SpatPlus64 & 0.0446 & 0.0279 & 0.0429 & 0.0272 & 0.0433 & 0.0267 \\
		& M-SpatPlus59 & 0.0419 & 0.0267 & 0.0407 & 0.0261 & 0.0415 & 0.0255 \\
		& M-SpatPlus54 & 0.0407 & 0.0264 & 0.0402 & 0.0262 & 0.0420 & 0.0259 \\
		& M-SpatPlus49 & 0.0399 & 0.0264 & 0.0398 & 0.0262 & 0.0419 & 0.0262 \\
		\hline
        \multirow{5}[2]{*}{Scenario 2} & M-Spatial & 0.0733 & 0.0401 & 0.0477 & 0.0265 & 0.0579 & 0.0307 \\
\hline
	\end{tabular}%
	\label{tab:addlabel}%
\end{table}%

\begin{table}[htbp]
	\centering
	\caption{Estimated standard errors ($s.e._{est}$) and simulated standard errors ($s.e._{sim}$) for $\beta_{2}$ based on $300$ simulated datasets for Simulation study 1 and Scenarios 1 to 6.}
	\begin{tabular}{lccccccc}
        \hline
		& & \multicolumn{2}{c}{ICAR} & \multicolumn{2}{c}{PCAR} & \multicolumn{2}{c}{BYM2} \\
		\hline
		& Model & $s.e._{est}$ & $s.e._{sim}$ & $s.e._{est}$ & $s.e._{sim}$  & $s.e._{est}$ & $s.e._{sim}$  \\
		\hline
		\multirow{5}[2]{*}{Scenario 1} & M-Spatial & 0.0596 & 0.0357 & 0.0407 & 0.0243 & 0.0462 & 0.0262 \\
		& M-SpatPlus64 & 0.0429 & 0.0268 & 0.0454 & 0.0267 & 0.0438 & 0.0259 \\
		& M-SpatPlus59 & 0.0408 & 0.0257 & 0.0432 & 0.0256 & 0.0423 & 0.0248 \\
		& M-SpatPlus54 & 0.0396 & 0.0252 & 0.0421 & 0.0252 & 0.0424 & 0.0251 \\
		& M-SpatPlus49 & 0.0393 & 0.0242 & 0.0417 & 0.0243 & 0.0426 & 0.0242 \\
		\hline
        \multirow{5}[2]{*}{Scenario 2} & M-Spatial & 0.0690 & 0.0382 & 0.0443 & 0.0252 & 0.0508 & 0.0273 \\
\hline
	\end{tabular}%
	\label{tab:addlabel}%
\end{table}%

\begin{table}[htbp]
	\centering
	\caption{Empirical $95\%$ coverage probabilities of the true value of $\beta_{1}$ and $\beta_{2}$ based on $300$ simulated datasets for Simulation study 1 and Scenarios 3, 4, 5 and 6.}
	\begin{tabular}{lccccccc}
        \hline
		& & \multicolumn{3}{c}{$\beta_{1}$} & \multicolumn{3}{c}{$\beta_{2}$} \\
		\hline
		& Model & ICAR & PCAR & BYM2 & ICAR & PCAR & BYM2  \\
		\hline
		\multirow{5}[2]{*}{Scenario 3} & M-Spatial & 19.3333 & 0.0000 & 6.0000 & 5.0000 & 0.0000 & 0.0000 \\
		& M-SpatPlus64 & 90.6667 & 87.3333 & 87.0000 & 96.0000 & 98.0000 & 97.3333 \\
		& M-SpatPlus59 & 94.6667 & 90.0000 & 89.3333 & 98.6667 & 99.3333 & 99.0000 \\
		& M-SpatPlus54 & 99.6667 & 99.6667 & 100.0000 & 96.6667& 97.3333 & 99.0000 \\
		& M-SpatPlus49 & 98.3333 & 98.6667 & 99.0000 & 78.0000 & 82.3333 & 86.0000 \\
		\hline
        \multirow{5}[2]{*}{Scenario 4} & M-Spatial & 70.6667 & 26.3333 & 56.0000 & 6.0000 & 0.0000 & 0.0000 \\
		& M-SpatPlus64 & 99.6667 & 98.3333 & 98.3333 & 96.6667 & 97.6667 & 97.3333 \\
		& M-SpatPlus59 & 99.6667 & 98.6667 & 98.3333 & 99.6667 & 99.6667 & 99.6667 \\
		& M-SpatPlus54 & 98.0000 & 96.6667 & 97.6667 & 94.0000 & 95.3333 & 97.6667 \\
		& M-SpatPlus49 & 95.3333 & 95.6667 & 97.0000 & 75.0000 & 79.6667 & 84.6667 \\
        \hline
        \multirow{5}[2]{*}{Scenario 5} & M-Spatial & 100.0000 & 100.0000 & 100.0000 & 9.0000 & 0.0000 & 0.0000 \\
		& M-SpatPlus64 & 64.3333 & 72.6667 & 73.6667 & 98.0000 & 98.3333 & 97.3333 \\
		& M-SpatPlus59 & 61.0000 & 72.6667 & 81.6667 & 100.0000 & 100.0000 & 99.3333 \\
		& M-SpatPlus54 & 12.0000 & 13.6667 & 20.3333 & 94.3333 & 96.6667 & 98.0000 \\
		& M-SpatPlus49 & 2.0000 & 3.0000 & 6.0000 & 76.6667 & 79.6667 & 83.0000 \\
        \hline
        \multirow{5}[2]{*}{Scenario 6} & M-Spatial & 100.0000 & 99.3333 & 100.0000 & 6.3333 & 0.0000 & 0.0000 \\
		& M-SpatPlus64 & 96.0000 & 95.6667 & 98.0000 & 97.6667 & 97.6667 & 98.3333 \\
		& M-SpatPlus59 & 96.3333 & 96.6667 & 100.0000 & 99.0000 & 100.0000 & 98.6667 \\
		& M-SpatPlus54 & 49.6667 & 49.3333 & 60.0000 & 94.6667 & 97.0000 & 98.0000 \\
		& M-SpatPlus49 & 34.0000 & 31.6667 & 42.0000 & 75.3333 & 82.3333 & 85.6667 \\
        \hline
	\end{tabular}
	\label{tab:addlabel}%
\end{table}

\begin{table}[htbp]
	\centering
    \scriptsize
	\caption{Posterior medians and $95\%$ credible intervals of estimated correlations between crime 1 and crime 2 based on $300$ simulated datasets for Simulation study 1 and Scenarios 3, 4, 5 and 6.}
    \begin{tabular}{lcccccccccc}
          \hline
          &       & \multicolumn{3}{c}{ICAR} & \multicolumn{3}{c}{PCAR} & \multicolumn{3}{c}{BYM2} \\
          \hline
          &  Model     & Median & \multicolumn{2}{c}{$95\%$ CI} & Median & \multicolumn{2}{c}{$95\%$ CI} & Median & \multicolumn{2}{c}{$95\%$ CI} \\
          \hline
          \multirow{5}[2]{*}{Scenario 3} & M-Spatial & 0.2568 & -0.2346 & 0.6783 & 0.2359 & -0.1913 & 0.6168 & 0.3250 & -0.2218 & 0.7265 \\
          & M-SpatPlus64 & 0.3574 & -0.0573 & 0.7085 & 0.2969 & 0.0897 & 0.5364 & 0.7132 & 0.4062 & 0.8911 \\
          & M-SpatPlus59 & 0.3508 & -0.0646 & 0.7041 & 0.2907 & 0.0839 & 0.5267 & 0.7183 & 0.4157 & 0.8902 \\
          & M-SpatPlus54 & 0.4754 & 0.1048 & 0.7708 & 0.4119 & 0.1556 & 0.6710 & 0.8001 & 0.5474 & 0.9283 \\
          & M-SpatPlus49 & 0.5076 & 0.1522 & 0.7859 & 0.4444 & 0.1718 & 0.7044 & 0.8170 & 0.5824 & 0.9340 \\
          \hline
          \multirow{5}[2]{*}{Scenario 4} & M-Spatial & 0.0977 & -0.3660 & 0.5370 & 0.1147 & -0.3026 & 0.5084 & 0.2039 & -0.3156 & 0.6323 \\
\hline
    \end{tabular}%
  \label{tab:addlabel}%
\end{table}%

\begin{table}[ht]
	\centering
	\caption{DIC and WAIC based on 300 simulated data sets for Simulation Study 1 and Scenarios 3, 4, 5 and 6.}
	\begin{tabular}{lcccccccc}
		\hline
		&  & \multicolumn{2}{c}{ICAR} & \multicolumn{2}{c}{PCAR} & \multicolumn{2}{c}{BYM2} \\
        \hline
        & Model & DIC & WAIC & DIC & WAIC & DIC & WAIC \\
        \hline
        \multirow{5}[2]{*}{Scenario 3} & M-Spatial & 953.3899 & 959.0199 & 944.7410 & 940.6010 & 940.0554 & 933.2278 \\
        & M-SpatPlus64 & 951.5502 & 952.1564 & 945.9650 & 940.6912 & 942.7260 & 932.5063 \\
        & M-SpatPlus59 & 951.5871 & 952.2058 & 945.7270 & 940.4662 & 942.6509 & 932.4621 \\
        & M-SpatPlus54 & 953.3013 & 951.8469 & 947.7106 & 939.4825 & 946.1554 & 935.1135\\
        & M-SpatPlus49 & 954.0929 & 952.0263 & 948.5981 & 939.6678 & 947.0794 & 935.5887 \\
        \hline
        \multirow{5}[2]{*}{Scenario 4} & M-Spatial & 956.4190 & 959.9196 & 948.3138 & 942.5435 & 943.8799 & 935.4106 \\
\hline
	\end{tabular}
\end{table}

\begin{table}[htbp]
	\centering
	\caption{Average value of mean absolute relative bias (MARB) and mean relative root mean prediction error (MRRMSE) of the relative risks based on $300$ simulated datasets for Simulation study 1 and Scenarios 1 to 6.}
	\begin{tabular}{lccccccc}
        \hline
		& & \multicolumn{2}{c}{ICAR} & \multicolumn{2}{c}{PCAR} & \multicolumn{2}{c}{BYM2} \\
		\hline
		& Model & MARB & MRRMSE & MARB & MRRMSE & MARB & MRRMSE \\
		\hline
		\multirow{5}[2]{*}{Scenario 1} & M-Spatial & 0.0916 & 0.1550 & 0.0823 & 0.1507 & 0.0805 & 0.1480 \\
		& M-SpatPlus64 & 0.0865 & 0.1553 & 0.0804 & 0.1539 & 0.0754 & 0.1513 \\
		& M-SpatPlus59 & 0.0862 & 0.1554 & 0.0794 & 0.1535 & 0.0740 & 0.1509 \\
		& M-SpatPlus54 & 0.0834 & 0.1574 & 0.0766 & 0.1557 & 0.0736 & 0.1544 \\
		& M-SpatPlus49 & 0.0828 & 0.1585 & 0.0761 & 0.1566 & 0.0722 & 0.1554 \\
		\hline
        \multirow{5}[2]{*}{Scenario 2} & M-Spatial & 0.0927 & 0.1717 & 0.0847 & 0.1673 & 0.0831 & 0.1640 \\
\hline
	\end{tabular}%
	\label{tab:addlabel}%
\end{table}%

\begin{table}[htbp]
	\centering
	\caption{Length of the $95\%$ credible intervals of $\beta_1$ and $\beta_2$ based on $300$ simulated datasets for Simulation study 1 and Scenarios 1 to 6.}
	\begin{tabular}{lccccccc}
        \hline
        & & \multicolumn{3}{c}{$\beta_{1}$} & \multicolumn{3}{c}{$\beta_{2}$} \\
        \hline
		& Model & ICAR & PCAR & BYM2 & ICAR & PCAR & BYM2\\
		\hline
		\multirow{5}[2]{*}{Scenario 1} & M-Spatial & 0.2654 & 0.1801 & 0.2104 & 0.2347& 0.1604 & 0.1820 \\
		& M-SpatPlus64 & 0.1755 & 0.1687 & 0.1703 & 0.1689 & 0.1786 & 0.1722 \\
		& M-SpatPlus59 & 0.1647 & 0.1600 & 0.1631 & 0.1604 & 0.1699 & 0.1663 \\
		& M-SpatPlus54 & 0.1601 & 0.1581 & 0.1652 & 0.1557 & 0.1655 & 0.1668 \\
		& M-SpatPlus49 & 0.1569 & 0.1563 & 0.1648 & 0.1545 & 0.1640 & 0.1676 \\
		\hline
        \multirow{5}[2]{*}{Scenario 2} & M-Spatial & 0.2887 & 0.1879 & 0.2283 & 0.2719 & 0.1747 & 0.2002 \\
\hline
	\end{tabular}%
	\label{tab:addlabel}%
\end{table}%

\begin{table}[htbp]
	\centering
	\caption{Average value of mean absolute relative bias (MARB) and mean relative root mean prediction error (MRRMSE) of $\beta_1$ based on $300$ simulated datasets for Simulation study 1 and Scenarios 1 to 6.}
	\begin{tabular}{lccccccc}
        \hline
		& & \multicolumn{2}{c}{ICAR} & \multicolumn{2}{c}{PCAR} & \multicolumn{2}{c}{BYM2} \\
		\hline
		& Model & MARB & MRRMSE & MARB & MRRMSE & MARB & MRRMSE \\
        \hline
		\multirow{5}[2]{*}{Scenario 1} & M-Spatial & 0.9972 & 1.0305 & 1.1079 & 1.1217 & 1.0322 & 1.0500 \\
		& M-SpatPlus64 & 0.2459 & 0.3084 & 0.2630 & 0.3196 & 0.2946 & 0.3444 \\
		& M-SpatPlus59 & 0.2121 & 0.2769 & 0.2421 & 0.2983 & 0.2878 & 0.3343 \\
		& M-SpatPlus54 & 0.1311 & 0.2195 & 0.1232 & 0.2137 & 0.1120 & 0.2057 \\
		& M-SpatPlus49 & 0.2394 & 0.2971 & 0.2322 & 0.2907 & 0.2255 & 0.2854 \\
		\hline
        \multirow{5}[2]{*}{Scenario 2} & M-Spatial & 0.8046 & 0.8479 & 0.7463 & 0.7670 & 0.7264 & 0.7547 \\
\hline
	\end{tabular}%
	\label{tab:addlabel}%
\end{table}%

\begin{table}[htbp]
	\centering
	\caption{Average value of mean absolute relative bias (MARB) and mean relative root mean prediction error (MRRMSE) of $\beta_2$ based on $300$ simulated datasets for Simulation study 1 and Scenarios 1 to 6.}
	\begin{tabular}{lccccccc}
        \hline
		& & \multicolumn{2}{c}{ICAR} & \multicolumn{2}{c}{PCAR} & \multicolumn{2}{c}{BYM2} \\
		\hline
		& Model & MARB & MRRMSE & MARB & MRRMSE & MARB & MRRMSE \\
        \hline
		\multirow{5}[2]{*}{Scenario 1} & M-Spatial & 0.9140 & 0.9312 & 1.0698 & 1.0767 & 0.9982 & 1.0067 \\
		& M-SpatPlus64 & 0.1292 & 0.1862 & 0.1266 & 0.1839 & 0.1411 & 0.1917 \\
		& M-SpatPlus59 & 0.0471 & 0.1369 & 0.0373 & 0.1331 & 0.0717 & 0.1431 \\
		& M-SpatPlus54 & 0.1685 & 0.2105 & 0.1725 & 0.2137 & 0.1615 & 0.2046 \\
		& M-SpatPlus49 & 0.2939 & 0.3178 & 0.2999 & 0.3236 & 0.2950 & 0.3189 \\
		\hline
        \multirow{5}[2]{*}{Scenario 2} & M-Spatial & 0.5761 & 0.6070 & 0.7782 & 0.7883 & 0.7016 & 0.7147 \\
\hline
	\end{tabular}%
	\label{tab:addlabel}%
\end{table}%

%%---------------------------------------------------------------------------------------
%% SUPPLEMENTARY MATERIAL: SIMULATION STUDY 2
%%---------------------------------------------------------------------------------------
\clearpage
\section{Supplementary material: Simulation study 2}
This section includes some additional scenarios to complement the analysis of Simulation study 2. The data is generated following the generating models (13)-(14) with $\aalpha=(\alpha_1, \alpha_2)^{'}=(0.12, 0.03)^{'}$, $\bbeta=(\beta_1, \beta_2)^{'}=(0.15, 0.20)^{'}$ and  $\X^{*}_1$ is simulated such that its correlation with spatial effects of crime 1, $\ttheta_1$, and spatial effects of crime 2, $\ttheta_2$, is the desired one. Below we define the correlations between $\X^{*}_1$, $\ttheta_1$ and $\ttheta_2$ chosen in each scenario:
\begin{itemize}
\item \textbf{Scenario 3:} $\rho=0.5$, $cor(\X^{*}_1, \ttheta_1)=0.5$ and $cor(\X^{*}_1, \ttheta_2)=0.7$ are chosen. This is similar to Scenario 1 but choosing medium correlation between the crimes.
\item \textbf{Scenario 4:} $\rho=0.3$, $cor(\X^{*}_1, \ttheta_1)=0.3$ and $cor(\X^{*}_1, \ttheta_2)=0.7$ are chosen.
\item \textbf{Scenario 5:} $\rho=0.3$, $cor(\X^{*}_1, \ttheta_1)=0.0$ and $cor(\X^{*}_1, \ttheta_2)=0.7$ are chosen.
\end{itemize}

\begin{figure}[h]
\centering
\includegraphics[width=0.8\textwidth]{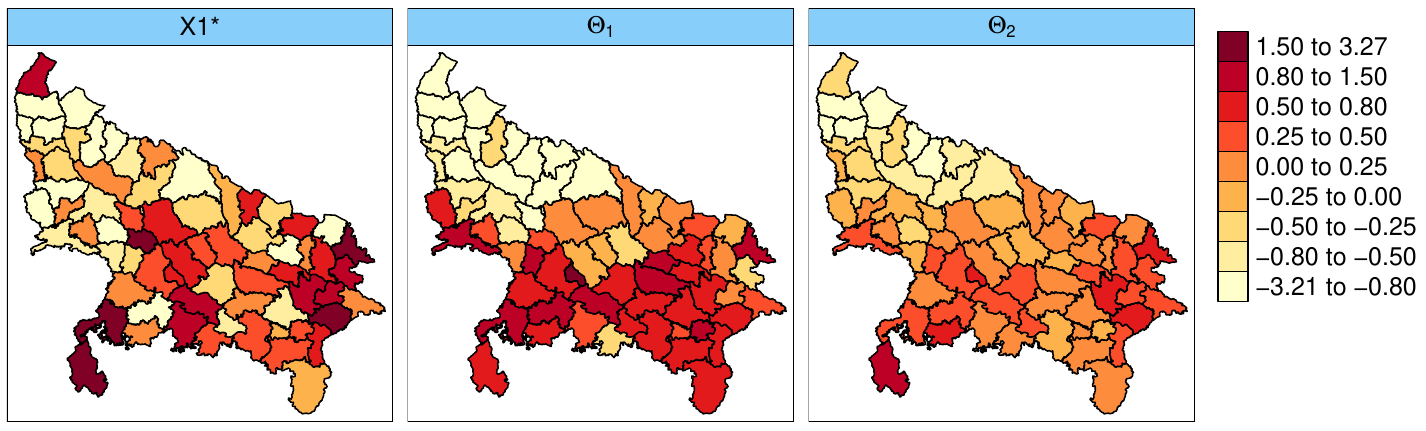}
\includegraphics[width=0.8\textwidth]{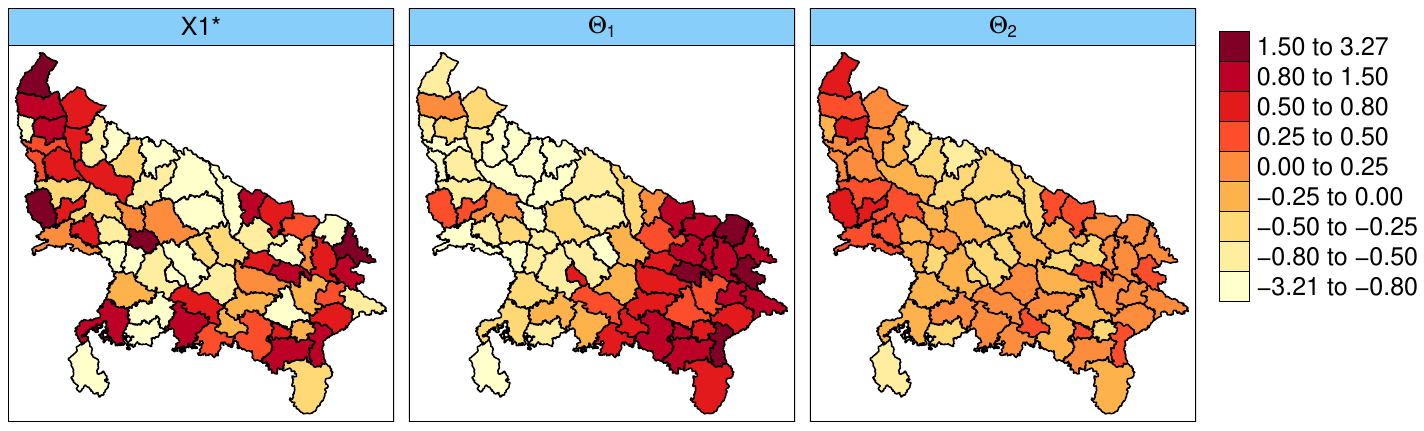}
\includegraphics[width=0.8\textwidth]{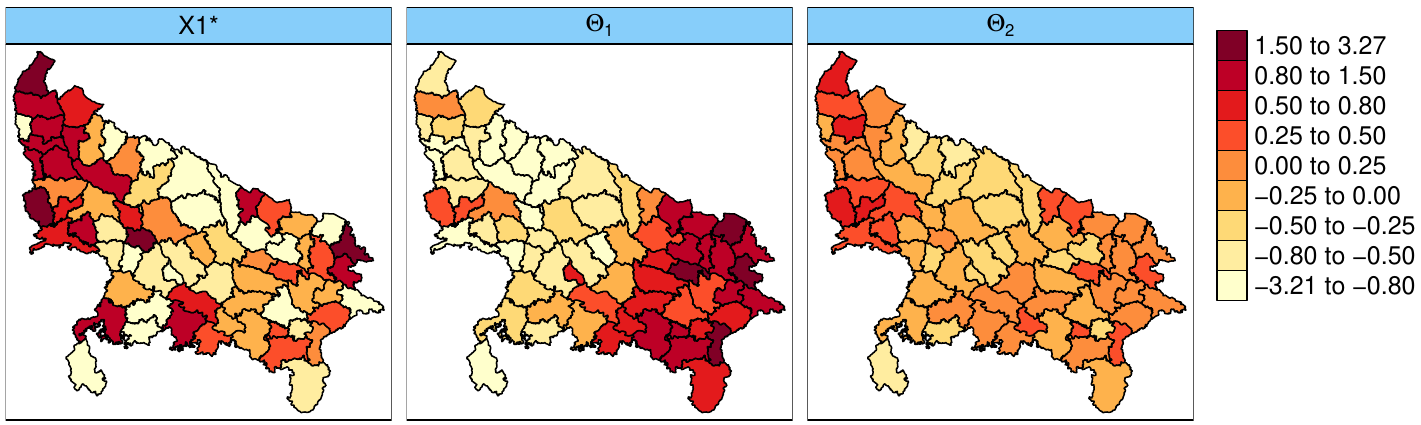}
\caption{From left to right, standardized simulated covariate $\X^{*}_1$, spatial effects $\ttheta_1$ for crime 1 and spatial effects $\ttheta_2$ for crime 2 in Simulation Study 2. The first row corresponds to Scenario 3, the second to Scenario 4 and the third to Scenario 5.}
\label{fig:SimuStudy1_sup}
\end{figure}

\begin{table}[htbp]
	\centering
    \scriptsize
	\caption{Posterior medians and $95\%$ credible intervals of estimated correlations between crime 1 and crime 2 based on $300$ simulated datasets for Simulation study 2 and Scenarios 3, 4 and 5.}
    \begin{tabular}{lcccccccccc}
          \hline
          &       & \multicolumn{3}{c}{ICAR} & \multicolumn{3}{c}{PCAR} & \multicolumn{3}{c}{BYM2} \\
          \hline
          & Model      & Median & \multicolumn{2}{c}{$95\%$ CI} & Median & \multicolumn{2}{c}{$95\%$ CI} & Median & \multicolumn{2}{c}{$95\%$ CI} \\
          \hline
          \multirow{5}[2]{*}{Scenario 3} & M-Spatial & 0.6203 & 0.3352 & 0.8260 & 0.6418 & 0.4166 & 0.8199 & 0.7990 & 0.5477 & 0.9377 \\
          & M-SpatPlus64 & 0.6126 & 0.3536 & 0.8014 & 0.6462 & 0.4428 & 0.8016 & 0.7140 & 0.4357 & 0.8897 \\
          & M-SpatPlus59 & 0.5834 & 0.3198 & 0.7797 & 0.6232 & 0.4125 & 0.7864 & 0.6801 & 0.3979 & 0.8674 \\
          & M-SpatPlus54 & 0.5828 & 0.3236 & 0.7758 & 0.6172 & 0.4091 & 0.7805 & 0.6834 & 0.3991 & 0.8716 \\
          & M-SpatPlus49 & 0.6073 & 0.3579 & 0.7920 & 0.6325 & 0.4258 & 0.7932 & 0.7046 & 0.4330 & 0.8810 \\
          & M-SpatPlus44 & 0.6092 & 0.3654 & 0.7900 & 0.6279 & 0.4240 & 0.7877 & 0.7168 & 0.4499 & 0.8896 \\
          & M-SpatPlus39 & 0.5939 & 0.3435 & 0.7810 & 0.6166 & 0.4060 & 0.7816 & 0.7076 & 0.4324 & 0.8870 \\
          \hline
          \multirow{5}[2]{*}{Scenario 4} & M-Spatial & 0.4241 & 0.1001 & 0.6840 & 0.2636 & -0.0474 & 0.5602 & 0.4675 & 0.1177 & 0.7309 \\
          & M-SpatPlus64 & 0.4637 & 0.1638 & 0.7025 & 0.3693 & 0.0808 & 0.6334 & 0.5192 & 0.2178 & 0.7557 \\
          & M-SpatPlus59 & 0.5060 & 0.2373 & 0.7170 & 0.4268 & 0.1553 & 0.6666 & 0.5622 & 0.2943 & 0.7680 \\
          & M-SpatPlus54 & 0.4995 & 0.2308 & 0.7115 & 0.4195 & 0.1455 & 0.6612 & 0.5566 & 0.2879 & 0.7644 \\
          & M-SpatPlus49 & 0.5190 & 0.2562 & 0.7248 & 0.4416 & 0.1662 & 0.6824 & 0.5738 & 0.3085 & 0.7777 \\
          & M-SpatPlus44 & 0.5437 & 0.2947 & 0.7387 & 0.4648 & 0.1944 & 0.6977 & 0.6016 & 0.3478 & 0.7931 \\
          & M-SpatPlus39 & 0.5535 & 0.3082 & 0.7448 & 0.4730 & 0.2012 & 0.7052 & 0.6144 & 0.3685 & 0.8000 \\
          \hline
          \multirow{5}[2]{*}{Scenario 5} & M-Spatial & 0.5283 & 0.2245 & 0.7629 & 0.3750 & 0.0832 & 0.6423 & 0.6028 & 0.2939 & 0.8212 \\
          & M-SpatPlus64 & 0.4963 & 0.2045 & 0.7218 & 0.4127 & 0.1107 & 0.6768 & 0.5293 & 0.2380 & 0.7546 \\
          & M-SpatPlus59 & 0.5199 & 0.2571 & 0.7243 & 0.4543 & 0.1797 & 0.6892 & 0.5594 & 0.2950 & 0.7638 \\
          & M-SpatPlus54 & 0.5011 & 0.2357 & 0.7091 & 0.4350 & 0.1562 & 0.6751 & 0.5404 & 0.2738 & 0.7488 \\
          & M-SpatPlus49 & 0.5049 & 0.2409 & 0.7125 & 0.4391 & 0.1613 & 0.6772 & 0.5447 & 0.2798 & 0.7527 \\
          & M-SpatPlus44 & 0.5287 & 0.2761 & 0.7275 & 0.4650 & 0.1943 & 0.6956 & 0.5704 & 0.3157 & 0.7660 \\
          & M-SpatPlus39 & 0.5217 & 0.2701 & 0.7214 & 0.4528 & 0.1773 & 0.6895 & 0.5742 & 0.3167 & 0.7730 \\
          \hline
    \end{tabular}%
  \label{tab:addlabel}%
\end{table}%

\begin{table}[ht]
	\centering
	\caption{Posterior means and standard deviations of $\beta_1$ based on 300 simulated datasets for Simulation study 2 and Scenarios 3, 4 and 5.}
	\begin{tabular}{lcccccccc}
		\hline
		&  & & \multicolumn{2}{c}{ICAR} & \multicolumn{2}{c}{PCAR} & \multicolumn{2}{c}{BYM2} \\
        \hline
        & Model & True value & Mean & SD & Mean & SD & Mean & SD  \\
		\hline
		\multirow{5}[2]{*}{Scenario 3} & M-Spatial & \multirow{5}[2]{*}{0.1500} & 0.2715 & 0.0807 & 0.2753 & 0.0813 & 0.2720 & 0.0822 \\
        & M-SpatPlus64 & & 0.1948 & 0.0657 & 0.1921 & 0.0653 & 0.1925 & 0.0670 \\
        & M-SpatPlus59 & & 0.1927 & 0.0626 & 0.1909 & 0.0624 & 0.1931 & 0.0641 \\
        & M-SpatPlus54 & & 0.1758 & 0.0595 & 0.1745 & 0.0593 & 0.1764 & 0.0611 \\
        & M-SpatPlus49 & & 0.1527 & 0.0576 & 0.1511 & 0.0575 & 0.1519 & 0.0595 \\
        & M-SpatPlus44 & & 0.1389 & 0.0554 & 0.1379 & 0.0554 & 0.1386 & 0.0578 \\
        & M-SpatPlus39 & & 0.1465 & 0.0538 & 0.1455 & 0.0539 & 0.1475 & 0.0563 \\
		\hline
        \multirow{5}[2]{*}{Scenario 4} & M-Spatial & \multirow{5}[2]{*}{0.1500} & 0.3082 & 0.0726 & 0.3111 & 0.0751 & 0.3149 & 0.0747 \\
        & M-SpatPlus64 & & 0.2722 & 0.0686 & 0.2725 & 0.0709 & 0.2743 & 0.0703 \\
        & M-SpatPlus59 & & 0.2213 & 0.0648 & 0.2193 & 0.0672 & 0.2193 & 0.0661 \\
        & M-SpatPlus54 & & 0.2205 & 0.0632 & 0.2190 & 0.0656 & 0.2194 & 0.0646 \\
        & M-SpatPlus49 & & 0.2018 & 0.0619 & 0.2004 & 0.0642 & 0.2005 & 0.0634 \\
        & M-SpatPlus44 & & 0.1758 & 0.0607 & 0.1743 & 0.0630 & 0.1743 & 0.0628 \\
        & M-SpatPlus39 & & 0.1536 & 0.0568 & 0.1527 & 0.0592 & 0.1529 & 0.0596 \\
        \hline
        \multirow{5}[2]{*}{Scenario 5} & M-Spatial & \multirow{5}[2]{*}{0.1500} & 0.2368 & 0.0776 & 0.2317 & 0.0799 & 0.2374 & 0.0792 \\
        & M-SpatPlus64 & & 0.2096 & 0.0703 & 0.2104 & 0.0724 & 0.2111 & 0.0718 \\
        & M-SpatPlus59 & & 0.1635 & 0.0660 & 0.1618 & 0.0682 & 0.1602 & 0.0674 \\
        & M-SpatPlus54 & & 0.1661 & 0.0640 &0.1648 & 0.0662 & 0.1639 & 0.0655 \\
        & M-SpatPlus49 & & 0.1572 & 0.0624 & 0.1559 & 0.0646 & 0.1555 & 0.0641 \\
        & M-SpatPlus44 & & 0.1312 & 0.0612 & 0.1300 & 0.0633 & 0.1288 & 0.0633 \\
        & M-SpatPlus39 & & 0.1202 & 0.0571 & 0.1194 & 0.0593 & 0.1188 & 0.0599 \\
        \hline
	\end{tabular}
\end{table}

\begin{table}[ht]
	\centering
	\caption{Posterior means and standard deviations of $\beta_2$ based on 300 simulated datasets for Simulation study 2 and Scenarios 3, 4 and 5.}
	\begin{tabular}{lcccccccc}
		\hline
		&  &  & \multicolumn{2}{c}{ICAR} & \multicolumn{2}{c}{PCAR} & \multicolumn{2}{c}{BYM2} \\
        \hline
        & Model & True value & Mean & SD & Mean & SD & Mean & SD  \\
		\hline
		\multirow{5}[2]{*}{Scenario 3} & M-Spatial & \multirow{5}[2]{*}{0.2000} & 0.3540 & 0.0423 & 0.3788 & 0.0446 & 0.3759 & 0.0442 \\
        & M-SpatPlus64 & & 0.2602 & 0.0381 & 0.2633 & 0.0425 & 0.2630 & 0.0412 \\
        & M-SpatPlus59 & & 0.2402 & 0.0375 & 0.2413 & 0.0415 & 0.2401 & 0.0405 \\
        & M-SpatPlus54 & & 0.2132 & 0.0368 & 0.2144 & 0.0407 & 0.2111 & 0.0399 \\
        & M-SpatPlus49 & & 0.1977 & 0.0357 & 0.1998 & 0.0394 & 0.1962 & 0.0389 \\
        & M-SpatPlus44 & & 0.1780 & 0.0359 & 0.1793 & 0.0392 & 0.1773 & 0.0397 \\
        & M-SpatPlus39 & & 0.1764 & 0.0350 & 0.1776 & 0.0384 & 0.1761 & 0.0389 \\
		\hline
        \multirow{5}[2]{*}{Scenario 4} & M-Spatial & \multirow{5}[2]{*}{0.2000} & 0.3474 & 0.0384 & 0.3578 & 0.0398 & 0.3648 & 0.0399 \\
        & M-SpatPlus64 & & 0.3157 & 0.0378 & 0.3177 & 0.0389 & 0.3259 & 0.0395 \\
        & M-SpatPlus59 & & 0.2603 & 0.0391 & 0.2612 & 0.0399 & 0.2623 & 0.0409 \\
        & M-SpatPlus54 & & 0.2479 & 0.0385 & 0.2486 & 0.0393 & 0.2491 & 0.0404 \\
        & M-SpatPlus49 & & 0.2333 & 0.0380 & 0.2333 & 0.0387 & 0.2344 & 0.0399 \\
        & M-SpatPlus44 & & 0.2049 & 0.0389 & 0.2051 & 0.0395 & 0.2066 & 0.0413 \\
        & M-SpatPlus39 & & 0.1620 & 0.0380 & 0.1623 & 0.0386 & 0.1635 & 0.0409 \\
        \hline
         \multirow{5}[2]{*}{Scenario 5} & M-Spatial & \multirow{5}[2]{*}{0.2000} & 0.3707 & 0.0392 & 0.3883 & 0.0404 & 0.3899 & 0.0405 \\
         & M-SpatPlus64 & & 0.3251 & 0.0379 & 0.3257 & 0.0388 & 0.3327 & 0.0393 \\
        & M-SpatPlus59 & & 0.2702 & 0.0394 & 0.2704 & 0.0402 & 0.2714 & 0.0411 \\
        & M-SpatPlus54 & & 0.2542 & 0.0387 & 0.2545 & 0.0394 & 0.2548 & 0.0404 \\
        & M-SpatPlus49 & & 0.2396 & 0.0382 & 0.2395 & 0.0389 & 0.2401 & 0.0400  \\
        & M-SpatPlus44 & & 0.2140 & 0.0390 & 0.2139 & 0.0397 & 0.2150 & 0.0412 \\
        & M-SpatPlus39 & & 0.1680 & 0.0384 & 0.1680 & 0.0390 & 0.1689 & 0.0411 \\
        \hline

	\end{tabular}
\end{table}

\begin{figure}[h]
\centering
\includegraphics[width=1.1\textwidth]{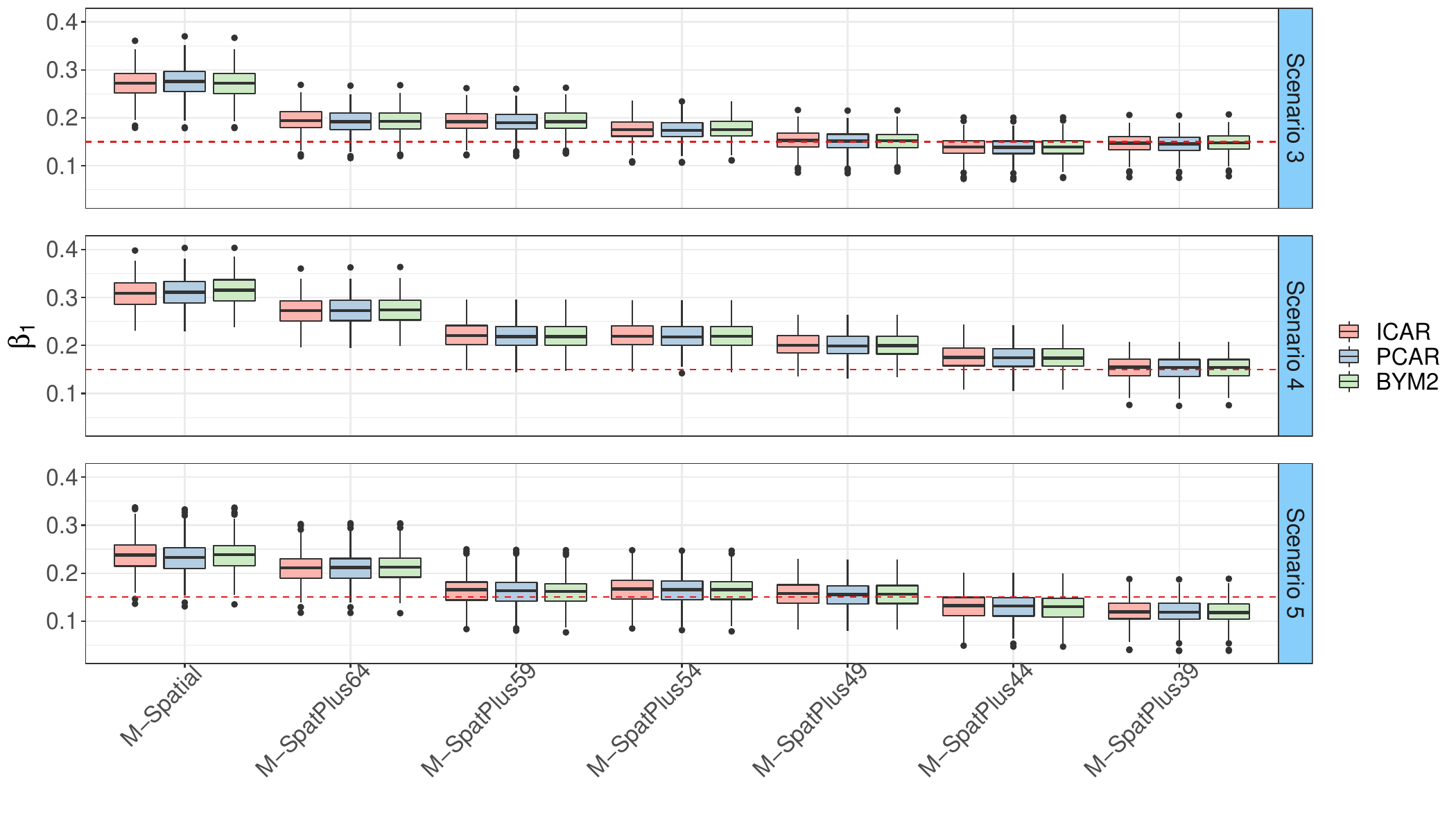}
\caption{Boxplots of the estimated means of $\beta_1$ based on 300 simulated datasets for Simulation Study 2, Scenarios 3, 4, 5 and 6 (ordered from top to bottom). Each color represents a different prior given to the columns of $\PPhi$, namely, red for ICAR, blue for PCAR and green for BYM2.}
\label{SimuStudy1_boxplots_rapes_sup}
\end{figure}

\begin{figure}[h]
\centering
\includegraphics[width=1.1\textwidth]{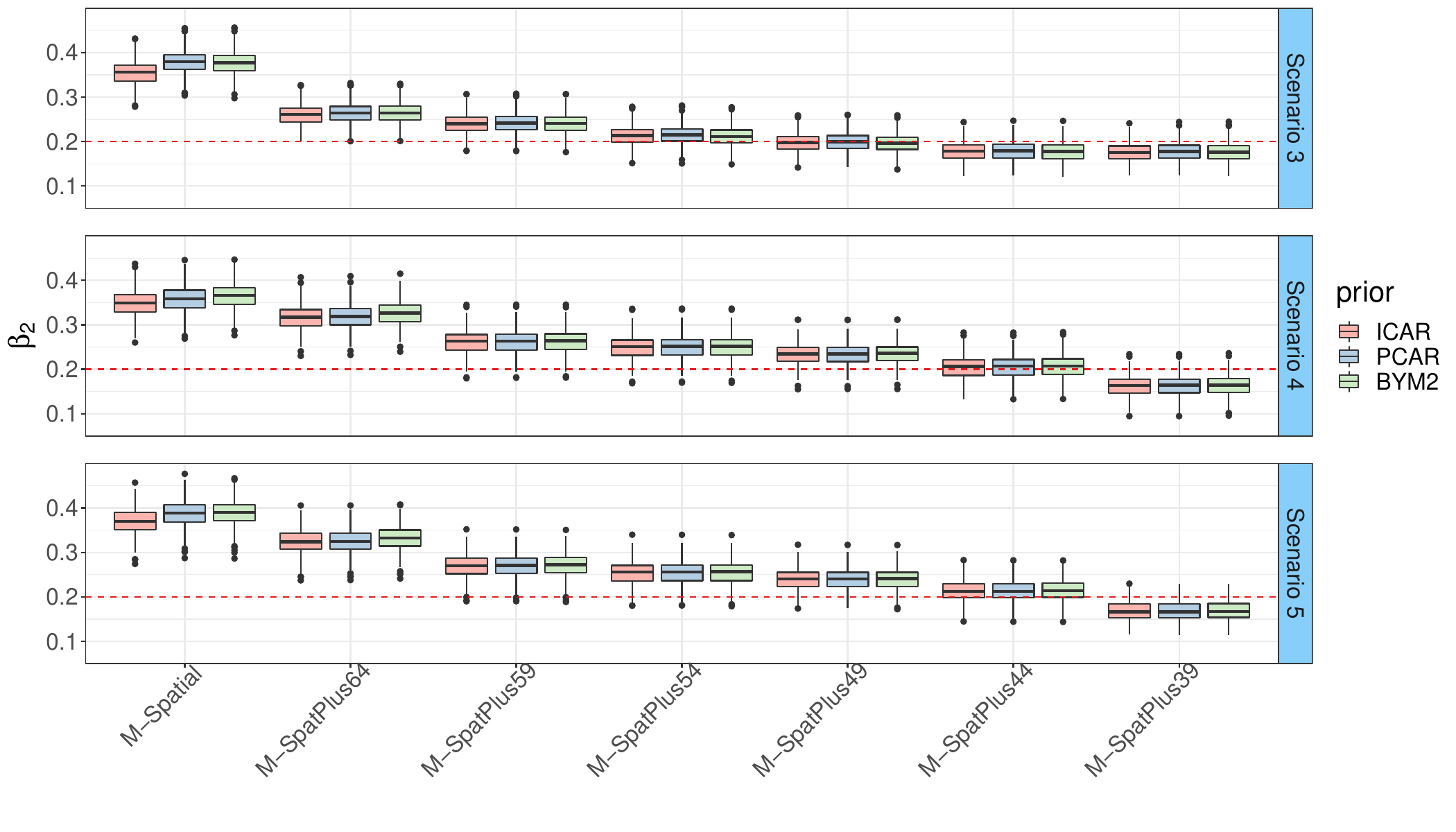}
\caption{Boxplots of the estimated means of $\beta_2$ based on 300 simulated datasets for Simulation Study 2, Scenarios 3, 4, 5 and 6 (ordered from top to bottom). Each color represents a different prior given to the columns of $\PPhi$, namely, red for ICAR, blue for PCAR and green for BYM2.}
\label{SimuStudy1_boxplots_dowry_sup}
\end{figure}

\begin{table}[htbp]
	\centering
	\caption{Estimated standard errors ($s.e._{est}$) and simulated standard errors ($s.e._{sim}$) for $\beta_{1}$ based on $300$ simulated datasets for Simulation study 2 and Scenarios 1 to 5.}
\begin{tabular}{lccccccc}
        \hline
		& & \multicolumn{2}{c}{ICAR} & \multicolumn{2}{c}{PCAR} & \multicolumn{2}{c}{BYM2} \\
		\hline
		& Model & $s.e._{est}$ & $s.e._{sim}$ & $s.e._{est}$ & $s.e._{sim}$ & $s.e._{est}$ & $s.e._{sim}$  \\
		\hline
		\multirow{5}[2]{*}{Scenario 1} &  M-Spatial & 0.0815 & 0.0338 & 0.0519 & 0.0290 & 0.0846 & 0.0338 \\
        & M-SpatPlus64 & 0.0612 & 0.0254 & 0.0404 & 0.0214 & 0.0630 & 0.0253 \\
        & M-SpatPlus59 & 0.0585 & 0.0249 & 0.0393 & 0.0206 & 0.0605 & 0.0250 \\
        & M-SpatPlus54 & 0.0580 & 0.0242 & 0.0390 & 0.0199 & 0.0599 & 0.0242 \\
        & M-SpatPlus49 & 0.0563 & 0.0225 & 0.0378 & 0.0191 & 0.0584 & 0.0225 \\
        & M-SpatPlus44 & 0.0547 & 0.0228 & 0.0555 & 0.0228 & 0.0574 & 0.0230 \\
\hline
	\end{tabular}%
	\label{tab:addlabel}%
\end{table}%

\begin{table}[htbp]
	\centering
	\caption{Estimated standard errors ($s.e._{est}$) and simulated standard errors ($s.e._{sim}$) for $\beta_{2}$ based on $300$ simulated datasets for Simulation study 2 and Scenarios 1 to 5.}
	\begin{tabular}{lccccccc}
		\hline
		& & \multicolumn{2}{c}{ICAR} & \multicolumn{2}{c}{PCAR} & \multicolumn{2}{c}{BYM2} \\
		\hline
		& Model & $s.e._{est}$ & $s.e._{sim}$ & $s.e._{est}$ & $s.e._{sim}$  & $s.e._{est}$ & $s.e._{sim}$  \\
		\hline
		\multirow{5}[2]{*}{Scenario 1} & M-Spatial & 0.0476 & 0.0272 & 0.0519 & 0.0290 & 0.0501 & 0.0275 \\
        & M-SpatPlus64 & 0.0380 & 0.0210& 0.0404 & 0.0214 & 0.0398 & 0.0211 \\
        & M-SpatPlus59 & 0.0370 & 0.0203 & 0.0393 & 0.0206 & 0.0388 & 0.0203 \\
        & M-SpatPlus54 & 0.0367 & 0.0196 & 0.0390 & 0.0199 & 0.0385 & 0.0197 \\
        & M-SpatPlus49 & 0.0355 & 0.0189 & 0.0378 & 0.0191 & 0.0374 & 0.0189 \\
        & M-SpatPlus44 & 0.0353 & 0.0199 & 0.0375 & 0.0201 & 0.0375 & 0.0201\\
\hline
	\end{tabular}%
	\label{tab:addlabel}%
\end{table}%

\begin{table}[htbp]
	\centering
	\caption{Empirical $95\%$ coverage probabilities of the true value of $\beta_{1}$ and $\beta_{2}$ based on $300$ simulated datasets for Simulation study 2 and Scenarios 3, 4 and 5.}
	\begin{tabular}{lccccccc}
        \hline
		& & \multicolumn{3}{c}{$\beta_{1}$} & \multicolumn{3}{c}{$\beta_{2}$} \\
		\hline
		& & ICAR & PCAR & BYM2 & ICAR & PCAR & BYM2  \\
		\hline
		\multirow{5}[2]{*}{Scenario 3} & M-Spatial & 88.3333 & 85.3333 & 89.3333 & 1.3333 & 0.0000 & 0.0000 \\
        & M-SpatPlus64 & 100.0000 & 100.0000 & 100.0000 & 73.6667 & 80.6667 & 78.0000 \\
        & M-SpatPlus59 & 100.0000 & 100.0000 & 100.0000 & 92.6667 & 95.6667 & 95.6667 \\
        & M-SpatPlus54 & 100.0000 & 100.0000 & 100.0000 & 99.0000 & 99.6667 & 99.6667 \\
        & M-SpatPlus49 & 100.0000 & 100.0000 & 100.0000 & 100.0000 & 100.0000 & 100.0000 \\
        & M-SpatPlus44 & 100.0000 & 100.0000 & 100.0000 & 98.6667 & 99.6667 & 99.6667 \\
\hline
	\end{tabular}
	\label{tab:addlabel}%
\end{table}

\begin{table}[ht]
	\centering
	\caption{DIC and WAIC based on 300 simulated data sets for Simulation Study 2 and Scenarios 3, 4 and 5.}
	\begin{tabular}{lcccccccc}
		\hline
		&  & \multicolumn{2}{c}{ICAR} & \multicolumn{2}{c}{PCAR} & \multicolumn{2}{c}{BYM2} \\
        \hline
        & Model & DIC & WAIC & DIC & WAIC & DIC & WAIC \\
        \hline
        \multirow{5}[2]{*}{Scenario 3} & M-Spatial & 967.9454 & 954.5795 & 967.5495 & 952.1199 & 967.9909 & 951.7634 \\
        & M-SpatPlus64 & 969.1749 & 951.7115 & 970.9422 & 951.6440 & 970.9431 & 950.8905 \\
        & M-SpatPlus59 & 970.2056 & 951.4510 & 972.0560 & 951.7777 & 972.2680 & 951.2902 \\
        & M-SpatPlus54 & 972.2309 & 953.1453 & 973.6655 & 952.8703 & 974.4395 & 953.3604 \\
        & M-SpatPlus49 & 973.2094 & 954.5434 & 974.3598 & 953.7926 & 975.3738 & 954.5668 \\
        & M-SpatPlus44 & 974.3461 & 955.1660 & 975.2471 & 954.0045 & 976.5248 & 955.1149  \\
\hline
	\end{tabular}
\end{table}

\begin{table}[htbp]
	\centering
	\caption{Average value of mean absolute relative bias (MARB) and mean relative root mean prediction error (MRRMSE) of the relative risks based on $300$ simulated datasets for Simulation study 2 and Scenarios 1 to 5.}
	\begin{tabular}{lccccccc}
        \hline
		& & \multicolumn{2}{c}{ICAR} & \multicolumn{2}{c}{PCAR} & \multicolumn{2}{c}{BYM2} \\
		\hline
		& Model & MARB & MRRMSE & MARB & MRRMSE & MARB & MRRMSE \\
		\hline
		\multirow{5}[2]{*}{Scenario 1} & M-Spatial & 0.0675 & 0.1550 & 0.0649 & 0.1567 & 0.0639 & 0.1557  \\
        & M-SpatPlus64 & 0.0652 & 0.1567 & 0.0632 & 0.1585 & 0.0626 & 0.1579 \\
        & M-SpatPlus59 & 0.0660 & 0.1579 & 0.0643 & 0.1597 & 0.0642 & 0.1592 \\
        & M-SpatPlus54 & 0.0657 & 0.1583 & 0.0638 & 0.1601 & 0.0639 & 0.1598 \\
        & M-SpatPlus49 & 0.0694 & 0.1608 & 0.0671 & 0.1623 & 0.0676 & 0.1622 \\
        & M-SpatPlus44 & 0.0695 & 0.1612 & 0.0671 & 0.1626 & 0.0677 & 0.1626 \\
\hline
	\end{tabular}%
	\label{tab:addlabel}%
\end{table}%

\begin{table}[htbp]
	\centering
	\caption{Length of the $95\%$ credible intervals of $\beta_1$ and $\beta_2$ based on $300$ simulated datasets for Simulation study 2 and Scenarios 1 to 5.}
	\begin{tabular}{lccccccc}
        \hline
        & & \multicolumn{3}{c}{$\beta_{1}$} & \multicolumn{3}{c}{$\beta_{2}$} \\
        \hline
		& Model & ICAR & PCAR & BYM2 & ICAR & PCAR & BYM2\\
		\hline
		\multirow{5}[2]{*}{Scenario 1} & M-Spatial & 0.3208 & 0.3293 & 0.3328 & 0.1871 & 0.2039 & 0.1972 \\
        & M-SpatPlus64 & 0.2407 & 0.2435 & 0.2478 & 0.1495 & 0.1587 & 0.1565 \\
        & M-SpatPlus59 & 0.2304 & 0.2331 & 0.2382 & 0.1455 & 0.1546 & 0.1527 \\
        & M-SpatPlus54 & 0.2280 & 0.2309 & 0.2358 & 0.1443 & 0.1534 & 0.1516 \\
        & M-SpatPlus49 & 0.2215 & 0.2245 & 0.2298 & 0.1395 & 0.1485 & 0.1469 \\
        & M-SpatPlus44 & 0.2154 & 0.2184 & 0.2259 & 0.1389 & 0.1474 & 0.1476 \\
\hline
	\end{tabular}%
	\label{tab:addlabel}%
\end{table}%

\begin{table}[htbp]
	\centering
	\caption{Average value of mean absolute relative bias (MARB) and mean relative root mean prediction error (MRRMSE) of $\beta_1$ based on $300$ simulated datasets for Simulation study 2 and Scenarios 1 to 5.}
	\begin{tabular}{lccccccc}
        \hline
		& & \multicolumn{2}{c}{ICAR} & \multicolumn{2}{c}{PCAR} & \multicolumn{2}{c}{BYM2} \\
		\hline
		& Model & MARB & MRRMSE & MARB & MRRMSE & MARB & MRRMSE \\
        \hline
		\multirow{5}[2]{*}{Scenario 1} & M-Spatial & 0.8006 & 0.8317 & 0.8439 & 0.8738 & 0.8399 & 0.8696 \\
        & M-SpatPlus64 & 0.2523 & 0.3038 & 0.2473 & 0.2997 & 0.2635 & 0.3130 \\
        & M-SpatPlus59 & 0.2762 & 0.3223& 0.2766 & 0.3227 & 0.3076 & 0.3500 \\
        & M-SpatPlus54 & 0.1391 & 0.2129& 0.1365 & 0.2114 & 0.1541 & 0.2233 \\
        & M-SpatPlus49 & 0.0534 & 0.1595 & 0.0527 & 0.1594 & 0.0663 & 0.1642 \\
        & M-SpatPlus44 & 0.0632 & 0.1645 & 0.0618 & 0.1642 & 0.0499 & 0.1611 \\
\hline

	\end{tabular}%
	\label{tab:addlabel}%
\end{table}%

\begin{table}[htbp]
	\centering
	\caption{Average value of mean absolute relative bias (MARB) and mean relative root mean prediction error (MRRMSE) of $\beta_2$ based on $300$ simulated datasets for Simulation study 2 and Scenarios 1 to 5.}
	\begin{tabular}{lccccccc}
        \hline
		& & \multicolumn{2}{c}{ICAR} & \multicolumn{2}{c}{PCAR} & \multicolumn{2}{c}{BYM2} \\
		\hline
		& Model & MARB & MRRMSE & MARB & MRRMSE & MARB & MRRMSE \\
        \hline
		\multirow{5}[2]{*}{Scenario 1} & M-Spatial & 0.5088 & 0.5267 & 0.6421 & 0.6583 & 0.5893 & 0.6051 \\
        & M-SpatPlus64 & 0.0274 & 0.1087 & 0.0255 & 0.1098 & 0.0345 & 0.1109 \\
        & M-SpatPlus59 & 0.0309 & 0.1061 & 0.0327 & 0.1081 & 0.0263 & 0.1050 \\
        & M-SpatPlus54 & 0.0854 & 0.1301 & 0.0856 & 0.1314 & 0.0850 & 0.1300 \\
        & M-SpatPlus49 & 0.1297 & 0.1604 & 0.1303 & 0.1616 & 0.1277 & 0.1587 \\
        & M-SpatPlus44 & 0.1976 & 0.2213 & 0.2003 & 0.2241 & 0.1936 & 0.2181 \\
\hline
	\end{tabular}%
	\label{tab:addlabel}%
\end{table}%

\end{document}